\documentclass{jfm}

\usepackage{graphicx}

\usepackage{subfig}
\usepackage{amsmath}
\usepackage{epstopdf}
\usepackage{multicol}
\usepackage{amssymb}
\usepackage{textcomp}
\usepackage{color}

 \newcommand\nb{\boldsymbol{n}}
 \newcommand\xb{\boldsymbol{x}}
 \newcommand\ub{\boldsymbol{u}}
 
 \newcommand\ubinf{\boldsymbol{u}_{\infty}}

 \newcommand\psiinf{\psi_{\infty}}
 \newcommand\umax{U_{\text{max}}}

 \newcommand\omegamax{\omega_\text{max}}
 
 \newcommand\grad{\nabla}
 \newcommand\lap{\Delta}
 \newcommand\xmin{x_{\text{min}}}
 \newcommand\xmax{x_{\text{max}}}
 \newcommand\ymax{y_{\text{max}}}
 \newcommand\ymin{y_{\text{min}}}
 \newcommand\zo{\textbf{x}_o}
 \newcommand\zbd{\mathbf{x}_{bd}}
 \newcommand\zb{\mathbf{x}}
 \newcommand\vs{\textit{vs.\ }}
 \newcommand\xhat{\widehat{x}}
 \newcommand\xhhat{\widehat{\widehat{x}}}
 \newcommand\yhat{\widehat{y}}
 \newcommand\yhhat{\widehat{\yhat}}
 \newcommand\that{\widehat{t}}
 \newcommand\thhat{\widehat{\that}}
 \newcommand\gamhat{\widehat{\Gamma}}
 \newcommand\gamhhat{\widehat{\gamhat}}
 \newcommand\fhat{\widehat{F_D}}
 \newcommand\fhhat{\widehat{\fhat}}
 \newcommand\flhat{\widehat{F_L}}
 \newcommand\flhhat{\widehat{\flhat}}
 \newcommand\what{\widehat{\omega}}
 \newcommand\whhat{\widehat{\what}}
 \newcommand\delhat{\widehat{\delta}}
 \newcommand\delhhat{\widehat{\delhat}}
 \newcommand\xchat{\widehat{x_c}}
 \newcommand\xchhat{\widehat{\xchat}}
 \newcommand\FD{F_{\tiny D}}

\title[]{Numerical study of viscous starting flow past a flat plate}
  \author[L. Xu and M. Nitsche]%
  {L\ls I\ls N\ls G\ns X\ls U$^1$%
\ns
  \and M\ls O\ls N\ls I\ls K\ls A\ns N\ls I\ls T\ls S\ls C\ls H\ls E$^2$
}
\affiliation{$^{1}$ Department of Mathematics and
Statistics, Georgia State University, Atlanta, GA 30303,USA\\
$^{2}$Department of Mathematics and Statistics, University of New Mexico,
Albuquerque, NM 87131, USA}

\begin{document}
\maketitle

\abstract{Viscous flow past a finite flat plate which is impulsively started 
in direction normal to itself is studied numerically
using a high order mixed finite difference and semi-Lagrangian scheme.
The goal is to resolve details of the vorticity generation at early 
times, and to determine the effect of viscosity on flow quantities
such as the core trajectory and vorticity, and the shed circulation.
Vorticity contours, streaklines and streamlines 
are presented for a range of Reynolds numbers 
$Re\in[250,2000]$ and a range of times $t\in[0.0002,5]$.
At early times, most of
the vorticity is attached to the plate.  
The paper proposes a definition for the shed circulation at
early as well as late times, and shows that 
it indeed represents vorticity
that separates from the plate without reattaching.
The contribution of viscous diffusion to the circulation shedding rate 
is found to be significant, but, 
interestingly, to depend only slightly on the value of the Reynolds number.
The shed circulation and the vortex core trajectories
follow scaling laws for inviscid self-similar flow
over several decades in time. 
Scaling laws describing the 
core vorticity, core dissipation, 
boundary layer thickness, drag and lift forces in time and Reynolds number
are also presented.
The simulations provide benchmark results 
to evaluate, for example, simpler separation models such as
point vortex and vortex sheet models.

\section{Introduction}


Vorticity separation in flow
past sharp edges is a fundamental process of
intrinsic interest in fluid dynamics. 
The boundary layer vorticity is convected around the edge, 
where it concentrates and forms a vortex. The vortex grows
in strength and size, eventually causing the boundary layers to separate
as a shear layer that rolls up in a spiral shape around the 
vortex core.  The starting vortex flow has been 
the focus of many experimental, analytical and numerical studies,
beginning with the work of Prandtl (see Lugt 1995). 
%
%
This paper concerns flow past a finite flat plate of zero thickness
which is impulsively started in direction normal to itself. 
Closely related laboratory experiments 
include the works of \cite{pierce61}, 
\cite{tanhon71}, 
\cite{pullinperry80}, 
\cite{lianhuang89}, 
and \cite{lepagelewekeverga05}. 
They visualize the rolled-up layer and yield data on 
the vortex size, core trajectory, core vorticity distribution,
and the onset of an instability along the outer spiral turns.
%
Related numerical results include
the simulations of 
\cite{wang00} and \cite{eldredge07}  
for viscous flow past thin rounded plates, 
and those of 
\cite{hudsondennis85} and
\cite{dennisetal93}, followed by
\cite{koushiels96},
and \cite{luchini02}, 
for flow past plates of zero thickness.
The first three of these consider finite plates, while
%
\cite{luchini02} 
compute flow past a semi-infinite plate. 
These works report vortex fields, vortex core trajectories and induced forces 
at intermediate to relatively large times.

The main goal of this paper is to complement these earlier
works with numerical results that yield new information
about the flow, in particular on quantities 
that may be more difficult to measure experimentally.
%
%
%
Our focus is to resolve the flow over several decades in time for 
a range of Reynolds numbers, show details of the vorticity 
generation and study the effect of viscosity on various flow quantities.
Specific quantities of interest include the vortex trajectory,
the forces induced by the wall vorticity, and the shed circulation.
Computing shed circulation 
requires defining the region of entrainment of the starting 
vortex, which is not clearly apparent in the early formation stages.
Once this region is determined, we investigate convective and
diffusive contributions to the circulation shedding rate, and 
thereby obtain detailed insight into how viscosity affects 
circulation shedding. 
The computations also yield several scaling laws that show the dependence
on time and Reynolds number for the corresponding quantities.
While the results pertain to an idealized flow past a plate of
zero thickness,
they can be used as a basis of comparison to evaluate
widely used low order models for separated flows, 
such as point vortex models
(eg., Cortelezzi and Leonard 1993, Michelin and Llewellyn Smith 2009,
Eldredge and Wang 2010,
Ysasi et al 2011),
or
vortex sheet models (eg., Krasny 1991, 
Nitsche \& Krasny 1994,
Jones 2003, Jones and Shelley 2005, Alben and Shelley 2008, 
Shukla and Eldredge 2007), which are all
based on simple approximations for the circulation shedding rate. 
The results also provide a basis of comparison 
to determine, for example, the effect of finite plate thickness,
the shape of the plate tip 
(Schneider {\it et al} 2014),
or the wedge angle in flow past wedges 
(Pullin \& Perry 1980).

Impulsively started flow past a plate of zero thickness is difficult to 
compute for several reasons. The fluid velocity and shed vorticity are
initially unbounded, requiring a fine mesh and small timesteps.
Velocity and vorticity gradients near the wall are large, causing 
numerical instabilities. 
Here, we use 
a split method in time in which advection is treated using a semi-Lagrangian
scheme, diffusion is treated with a 3-level Crank-Nicholson method,
and all finite difference and interpolation approximations are of 4th order.
The method uses ideas from several previous works, including
\cite{eliu96a},
\cite{luchini02},
\cite{staniforth91},
\cite{seaid02},
\cite{johnston99},
\cite{nitschetal03}.
The method is 
of 2nd order for the present highly singular flow.

The simulations yield well resolved vorticity evolution profiles over
a large range of times. 
%
Results are first presented for fixed Reynolds number $Re=500$, 
including details of the vorticity near the boundary. 
Following results using 
$Re\in [250,2000]$ show the dependence on $Re$. 
Based on the computed profiles, we define 
the separated vorticity at early times, when it is not clearly
differentiated from boundary vorticity, and use this definition
to compute shed circulation, as well as convective and diffusive
components of the vorticity flux into the starting vortex. 
The results  show that the chosen vortex boundary indeed 
bounds separated vorticity from vorticity that remains attached.
They also show that viscous diffusion contributes significantly 
to the circulation shedding rate, but its contribution 
depends surprisingly little on the value of the Reynolds number.
The shed circulation and the vortex core trajectory are found to follow 
inviscid scaling laws (Kaden 1931, Pullin 1978) over several decades in time. 
The computed trajectory is also in good agreement
with experimental results of \cite{pullinperry80}.
The core vorticity and dissipation, and the induced drag and
lift forces, are found to follow scaling laws 
that define their dependence on time and on $Re$.
%


The paper is organized as follows.
Section \ref{sec:problem} describes the problem of interest and the
governing equations.
Section \ref{sec:method} 
presents
the numerical method, its accuracy, and the
resolution obtained.
Section \ref{sec:results} presents the numerical 
results, including the evolution in time for fixed $Re$,
the dependence at a fixed time on $Re$, the core trajectory and vorticity,
the circulation and circulation shedding rates, and the induced drag and lift forces,
in that order.
The results are summarized in section \ref{sec:conc}.

\section{Problem Formulation}\label{sec:problem}

\begin{figure}
\centering
\includegraphics[height=0.38\textwidth]{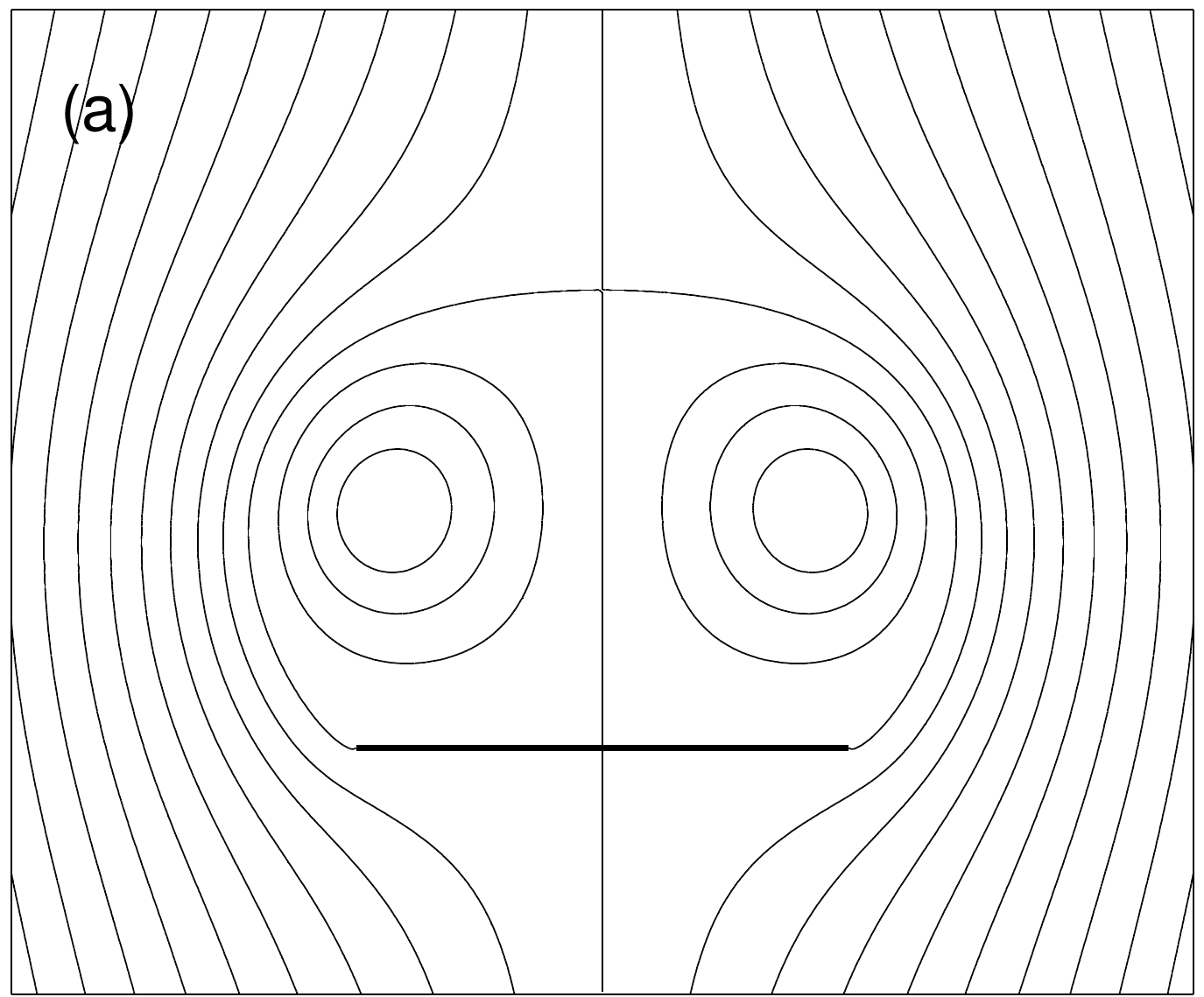}
\includegraphics[height=0.38\textwidth]{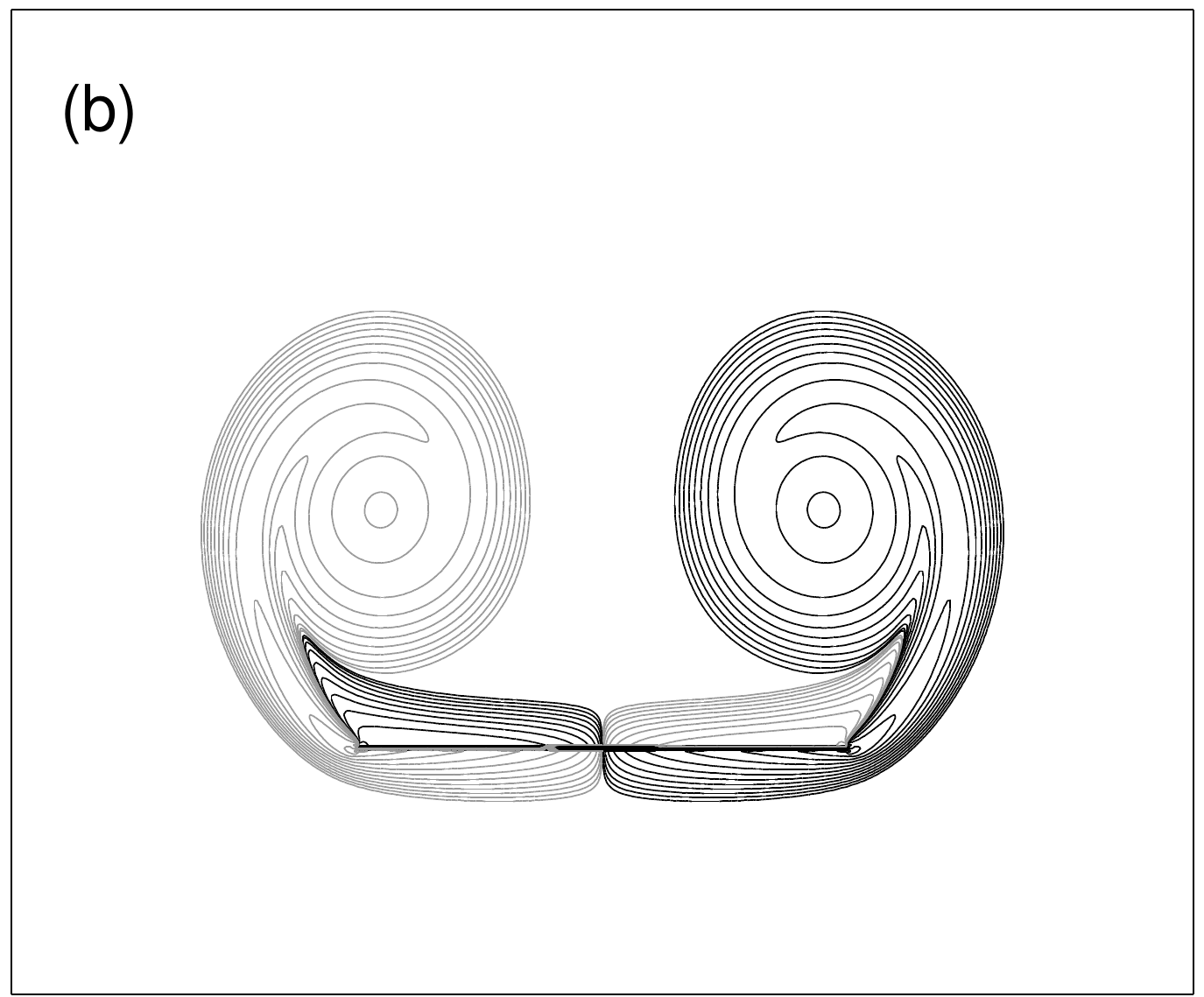}
 \caption{Sample solution at a relatively large time. 
(a) Streamlines.  (b) Vorticity field.
Positive vorticity contours are shown in black, negative ones in a lighter shade of grey.}
\label{F:sample}
\end{figure}

\subsection{Problem description}
A finite plate of length $L$ and zero thickness is immersed 
in viscous fluid and impulsively started to move from zero velocity to
a constant velocity $U>0$ in direction normal to itself.
The flow is nondimensionalized
using the plate length $L$ as the characteristic length scale,
and $U$ as the characteristic velocity.
An alternative nondimensionalization, appropriate in the absence of a length scale,
or at very early times,
is given by using $\nu/U$ instead of $L$ as the characteristic length scale,
where $\nu$ is the kinematic fluid viscosity. This alternative is discussed in 
the Appendix.

The flow is assumed to be two dimensional.
It is described in nondimensional Cartesian 
coordinates $\xb=(x,y)$, and time $t$,
with fluid velocity $\ub(\xb,t)=\big\langle u(x,y,t),v(x,y,t)\big\rangle $.
We choose a reference frame moving with the plate, in 
which the plate is positioned horizontally on the x-axis, 
centered at the origin, 
at
\begin{equation}
 S=\Big\{(x,y): x \in\big[-\frac{1}{2},\frac{1}{2}\,\big]~,~y=0~\Big\}~,
\label{E:plate}
\end{equation}
the plate velocity is zero, and 
the far field velocity points upwards,
\begin{equation}
\ubinf(t)=\big\langle 0,1\big\rangle~.
\label{E:farfield}
\end{equation}
To illustrate, figure \ref{F:sample} plots the streamlines and vorticity
field at some relatively large time past the start of the motion. 
Here and throughout the paper, positive vorticity contours are shown in black,
negative contours are shown in a lighter grey scale.
The flow is assumed to remain symmetric about $x=0$, 
since at the times considered here
symmetry breaking instabilities are not expected to significantly 
affect the flow.
Hereafter, results are only shown for the right half plane, $x\ge 0$.

The flow is driven by the potential flow past the plate induced by the far field velocity.
The corresponding stream function,
$\psi_\infty(x,y)$, is given by the complex potential
\begin{eqnarray}
 W_\infty(x,y) = \sqrt{{1\over4}-(x+iy)^2} = \phi_\infty +
i\psiinf~.
\label{E:psiinf}
\end{eqnarray}
This flow is induced by a vortex sheet in place of the plate
whose strength is such that no flow passes through the plate.
For higher generality, instead of using equation (\ref{E:psiinf}), we approximate 
$\psiinf$ using a sufficiently fine discretization of the vortex sheet,
following the approach taken in Nitsche \& Krasny (1994), 
which
can be applied to other geometries as well, even if the analytic
expression for the potential is not known.

\subsection{Governing equations}
The fluid flow is modeled by the 
incompressible Navier Stokes equations
with constant density.
The governing equations,
given in terms of the fluid vorticity $\omega(\xb,t)=v_x-u_y$ and
stream function $\psi(\xb,t)$, are
\begin{subequations}
\begin{eqnarray}
&\text{(a)} &\quad \frac{\partial \omega}{\partial t} + (\ub\cdot\nabla) \omega =
 \frac{1}{Re}\nabla^2\omega, 
\label{E:vortevol}
\\ &\text{(b)} &\quad  \nabla^2\psi = -\omega ,
~~ \text{with}~~ 
\psi=0~\text{on}~ S~ \text{and}~ \psi\to\psi_{\infty} ~\text{as}~ \mathbf{x}\to\infty,
\label{E:poisson} \\ 
&\text{(c)} &\quad
\ub = \nabla^{\perp}\psi ,
~~ \text{with}~~ \ub=0~\text{on}~ S ~,
\label{E:ufrompsi} 
\end{eqnarray}
\label{E:nse}
\end{subequations}
where
$\nabla^{\perp}\psi = 
\big\langle
\frac{\partial \psi}{\partial y}, -\frac{\partial \psi}{\partial x} 
\big\rangle
$,
$Re=LU/\nu~,$
and $\nu$ is the kinematic fluid viscosity.


\section{Numerical Approach} 
\label{sec:method}

\subsection{Numerical Method}
The numerical method is based on fourth order finite difference 
approximations of the governing equations on a regular grid.
The computational domain is the rectangular region
\begin{eqnarray}
 D=[0,\xmax]\times[\ymin,\ymax]~,
\label{D}
\end{eqnarray}
with symmetry imposed across $x=0$.
The interior of the domain is given by the interior of $D\backslash S$, 
where $S$ is the plate position given in (\ref{E:plate}).
The computational boundary consists of
$\partial D\cup S$.
Here $\xmax>1/2$, $\ymin<0$, $\ymax>0$ are chosen sufficiently large
so that 
the vorticity $\omega$ effectively vanishes 
on $\partial D$ for
all the times computed. 
The domain is discretized by $(N_x+1)\times (N_y+1)$ 
equally spaced gridpoints $(x_i,y_j)$, where
\begin{subequations}
\begin{eqnarray} 
x_i &=  ih~,\hskip0.8truecm\qquad i&=0,\dots,N_x~,\quad h=\xmax/N_x~,\\
y_j &=  \ymin+jk~,\quad j&=0,\dots,N_y~,\quad k=(\ymax-\ymin)/N_y~,
\end{eqnarray}
and $N_x,N_y$ are chosen so that $h=k$.
Similarly, time is discretized as 
\begin{eqnarray}
t_n &=  n\Delta t~,\hskip1.2truecm n&=0,\dots,N~,\quad \Delta t=T_{fin}/N~,\hskip1.7truecm
\end{eqnarray}
\end{subequations}
where $T_{fin}=t_N$ is the final time.
Streamfunction, velocity and vorticity are carried on the gridpoints, with
$\psi_{i,j}^n$, 
$\langle u,v\rangle_{i,j}^n$, 
$\omega_{i,j}^n$ 
approximating 
$\psi(x_i,y_j,t_n)$,
$\langle u,v\rangle(x_i,y_j,t_n)$,
and $\omega(x_i,y_j,t_n)$.

The boundary stream function at time $t^n$ is $\psi^n_{bd}$.
The values of $\psi_{bd}$ on the plate $S$ are zero,
which ensures that on the plate, $v=0$.
The values of $\psi_{bd}$ on the remaining boundaries of $\partial D$
are obtained numerically, as explained below.
The boundary vorticity at time $t^n$ is $\omega^n_{bd}$.
The values of $\omega_{bd}$ on $\partial D$ are zero.
The values of $\omega_{bd}$ on the upstream 
and downstream sides of the plate $S$,
denoted by $\omega_+^{n}$ and $\omega_-^{n}$, respectively,
are obtained by enforcing that $u=0$ on the plate, see below.
%
The initial conditions are given by zero vorticity in the interior of the domain.

The vorticity at time $t_n$ is updated to time $t_{n+1}$ by solving equation 
(\ref{E:nse}) in two steps:

\begin{description}
\item[Step 1:]~
The interior vorticity is convected by solving the equation
\begin{subequations}
\begin{equation}\label{E:convec}
 \frac{DQ}{Dt}=0~\quad \text{subject to}\quad Q(t_n)=\omega^n~,
\end{equation}
for one timestep and 
setting $\omega^*=Q(t_{n+1})$.
The values of $\omega^*$ are then used to obtain
updated interior and boundary values of the stream function, 
velocity and vorticity, 
 $\psi^{n+1}$, $\langle u,v\rangle^{n+1}$ and $\omega_{bd}^{n+1}$.

\item[Step 2:]~
The interior vorticity is diffused by solving the equation
\begin{equation}\label{visc}
 \frac{\partial Q}{\partial t} =  {1\over Re} \bigtriangledown^2 Q\quad
\text{subject to}\quad 
Q(t_n)=\omega^*~,\quad
Q_{bd}(t^{n+1})=\omega_{bd}^{n+1}~,
\end{equation}
\end{subequations}
for one timestep and setting
$\omega^{n+1}=Q(t_{n+1})$.
\end{description}

Several details in each of the two steps above remain to be explained. 
\begin{description}
\item[Step 1a:]~
Equation \eqref{E:convec} is solved using a 
semi-Lagrangian scheme which is
second order in time and fourth order in space, as follows.
For each interior grid point $(x_i,y_j)$, 
first find the location of a particle at $t_n$ 
that travels with the fluid velocity, and ends up at $(x_i,y_j)$
at $t_{n+1}$.
This is equivalent to solving 
\begin{eqnarray}
\frac{d \zb}{dt} = \ub(\zb,t)~, \quad
\zb(t_{n+1}) = (x_i,y_j)
\label{E:semilag}
\end{eqnarray}
for $\zb(t_n)$, 
where $\zb = (x,y)$. Equation (\ref{E:semilag}) is solved to
second order in time using velocity values at the current and
previous timestep,
$\langle u,v\rangle^{n-1}$ and $\langle u,v\rangle ^{n}$. 
Then, obtain
the vorticity of the particle at $t_n$, $\omega(\zb(t_n),t_{n})$,
from vorticity values at nearby grid points using a fourth
order bi-cubic interpolant. This step uses interior and 
boundary values of vorticity at $t_n$.
Finally, set $\omega^{*}_{i,j} =
\omega(\zb(t_n),t_{n})$. Details can be found in the paper by \cite{xu12}.
\item[Step 1b:]~
Updated interior values of the stream function at $t_{n+1}$ 
are obtained by solving 
\begin{equation}
\Delta\psi^{n+1}=\omega^{*} \text{ in interior}~, \quad
\psi=\psi^{n+1}_{bd} \text{ on } \partial D \cup S~.
\label{E:discpoisson}
\end{equation}
Here, the Laplace operator 
is approximated by a compact
fourth order finite difference
scheme (Strikwerda 1989, 
equation 12.5.6).
The boundary values $\psi_{bd}$ are given by
$$\psi_{bd}=0 \text{ on } S\cup\{x=0\}~.$$
On the remaining three sides of $\partial D$, $\psi_{bd}$ 
is computed using an integral formulation. 
We use the domain specific Green's function $G_{S}$
\begin{subequations}
\begin{eqnarray} \label{fpGreen}
   G_{S}(\zb,\zo) =
\log\left|\frac{\sqrt{\frac{x+iy-1/2}{x+iy+1/2}} -
\sqrt{\frac{x_o+iy_{o}-1/2}{x_o+iy_{o}+1/2}}}{\sqrt{\frac{x+iy-1/2}{
x+iy+1/2}}    -
\sqrt{\frac{x_o+iy_{o}-1/2}{x_o+iy_{o}+1/2}}
^{*}}\right|~,
  \end{eqnarray}
where $\zb=(x,y)$, $\zo=(x_o,y_o)$, 
and $*$ denotes the complex conjugate,
and compute
\begin{equation} \label{E:biot}
\psi_{bd}(\zbd,t) = \psi_{\infty}(\zbd,t) +
\int_{D/S} \omega(\zo,t) G_{S}(\zbd,\zb)d\zb, 
\end{equation}
\end{subequations}
where $\zbd\in\partial D$.
Alternatively, one can use the free space Green's function $G_{\infty}$,
and simulate the effect of the plate by a vortex sheet in its place.
In either of these approaches,
one needs to compute an 
area integral $\int \omega(\zb,t)G(\zbd,\zb)d\zb$. 
In practice, we only integrate
over the region in which $|\omega| \ge 10^{-9}$, and
use the fourth order Simpson's method. 
The linear system for $\psi_{ij}^{n+1}$ obtained by discretizing
(\ref{E:discpoisson}) is solved using the conjugate gradient method. 

\item[Step 1c:]~
Updated interior values of velocity at $t_{n+1}$ are obtained
by solving
\begin{equation} 
\langle u,v\rangle^{n+1} =\grad^{\perp}\psi^{n+1}
\end{equation} 
using fourth order centered difference approximations. 
Boundary values of velocity are only needed 
on the plate, where
they vanish, and on the axis $x=0$, where they are obtained by 
centered differences from $\psi^{n+1}$ and use of symmetry.
The updated velocity is used at the next timestep, in
Step 1a.

\item[Step 1d:]~
Updated boundary vorticity values at $t_{n+1}$
are obtained from the updated stream function by enforcing 
the no-slip boundary condition on the plate.
The boundary condition $\phi=0$ ensures that $v=0$ on the walls.
The boundary vorticity is
$$ \omega_{bd}^{n+1}=-{\partial^2\psi^{n+1}\over\partial y^2}~.$$
This equation is discretized so that 
$\psi=0$ and $\partial\psi/\partial y=0$ on the wall, and thus $u=0$. 
Here, we use a fourth order version of the 
Thomas formula, known as Briley's formula, following 
\cite{eliu96a} (their equation 2.11). 
The updated vorticity values are used in Step 2, 
below, as well as at the next timestep, in Step 1a.

\item[Step 2:]~
Equation \eqref{visc} is solved
by
discretizing the Laplace operator with the 4th order 
compact finite difference
scheme also used for equation (\ref{E:discpoisson}) and then 
applying an implicit Crank-Nicolson method which is second order in time and
fourth order in space (Fletcher 1991, page 255ff).  
The resulting linear system for the
vorticity $\omega^{n+1}$ is solved using the conjugate gradient method.
\end{description}

This completes the description of the numerical method.
In order to visualize streaklines as may be observed in 
laboratory experiments, particles are also initially 
placed near the plate and passively transported by 
the fluid flow. Their position $\zb(t)$ is given by 
\begin{eqnarray}\label{ode}
 \centering
 \frac{d\zb}{dt} = \ub(\zb,t)~, \quad \zb(0)=\zo~,
\end{eqnarray}
where $\ub$ is the fluid velocity.
The velocity at the current particle position is obtained
by interpolation, and 
the equation is solved using the second order explicit 
Adam-Bashforth scheme, for a range of initial positions.

\subsection{Resolution and Convergence}

To test this numerical scheme, \cite{xu12} applied it to 
the driven cavity problem of \cite{eliu96a}, and reproduced their results.
For smooth cavity lid motion (see also Johnston 1999), 
the method was confirmed to converge to 4th order in space, and to first order in time.
The slow convergence in time is a property of standard splitting schemes.
Here, we discuss the performance of the method applied to 
the more singular case of impulsively started flow past a sharp edge.


\begin{figure}
\centering
\includegraphics[height=0.39\textwidth]{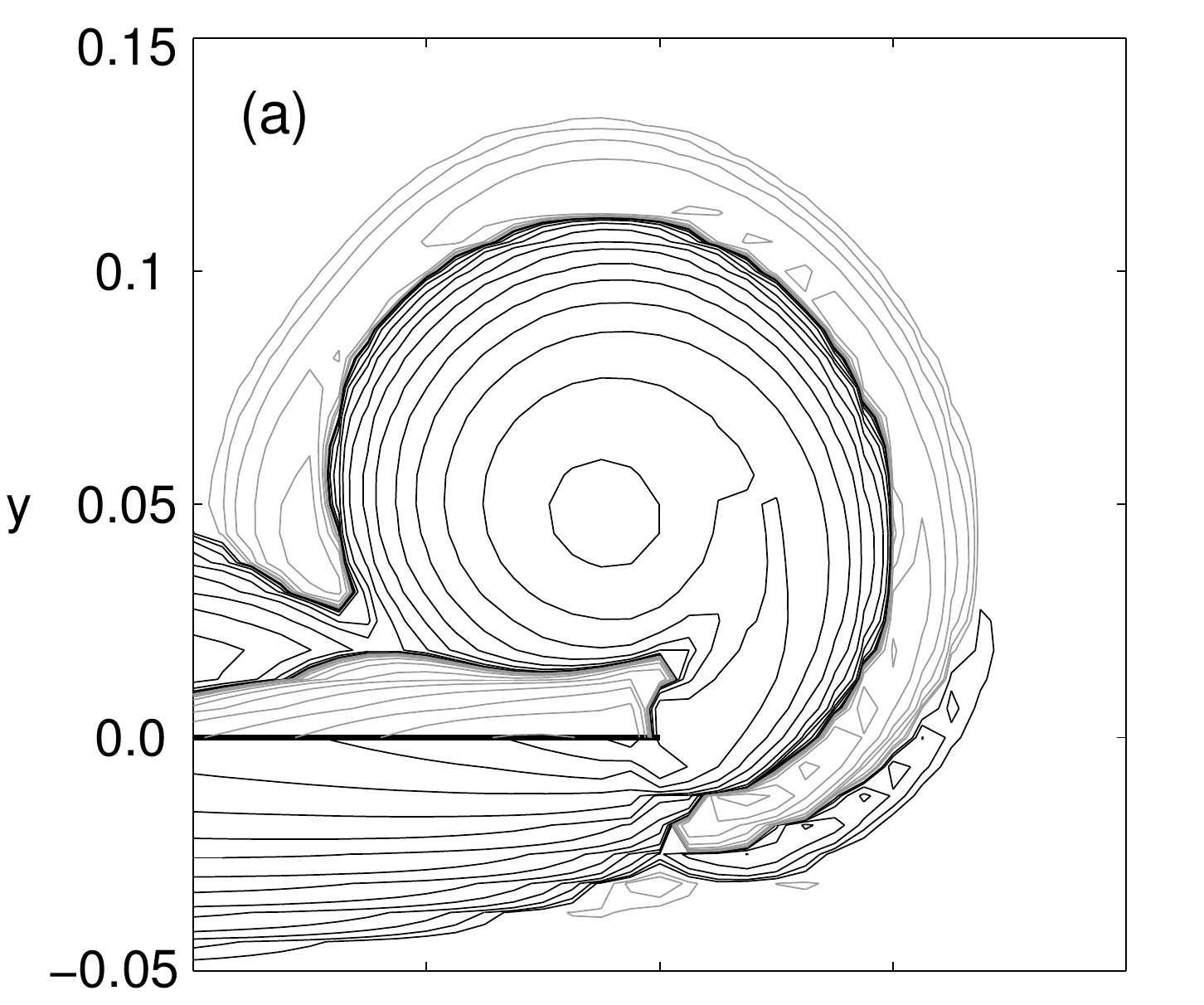}
\includegraphics[height=0.39\textwidth]{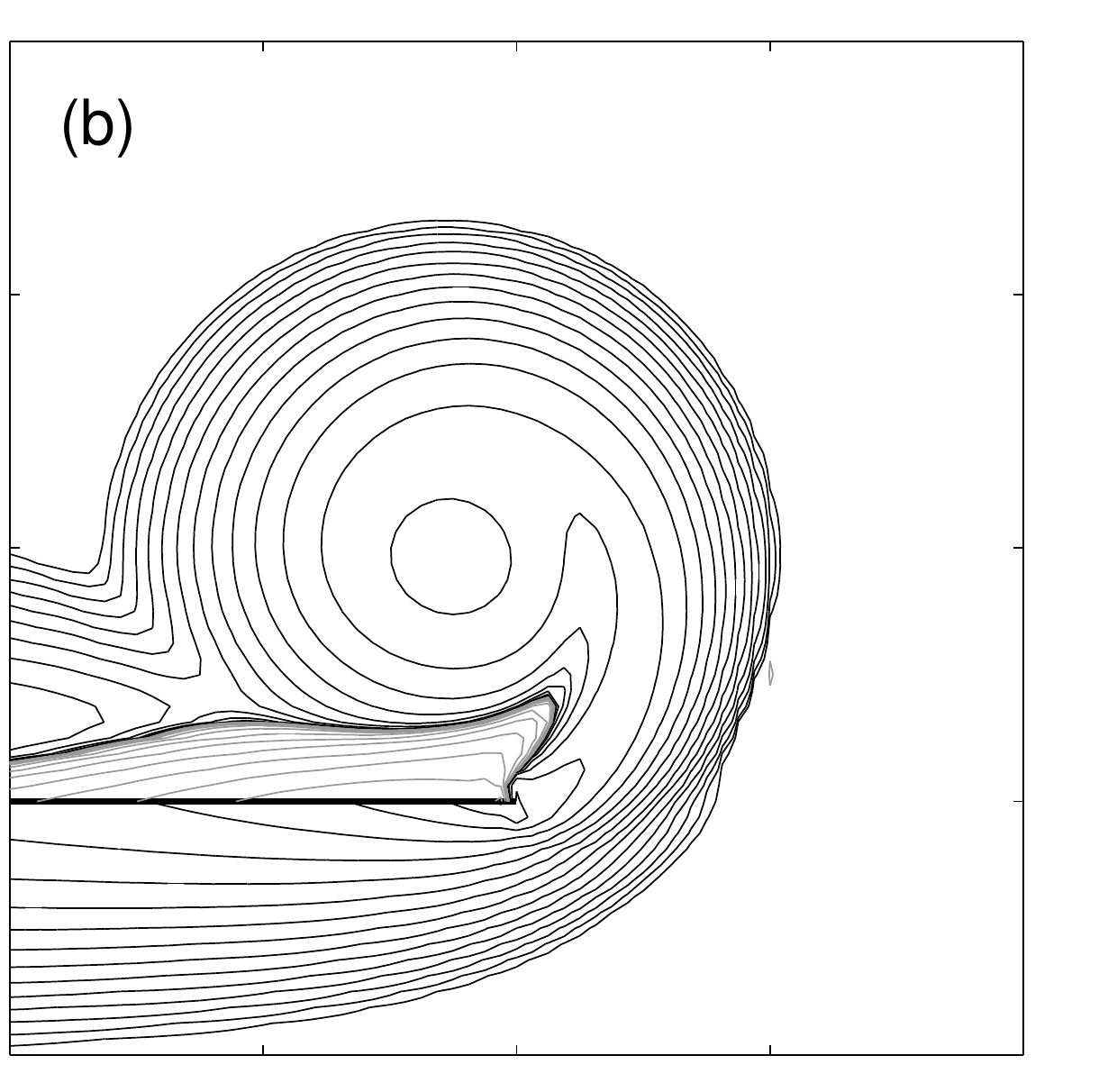}\\
\includegraphics[height=0.425\textwidth]{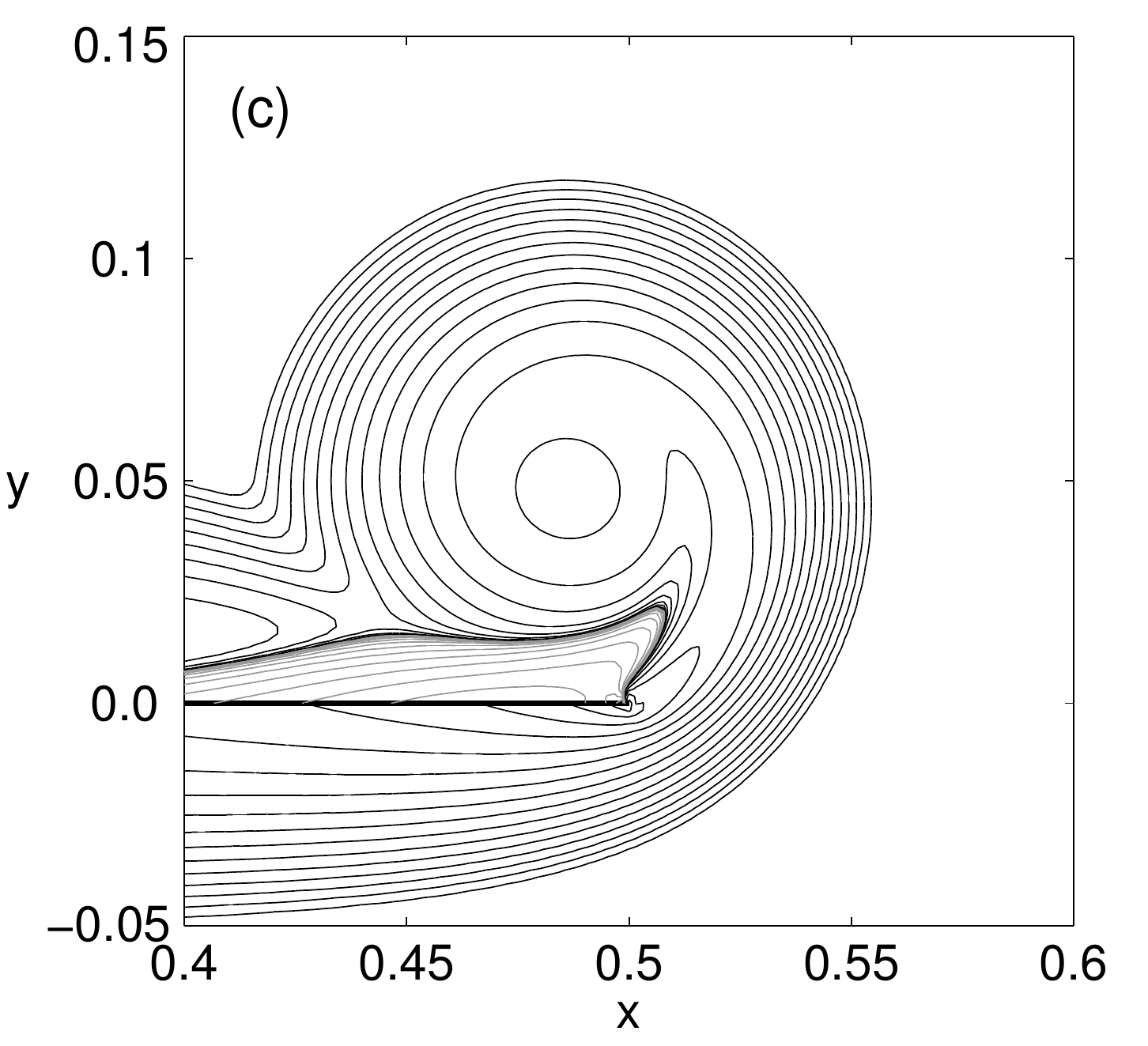}
\includegraphics[height=0.425\textwidth]{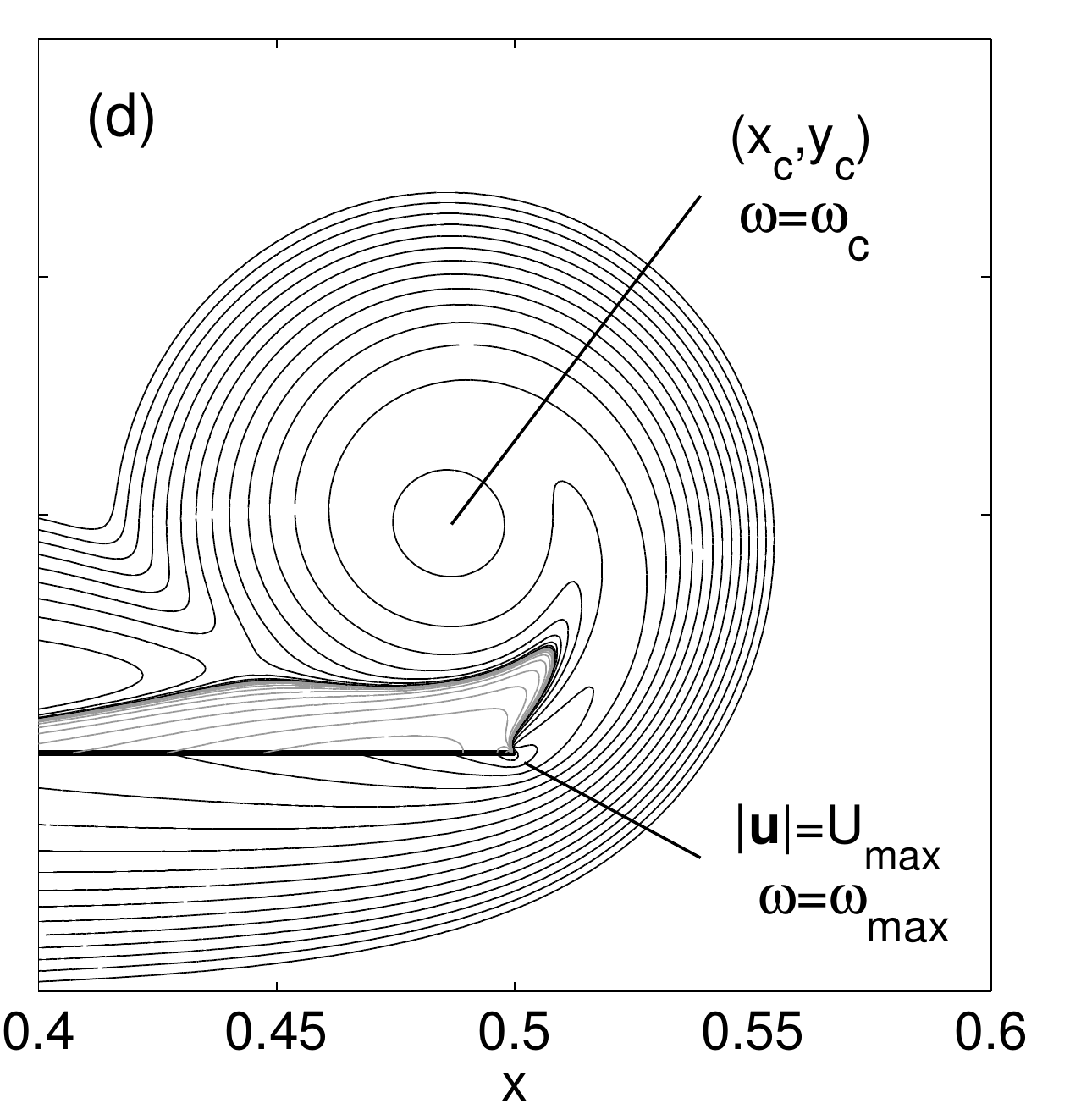}
 \caption{Vorticity contours $\omega=\pm 2^{-5:12}$ at $t$=0.05 for $Re$=500 for the problem of
impulsively started flow past a finite plate. The
mesh size $h$ and time step $\Delta t$ for each figure are 
\textit{(a)} $h$=1/160, $\Delta t$=$4\times10^{-4}$, 
\textit{(b)} $h$=1/320, $\Delta t$=$2\times10^{-4}$, 
\textit{(c)} $h$=1/640, $\Delta t$ = $1\times10^{-4}$, 
\textit{(d)} $h$=1/1280, $\Delta t$ = $5\times10^{-5}$. 
}
\label{F:conv}
\end{figure}

To illustrate the effect of resolution in space and time, figure 
\ref{F:conv} plots vorticity contours computed for $Re=500$ at
$t=0.05$, with various values of the meshsize $h$ and timestep $\Delta t$.
The resolution is coarsest in 
figure \ref{F:conv}(a), finest 
in figure \ref{F:conv}(d), 
as given in the caption. 
The figure shows contours $\omega=\pm2^{-5:12}$
in a region close to the tip of the plate, 
with positive vorticity in black, 
negative vorticity in a lighter shade of grey, and the plate 
as a black line. The zero vorticity contour level appears as a thick dark
curve which in fact consists of many positive and negative vorticity contour
levels of small magnitude.
We first describe the well-resolved result in figure \ref{F:conv}(d).
Recall that the background driving velocity flows from bottom to top. 
This causes the formation of 
a boundary layer of positive vorticity around the plate. 
The maximum vorticity and velocity magnitude, $\omega_{max}$ and
$U_{max}$, occur at the tip of the plate, one gridpoint away from it,
at all times. 
At the time
shown here, positive vorticity has moved upward to form a concentrated
vortex on the downstream side of the plate,
with a second local maximum in the core vorticity,
at $(x_c,y_c)$ with magnitude $\omega_c$. 
This vortex induces positive flow on the dowstream plate wall,
which causes the formation of
a thin region of opposite signed, negative vorticity 
along the wall. The negative 
vorticity
is entrained into the leading vortex.

If $h,\Delta t$ are too large, as in figure \ref{F:conv}(a), the lack of resolution
is evidenced by alternate layers of positive and negative vorticity 
that form outside the leading vortex.
If the flow is only slightly underresolved, 
as in figure \ref{F:conv}(b), 
ripples in the vorticity 
are first visible below and to the right of the tip.
As the resolution increases,
as in figure \ref{F:conv}(c), the ripples disappear and the vorticity is smooth.
Finer resolution, as in figure \ref{F:conv}(d), leaves the results practically unchanged.

We found this to be the case at all times computed: at all times, 
an instability is apparent for large enough values of $h,\Delta t$. 
If the resolution is sufficiently fine, the results are smooth and remain unchanged
to the eye under further refinement. The values of $h, \Delta t$ 
required for smooth results are smaller at earlier times. We thus 
take the following approach: results at a given time 
are computed with a value of $h,\Delta t$ sufficiently small so 
that the vorticity contours with $|\omega|\ge 2^{-5}$ appear resolved.
Table 1 lists the meshsizes and timesteps used in time intervals
$[0,T_{fin}]$, for different values of $T_{fin}$.
The range given for $\Delta t$ and $T_{fin}$ reflects values 
used for different Reynolds numbers $Re$. 
For larger Re, a given time requires a smaller value of $h$ and $\Delta t$.
For example, $h=1/320$ is used for the runs with $T_{fin}=0.5$ for $Re=250$,
but is required for much larger $T_{fin}=3$ for $Re=2000$.

An estimate of the order of convergence of the method 
for impulsively started flow 
is obtained from Table 2.
The table lists the 
errors in the position $(x^h_c,y^h_c)$ and vorticity magnitude $\omega^h_c$
of the vortex 
core at $t=0.05$ 
(see figure \ref{F:conv}(d)),
as well as errors in the stream function along horizontal and vertical
lines near the vortex core.
The errors are computed relative to the results 
with $h=1/1280$, as follows
\begin{subequations}
\begin{eqnarray}
e^h_{\omega}& =& \frac{|\omega^h_c-\omega^{1/1280}_c|}{|\omega^{1/1280}_c|}~,
\quad  
e^h_c = \frac{||\zb^h_c-\zb^{1/1280}_c||_2}{||\zb^{1/1280}_c||_2}\\
e^h_{\psi,x}& =& ||\psi^h-\psi^{1/1280}||_{\infty} \text{ along } x=0.5\\
e^h_{\psi,y}& =& ||\psi^h-\psi^{1/1280}||_{\infty} \text{ along } y=0.048
\end{eqnarray}
\end{subequations}
where $\zb_c = \langle x_c, y_c \rangle$.
The data in table 2 is summarized in figure \ref{F:errors}, together with 
a line with slope $m=2$. Even though the amount of 
data points is rather limited,
the data is consistent
with second order or better rate of convergence,
with faster convergence away from the tip.

\begin{table}
 \centering
 \begin{tabular}{lrll}
 \multicolumn{4}{c}{}\\
 ~ $h$ & 
$\Delta t$ ~ ~ ~& 
~ $T_{fin}$ &
[$\xmin, \xmax$]$\times$[$\ymin, \ymax$]\\
 1/160&  (4-5)$\times 10^{-4}$ &~ ~ 5       &~ ~ ~ [0,2]$\times$[-0.50,5.50]\\
 1/320&~  ~ 2 $\times 10^{-4}$ &~ 0.5-3     &~ ~ ~ [0,1]$\times$[-0.25,0.75]\\
 1/640&(0.5-2)$\times 10^{-4}$ &~ 0.2-0.6   &~ [0,0.75]$\times$[-0.25,0.50]\\
1/1280&  (2-5)$\times 10^{-5}$ &~0.05-0.1   &~ [0,0.75]$\times$[-0.25,0.50]\\
1/2560&  (4-5)$\times 10^{-6}$ &0.005-0.038 &~ [0,0.55]$\times$[-0.05,0.10]\\
1/5120&~  ~ 2 $\times 10^{-6}$ &0.001-0.005 &~ [0,0.55]$\times$[-0.05,0.10]\\
\hline
 \end{tabular}
\caption{\label{comp} The mesh size $h$, time step $\Delta t$, 
final time $T_{fin}$, computational domain 
used for the range of $Re=\in[250,2000]$ used in the computations. 
For larger values of $Re$, a given value of 
$h$ is used with the smaller values of $\Delta t$
and to smaller final times $T_{fin}$.
%
}
\end{table}

\begin{table}
 \centering
 \begin{tabular}{lcccc}
$h$  & $e^h_c$ & $e^h_{\omega}$ & $e^h_{\psi,x}$ & $e^h_{\psi,y}$\\
     &       &              & ~ along $x=0.5$ ~ & ~ along $y=0.048$ \\
1/160 & $3.51\times10^{-3}$& $8.30\times10^{-3}$& 0.0155 & 0.005811 \\
1/320 & $1.80\times10^{-3}$& $1.83\times10^{-3}$& 0.0060 & 0.002386 \\
1/640 & $1.86\times10^{-4}$& $4.19\times10^{-5}$& 0.0010 & 0.000683 \\
\end{tabular}
\caption{Errors 
$e_{\omega}^h$, $e_c^h$, 
$e^h_{\psi,x}$, 
and $e^h_{\psi,y}$, 
at $t=0.05$, computed with $Re$=500. }
\end{table}

\begin{figure}
 \centering
\includegraphics[width=0.55\textwidth]{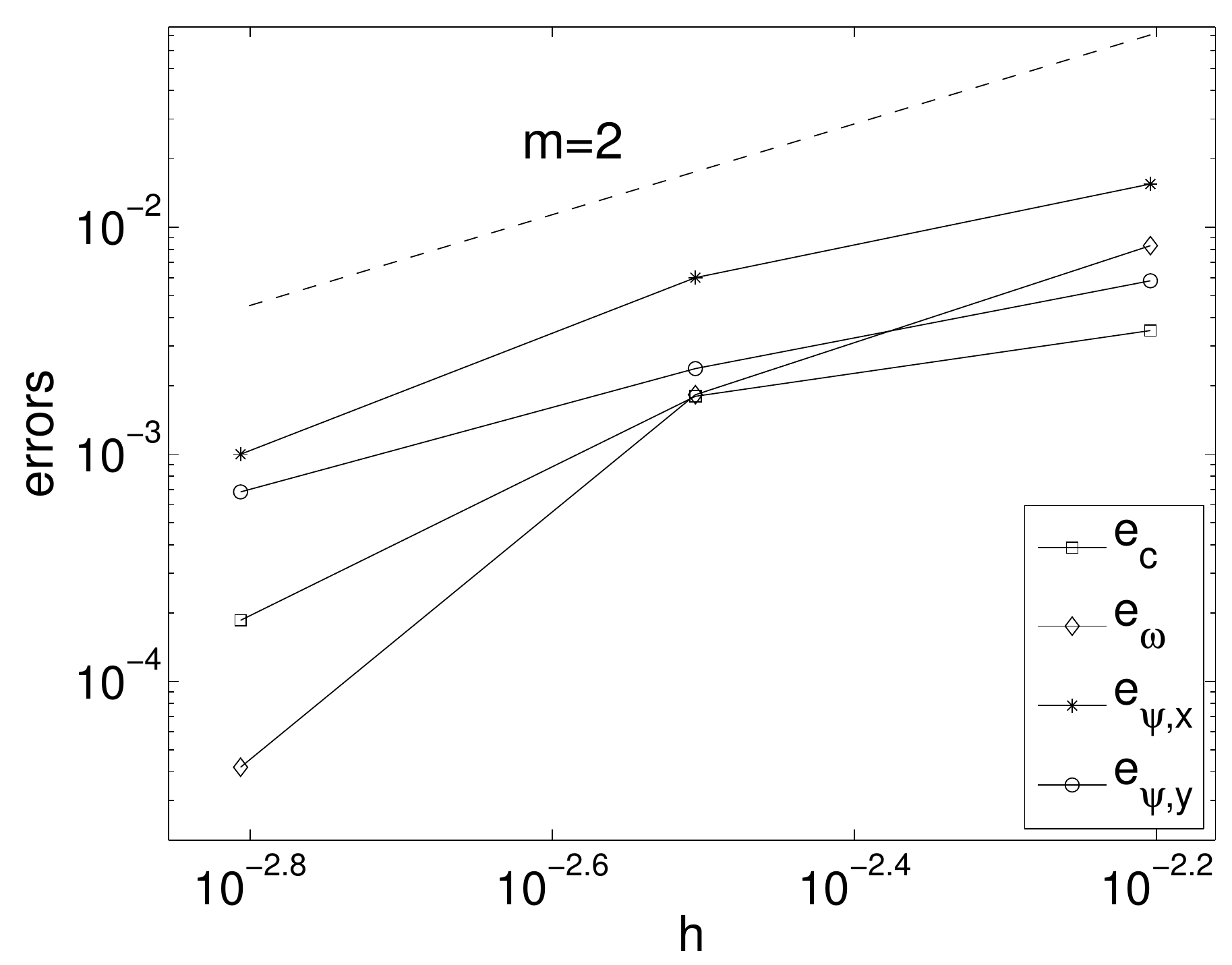}
\caption{Errors $e^h_c$, $e^h_{\omega}$, $e^h_{\psi,x}$,
$e^h_{\psi,y}$, \vs $h$, as indicated.
A line of slope $m=2$ is also shown.
}
\label{F:errors}
\end{figure}

\subsection{Singular initial flow}
The difficulty in resolving the flow 
is largely due to the singular nature  
of the initial flow. 
To illustrate, 
figure \ref{F:sing} 
plots the maximum velocity $\umax$,
and the maximum absolute vorticity $\omegamax$, \vs time $t$.
In each case, results for all values of $h$ used are plotted, 
as indicated. 

\begin{figure}
 \centering
\includegraphics[width=0.465\textwidth]{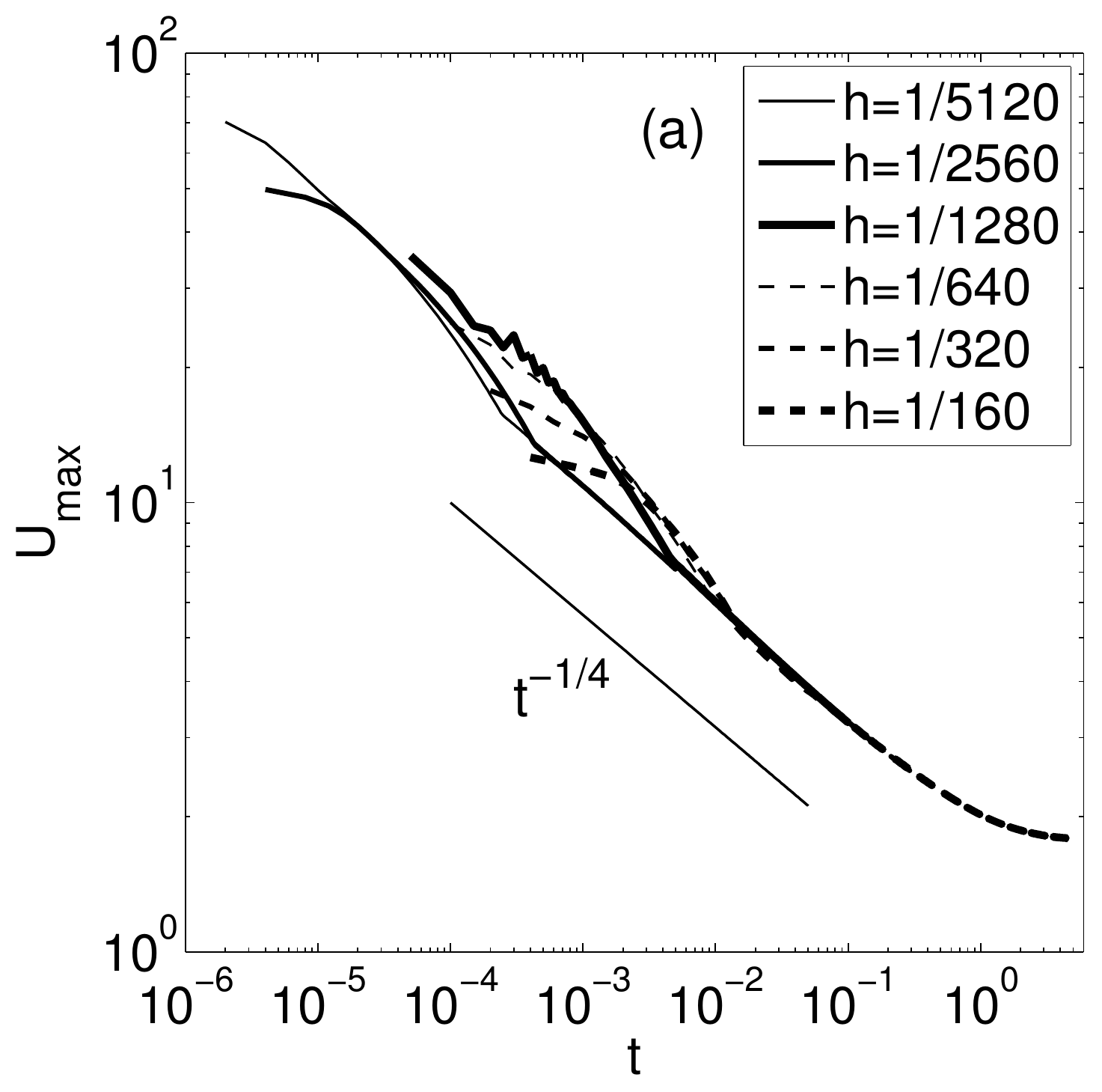}
\includegraphics[width=0.465\textwidth]{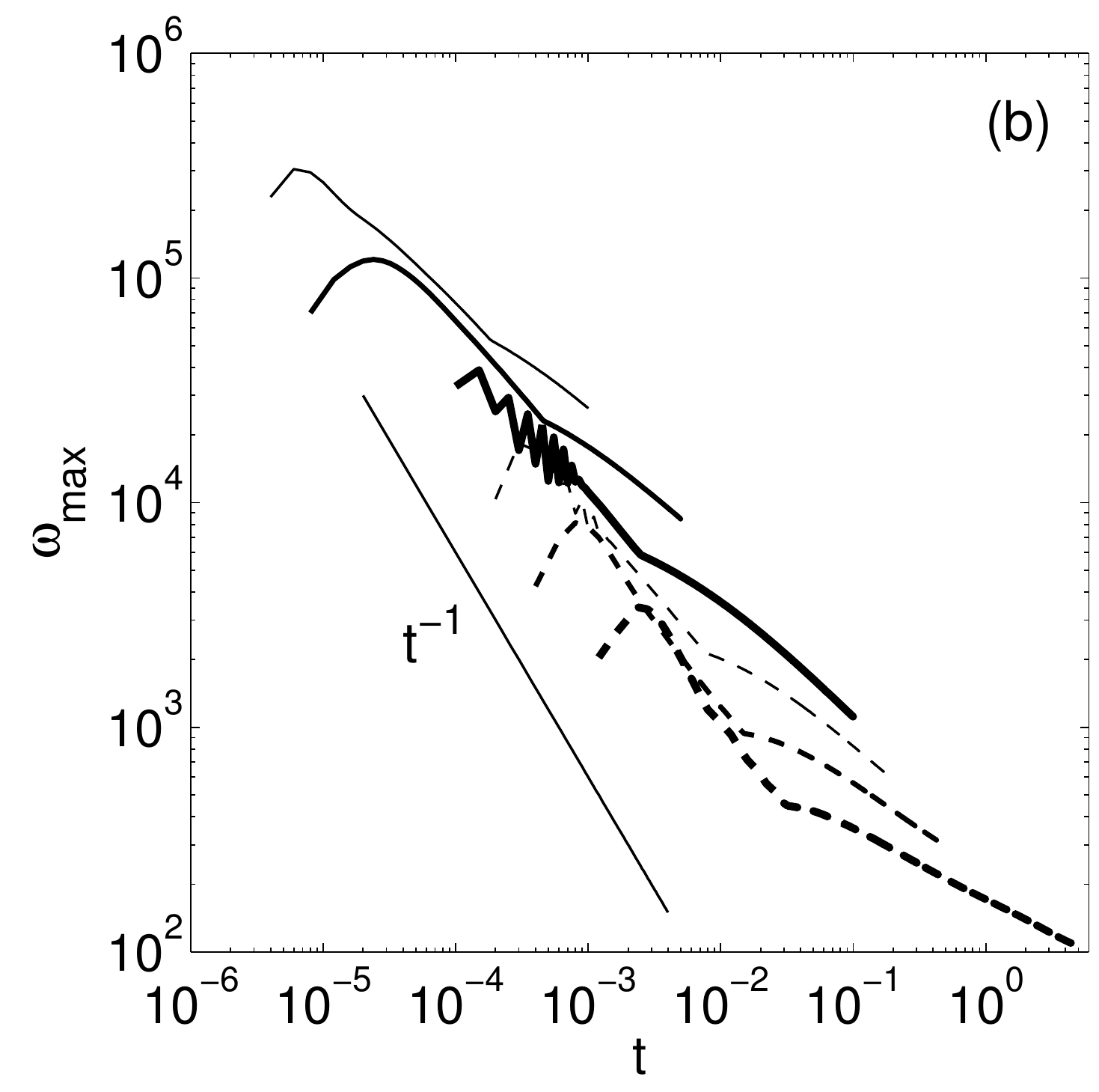}
\caption{(a) Maximum velocity $U_\text{max}$ and
(b) maximum absolute vorticity $\omegamax$, \vs $t$,
for $Re$=500, computed with the indicated values of $h$.}
\label{F:sing}
\end{figure}

The maximal absolute velocity $\umax$, plotted in figure \ref{F:sing}(a), 
becomes unbounded as $t$ approaches 0. Recall that the 
initial potential flow has unbounded velocity at 
the tip of the plate.
The maximum velocity $\umax$ is bounded at all positive times,
and decays fast initially.
However, because of the initial singularity, it is not possible
to resolve the flow until after some small initial times.
Figure \ref{F:sing}(a) 
shows that with smaller values of $h$,
$\umax$ can be computed smoothly during earlier times,
with $U_{\text{max}}\approx 200$
at $t\approx 10^{-5}$.
The results with varying $h$
appear to converge
to a line with slope $-1/4$, indicating that
$\umax$ decays as
\begin{eqnarray}
\umax\sim t^{-1/4}.
\end{eqnarray}

Figure \ref{F:sing}(b) plots the
maximum absolute vorticity $\omegamax$.
Its values are of order $10^6$ at the earliest 
time shown, and are fairly well
resolved with the smallest values of $h$ shown.
The maximum vorticity decreases in time approximately as
\begin{eqnarray} 
\omegamax\sim t^{-1}.  
\end{eqnarray}

We note that the maximum velocity and vorticity occur one 
gridpoint away from the tip of the plate. Thus, unlike
figure 3, the results 
in figure \ref{F:sing} do not represent a
study of pointwise convergence at a fixed point. 
They do
illustrate the singular nature of the flow and the extent
to which it is recovered by the finite numerical resolution.

\section{Numerical Results}
\label{sec:results}

\subsection{Vorticity, streaklines, streamlines, $Re=500$}
This section describes the evolution of the flow
near the plate, for the case of fixed $Re=500$, unless noted.
Figure \ref{F:evol}
shows vorticity contours (left column), 
streaklines (middle column) and streamlines (right column),
at the indicated times.
At each time, the results shown are computed with 
the finest resolution listed in table 2 for that time. 
The vorticity contours are $\omega=\pm 2^{[-5:12]}$, 
with positive contours in black, negative ones in grey. 


\begin{figure}
 \centering
\includegraphics[width=0.3505\textwidth]{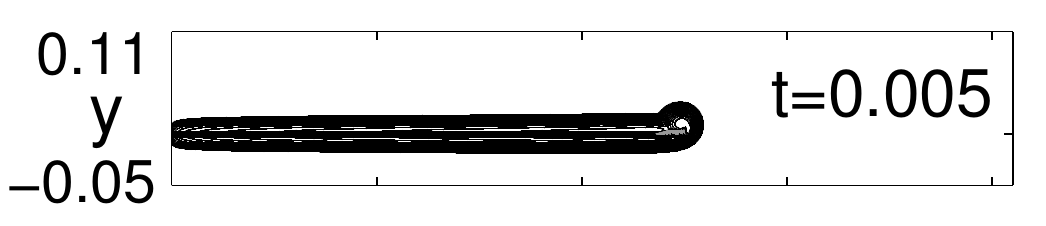}
\includegraphics[width=0.298\textwidth]{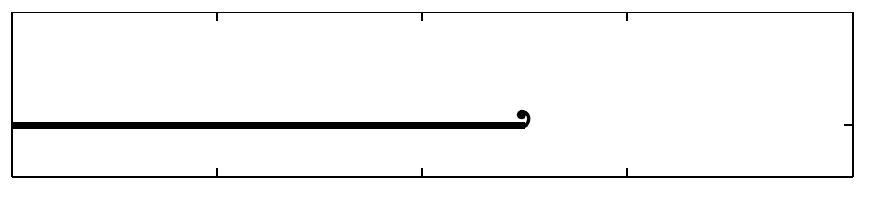}
\includegraphics[width=0.299\textwidth]{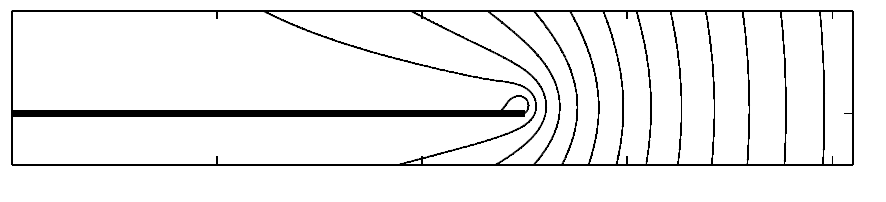}\\

\includegraphics[width=0.3505\textwidth]{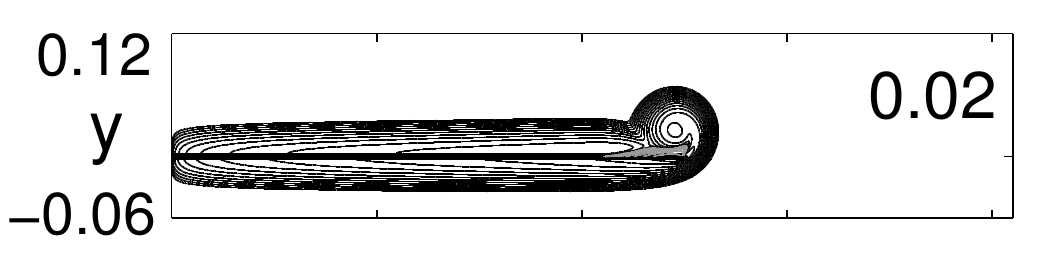}
\includegraphics[width=0.298\textwidth]{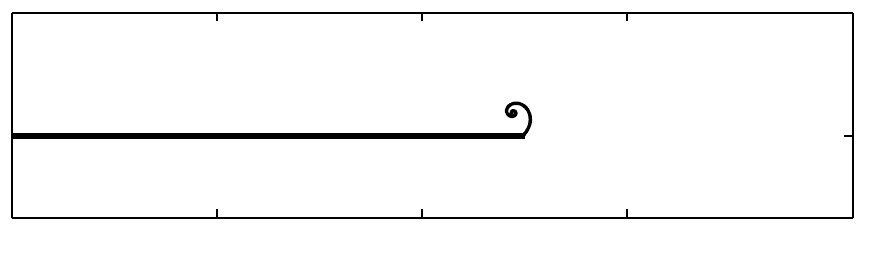}
\includegraphics[width=0.298\textwidth]{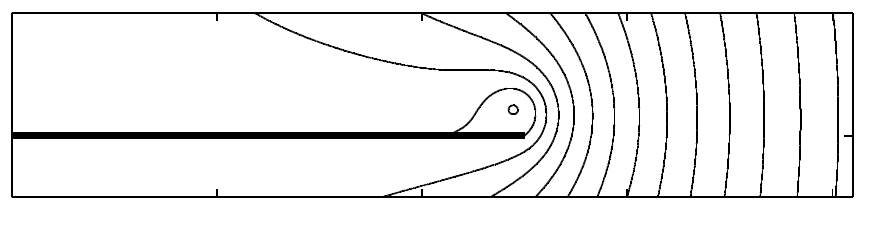}\\

\includegraphics[width=0.3505\textwidth]{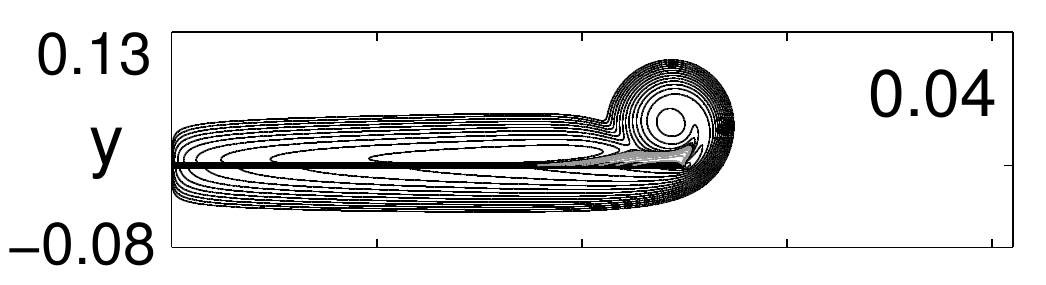}
\includegraphics[width=0.298\textwidth]{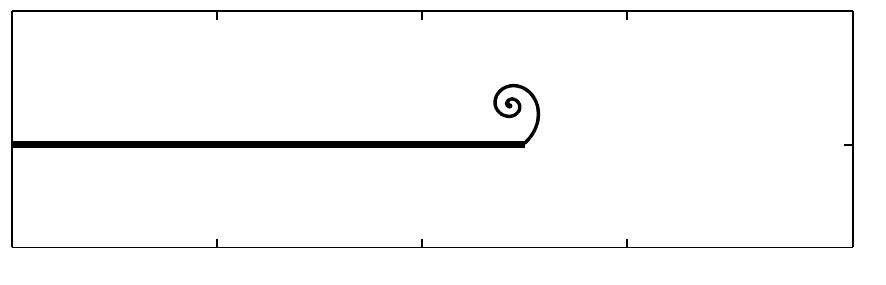}
\includegraphics[width=0.298\textwidth]{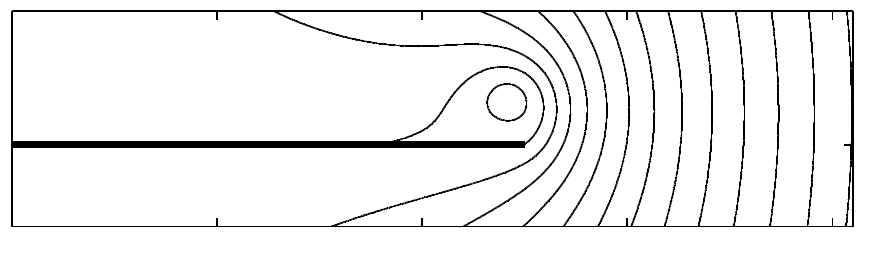}\\

\includegraphics[width=0.3505\textwidth]{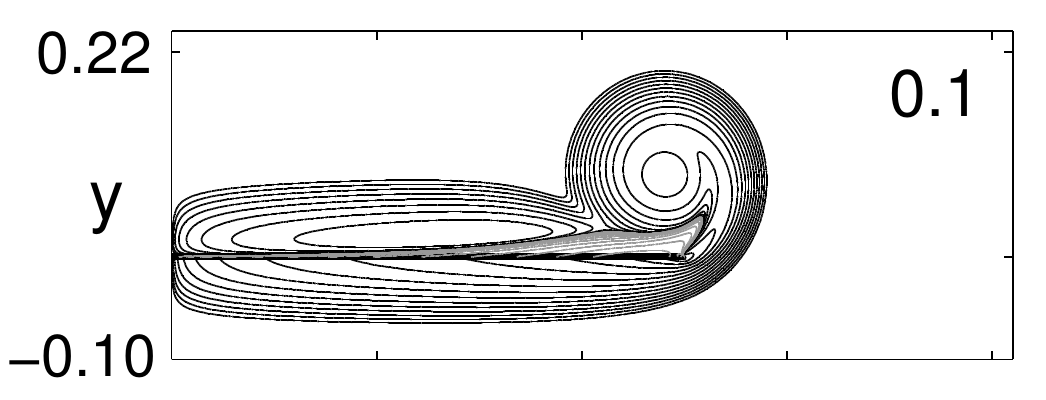}
\includegraphics[width=0.298\textwidth]{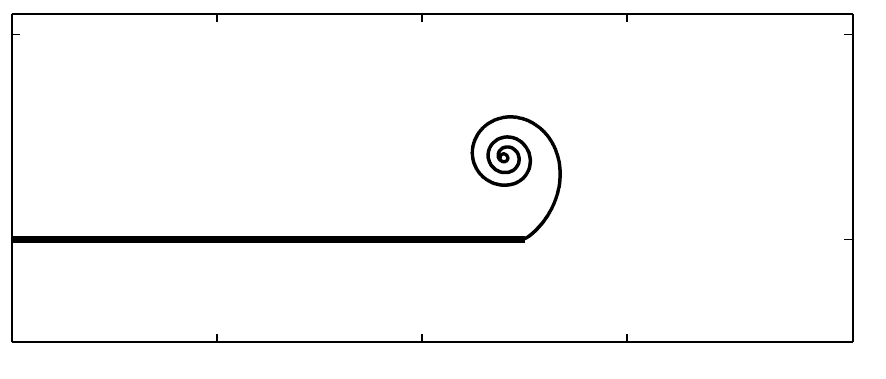}
\includegraphics[width=0.298\textwidth]{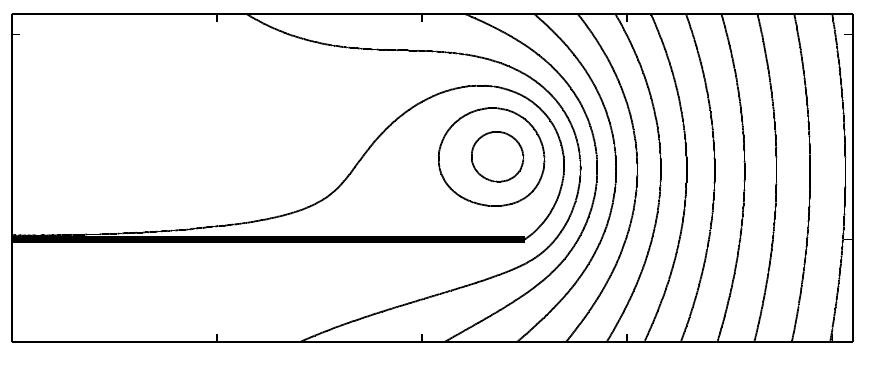}\\

\includegraphics[width=0.3505\textwidth]{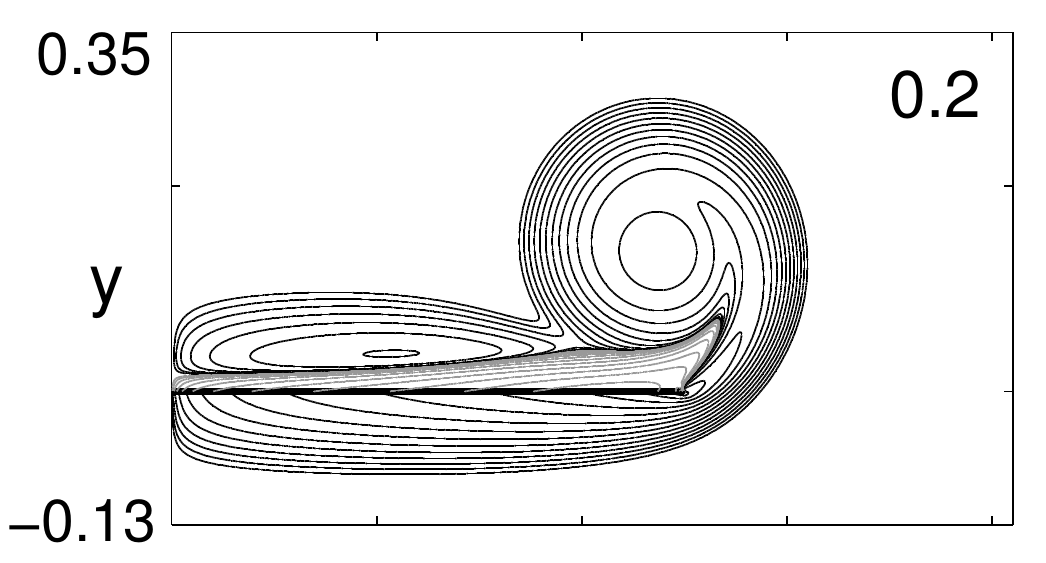}
\includegraphics[width=0.298\textwidth]{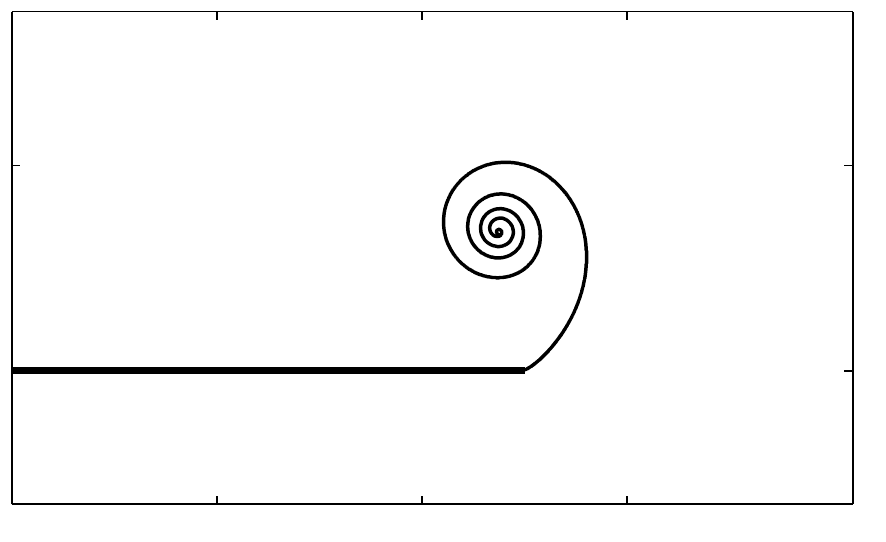}
\includegraphics[width=0.298\textwidth]{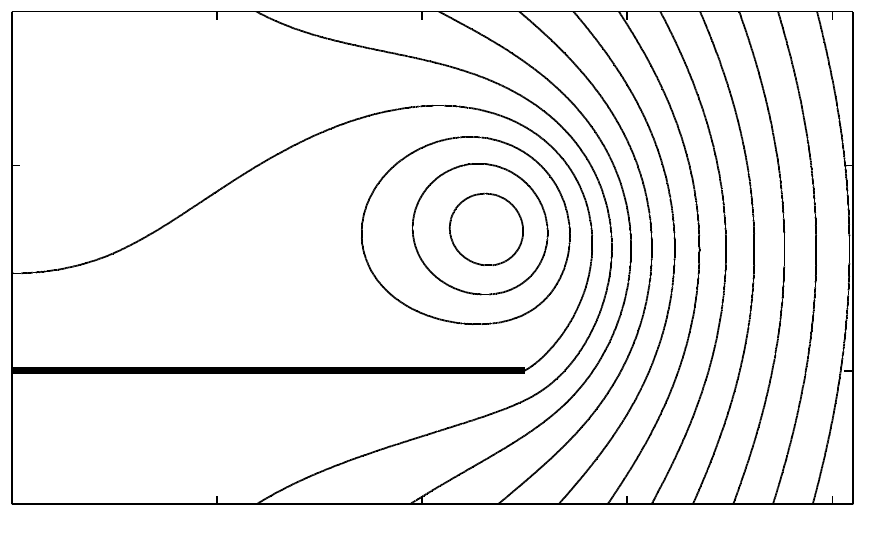}\\

\includegraphics[width=0.3505\textwidth]{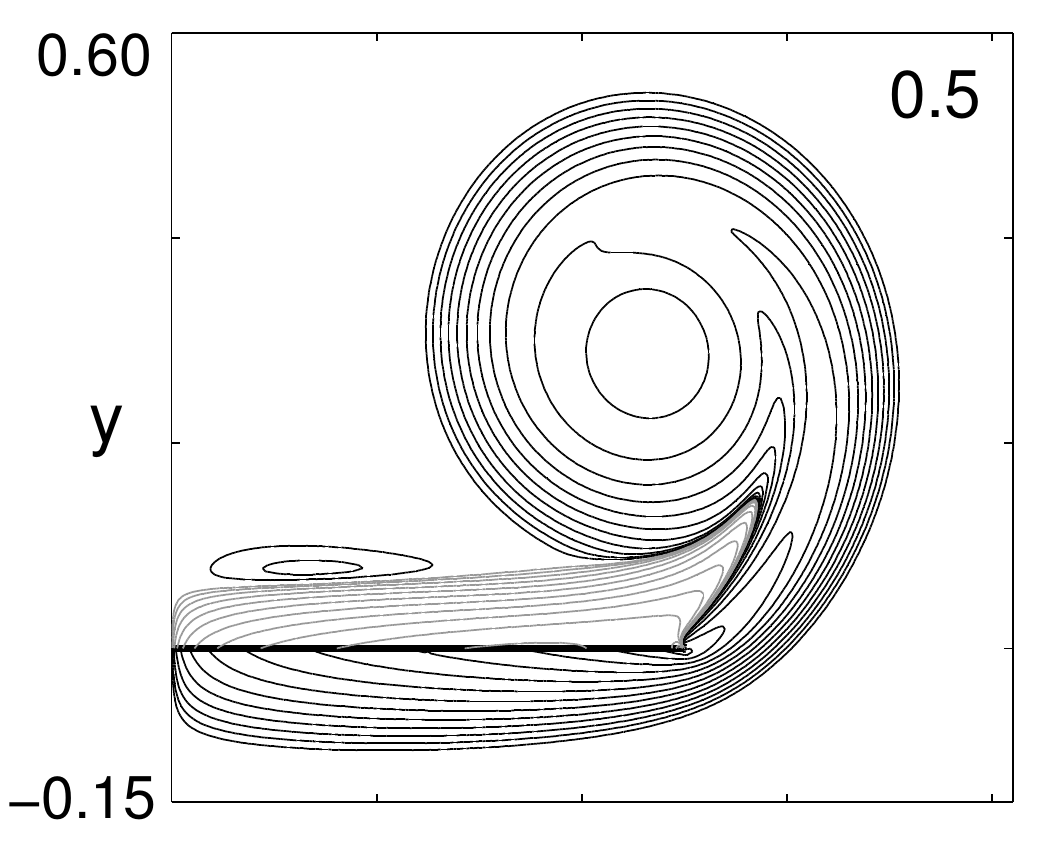}
\includegraphics[width=0.298\textwidth]{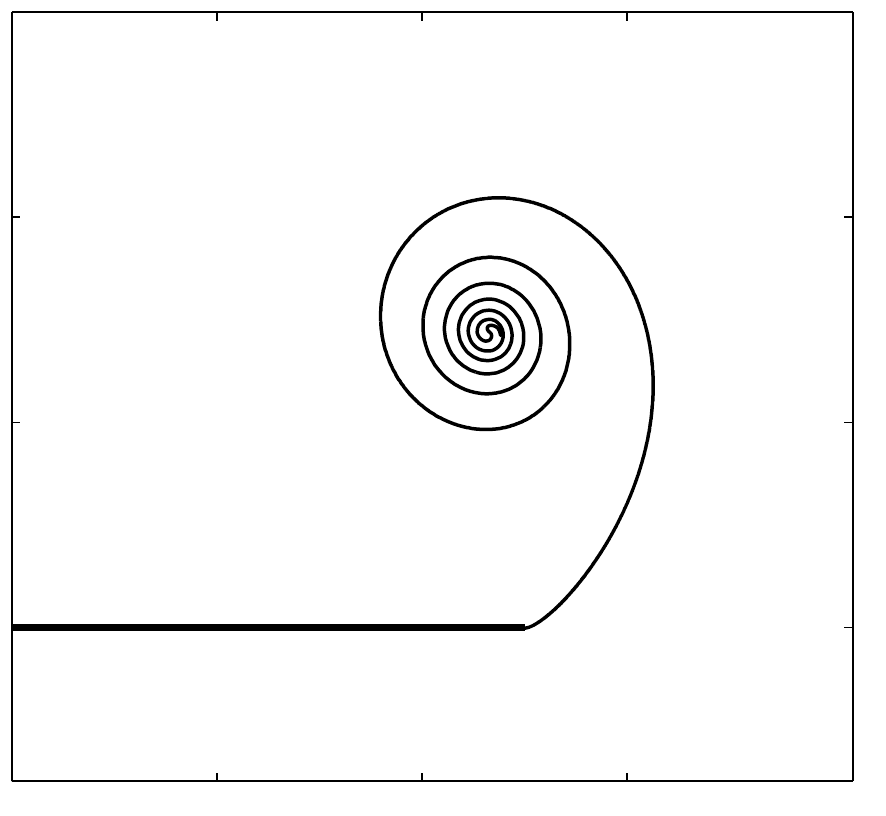}
\includegraphics[width=0.298\textwidth]{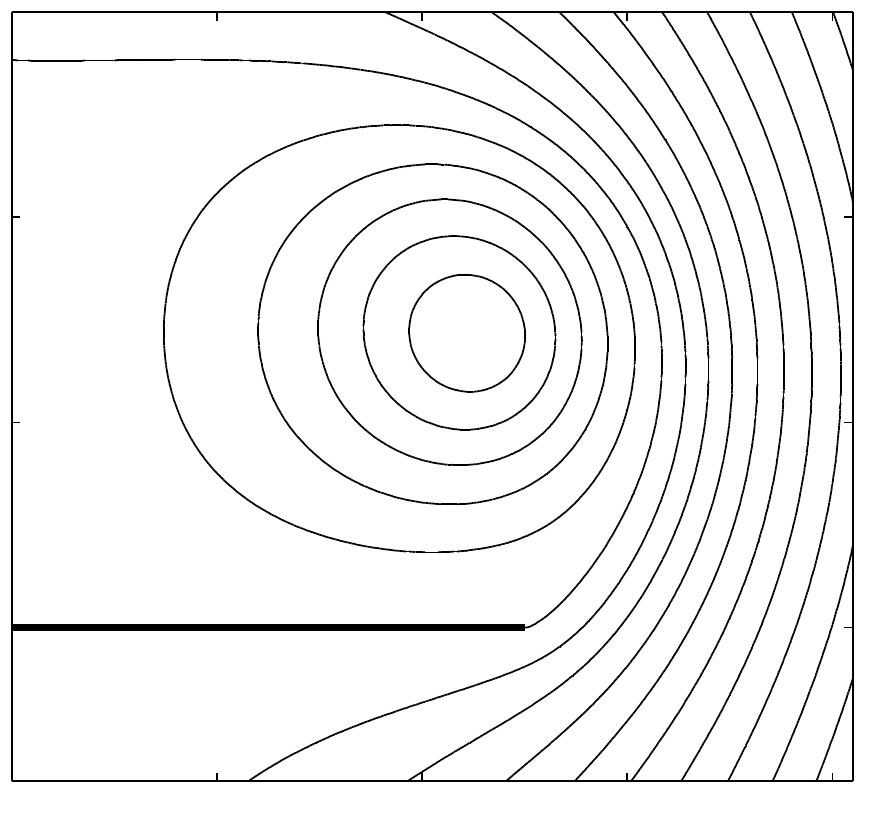}\\

\includegraphics[width=0.3505\textwidth]{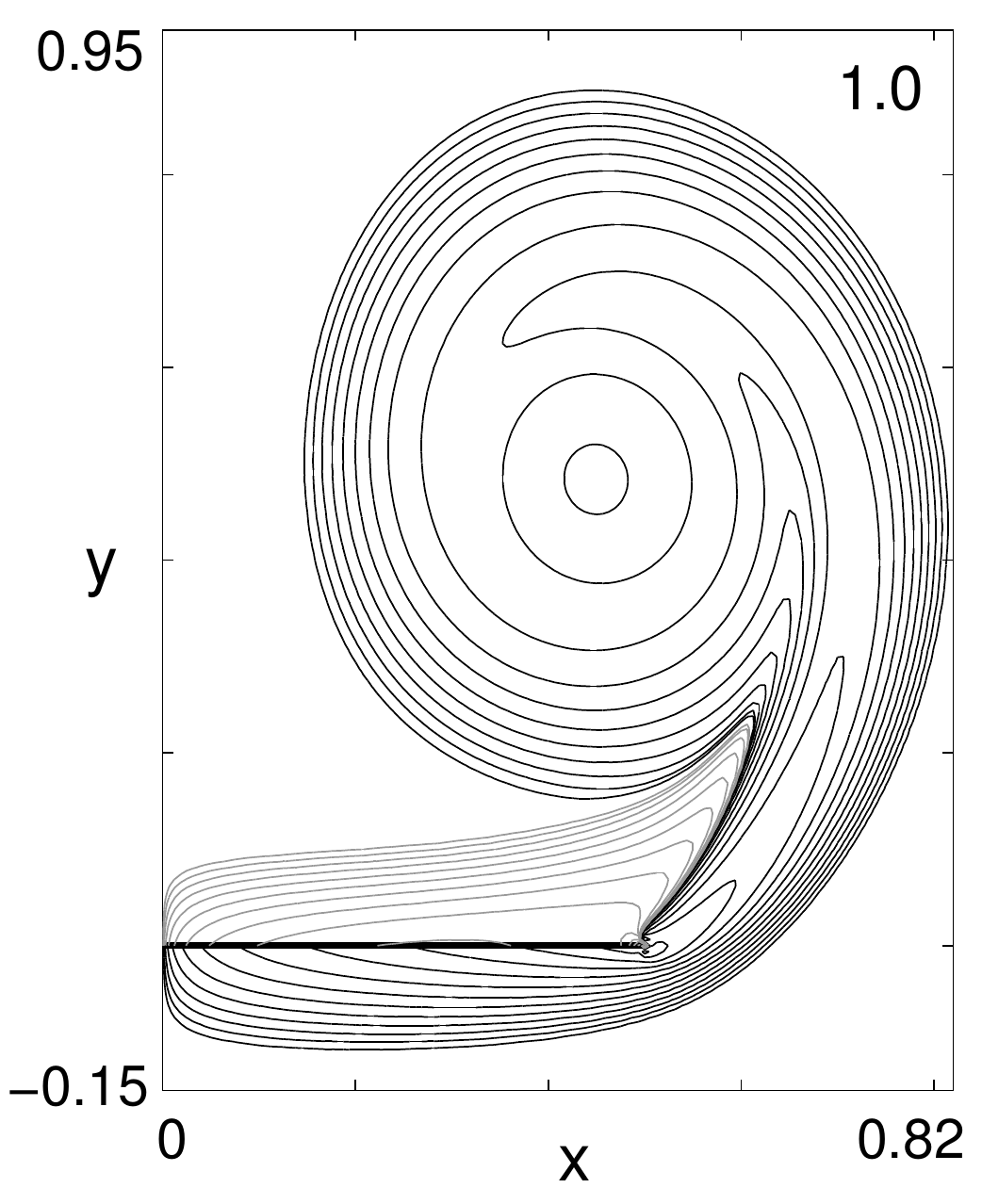}
\includegraphics[width=0.298\textwidth]{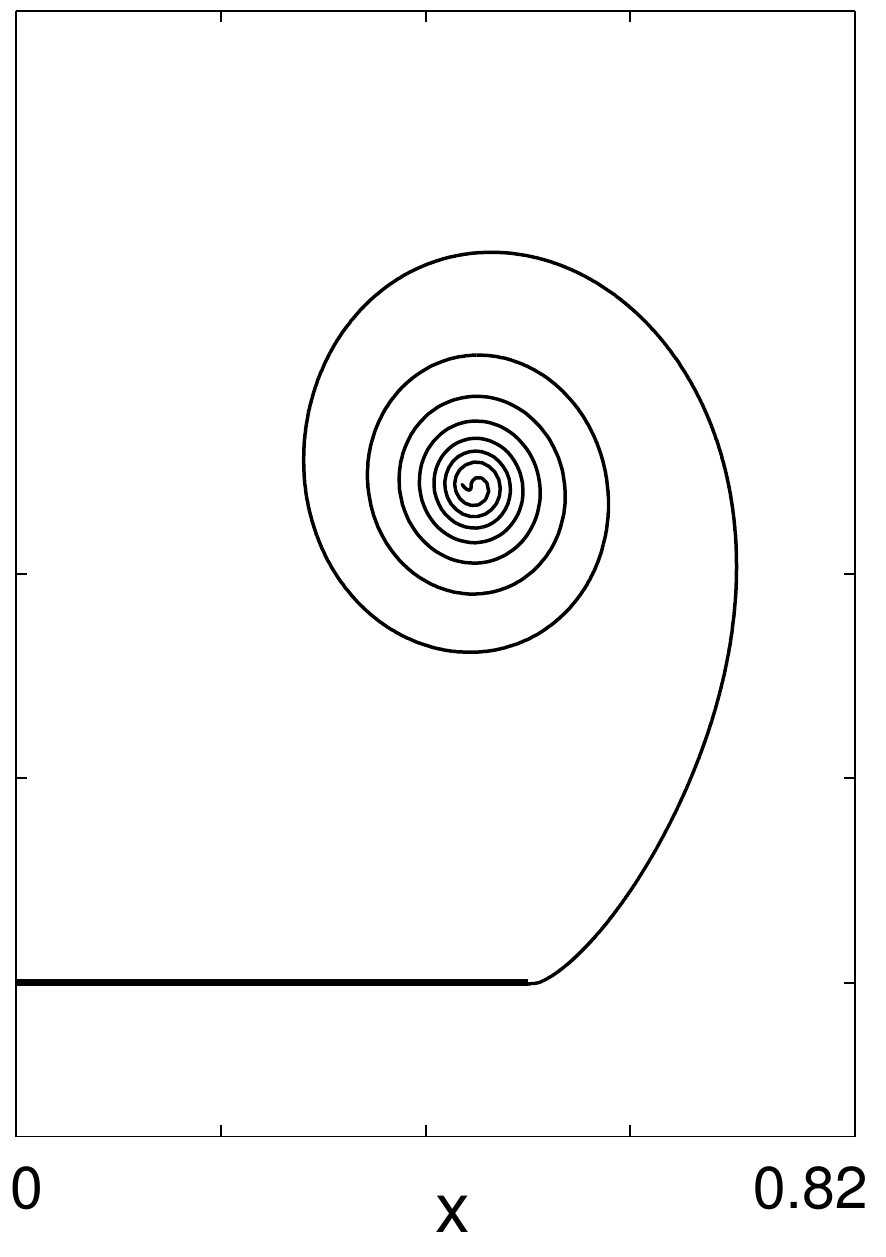}
\includegraphics[width=0.298\textwidth]{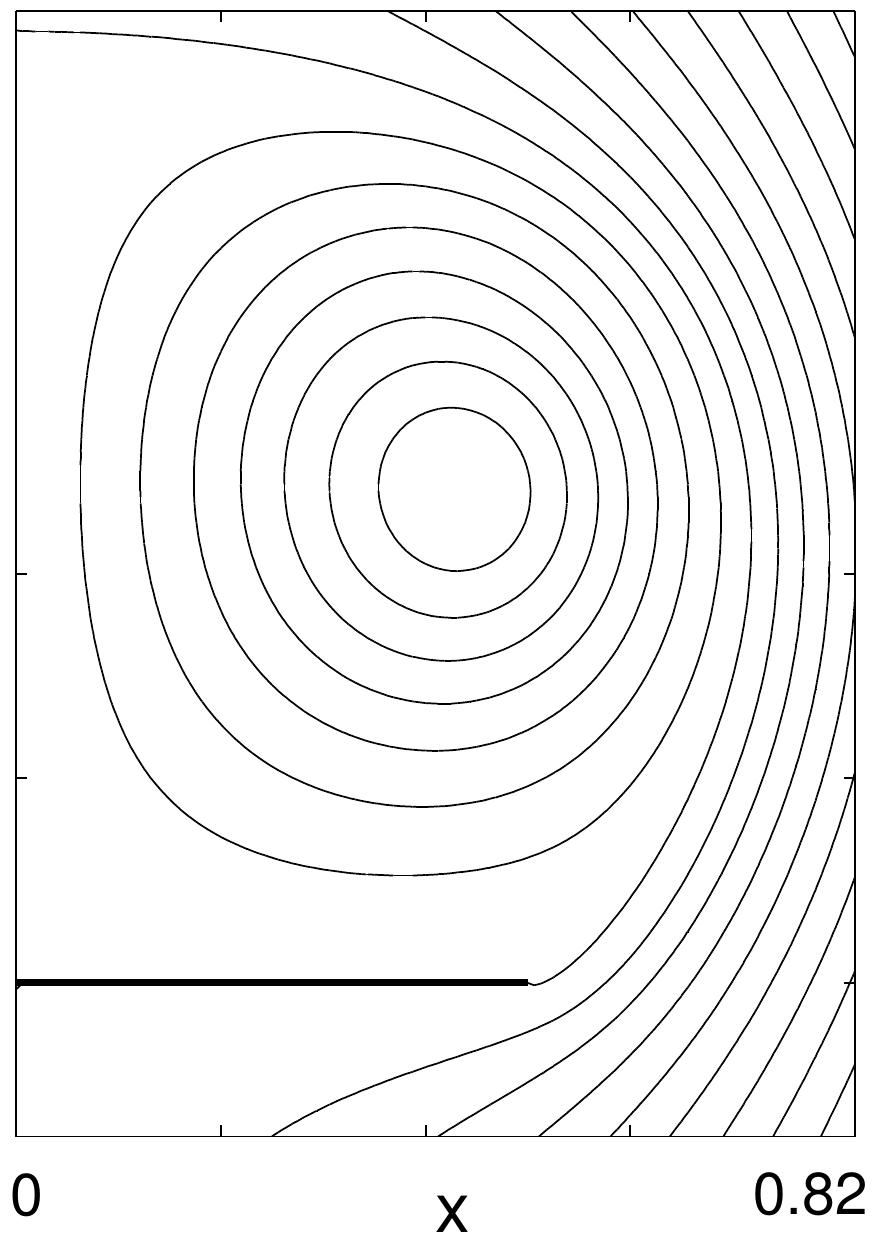}\\

 \caption*{
\centering (for caption see next page)
}
\end{figure}


\begin{figure}
 \centering

\includegraphics[width=0.3075\textwidth]{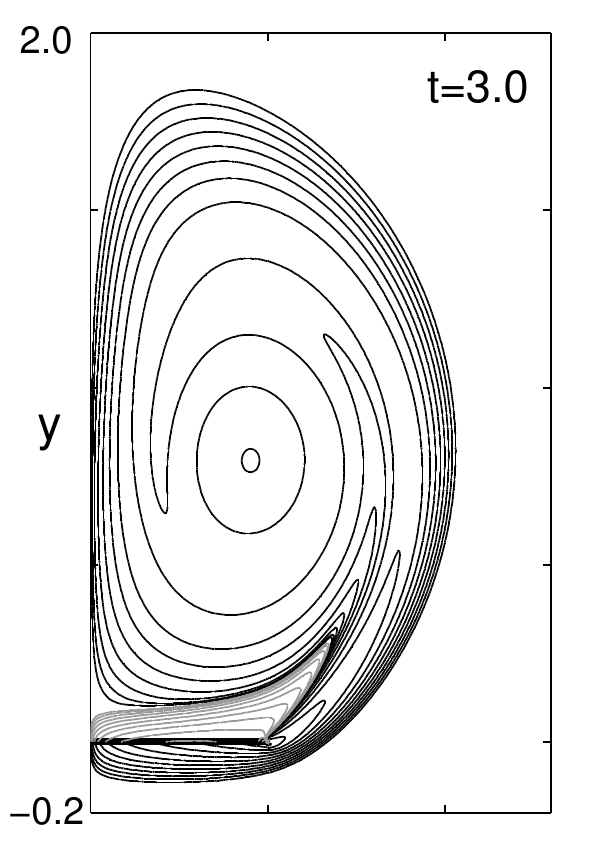}
\includegraphics[width=0.27\textwidth]{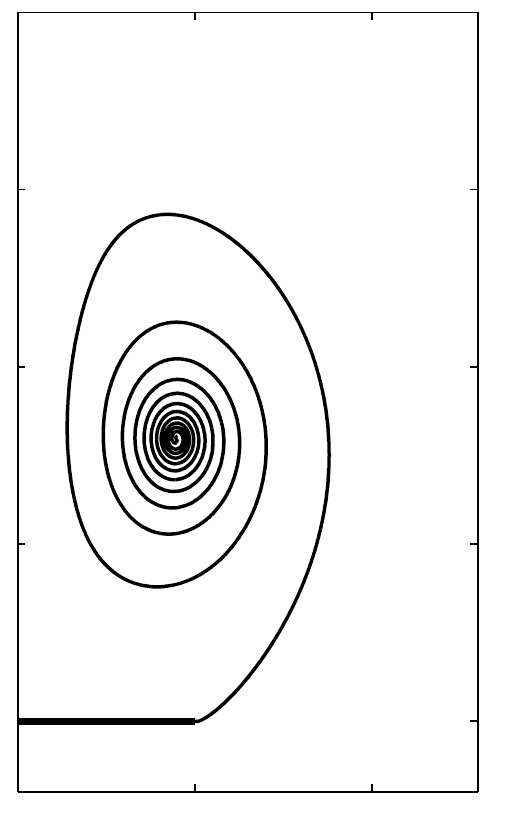}
\includegraphics[width=0.27\textwidth]{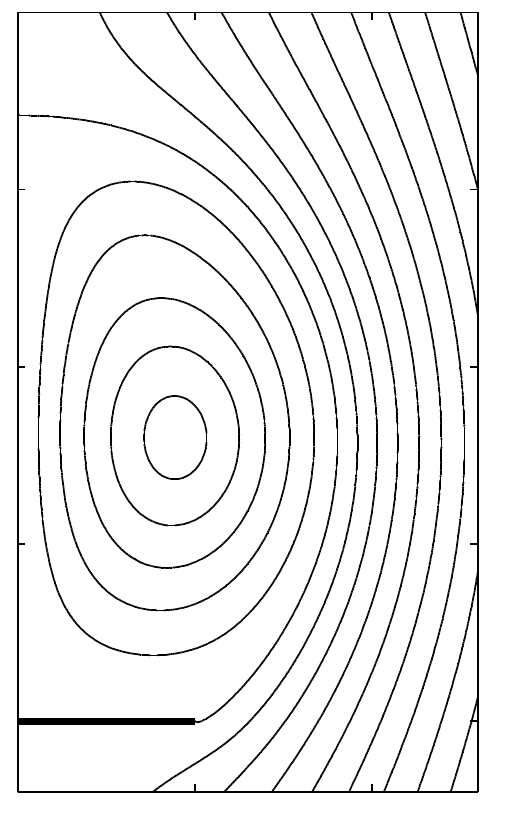}\\

\includegraphics[width=0.3075\textwidth]{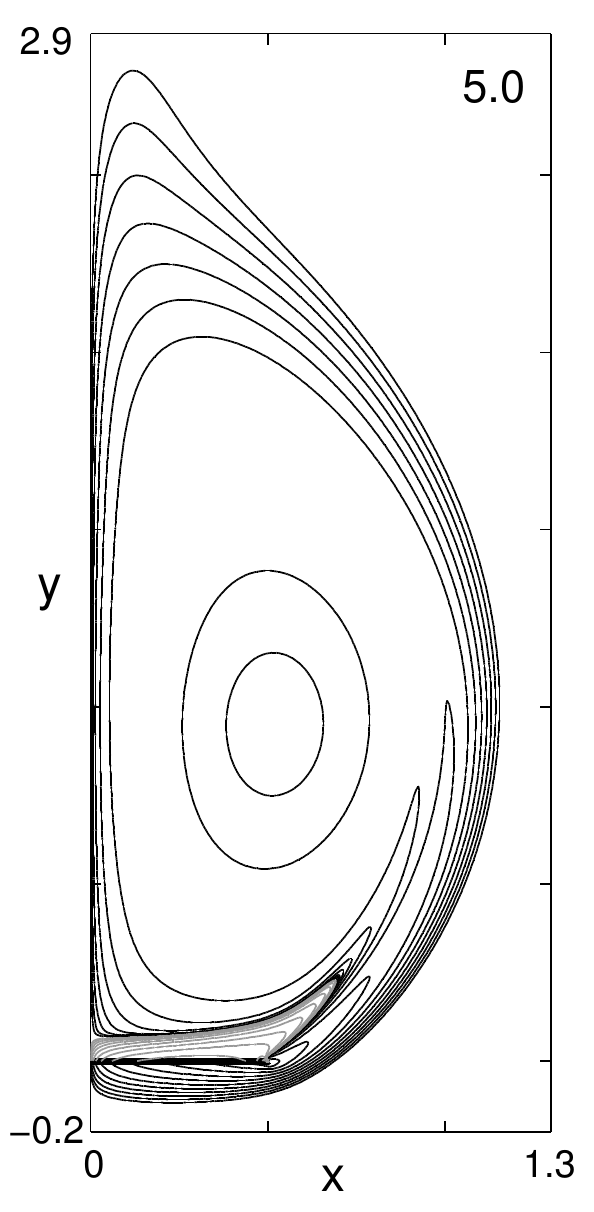}
\includegraphics[width=0.27\textwidth]{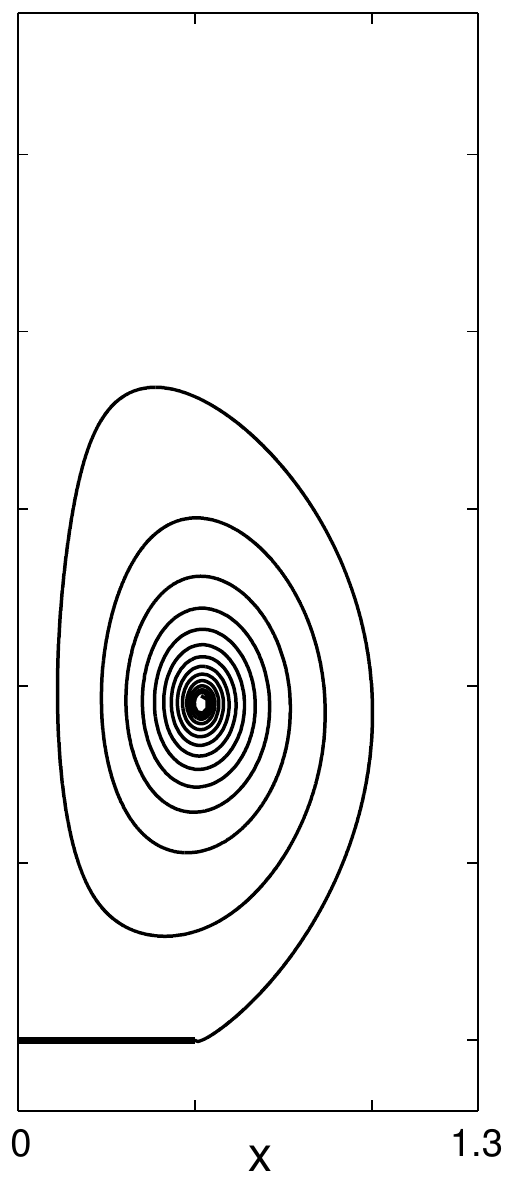}
\includegraphics[width=0.27\textwidth]{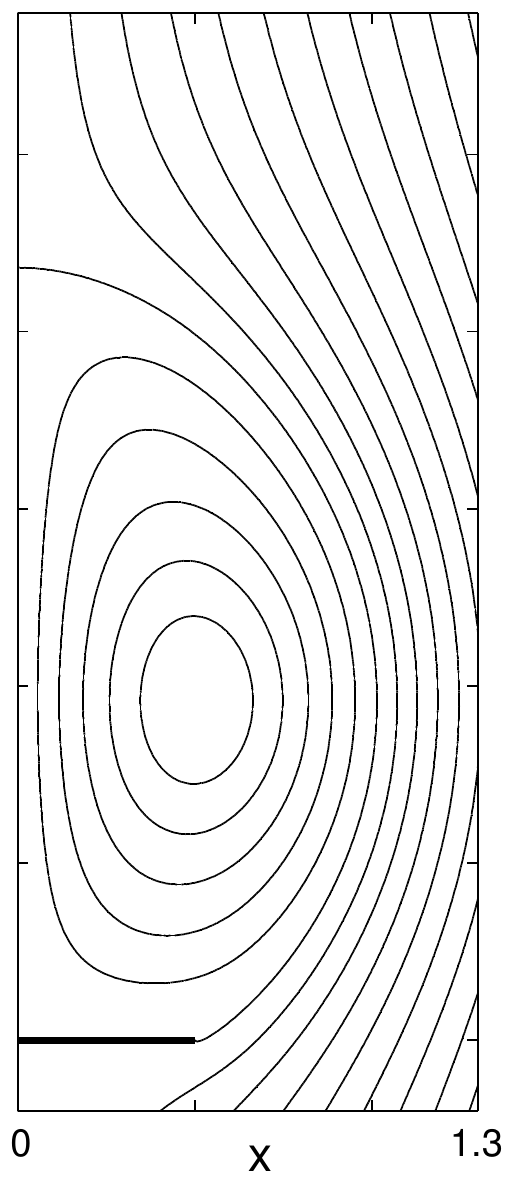}\\

\caption{
Vorticity, streaklines, and streamlines, for $Re=500$ at a sequence of 
times, $t=0.005,0.02,0.04,0.1,0.2,0.5,1.0,3.0,5.0$, as indicated.
The vorticity contour levels are 
$\omega = \pm 2^{[-5:12]}$. The stream function levels are 
$\psi = [-1:0.05:1]$ for $t\le 1$ (previous page) and
$\psi = [-1:0.1:1]$ for $t=3,5$ (this page).}
\label{F:evol}
\end{figure}



The driving far field flow $\ubinf$ is the potential flow
moving upwards past and around the plate. 
Initially, the flow generates a boundary layer of positive vorticity 
along 
both the upstream and downstream sides of the right half-plate.
Upstream vorticity is convected downstream,
concentrating near the tip as a vortex that grows in time.
The vortex entrains nearby vorticity, while vorticity  further away
is swept from the vortex towards the axis, thus depleting the region
in between.
As a result, the leading vorticity,
which initially is connected to the downstream boundary layer,
begins to separate from it.
At some time between t=0.2 and 0.5,
the positive vorticity in the leading vortex has completely 
separated from the positive boundary layer vorticity,
resulting in a more clearly defined starting vortex. 


The leading vortex 
induces a region of recirculating flow that can be seen 
in the corresponding streamlines.
The region of recirculating flow 
forms immediately after the motion begins. The fluid within this 
region, below the center of rotation, flows in direction 
opposite to the starting flow, and generates negative
vorticity attached to the wall.
In the computations, the negative vorticity is observed 
already after a few timesteps. 
It is barely visible in figure \ref{F:evol}, at t=0.005, but grows in time and
is clearly discernible by the grey contours at later times.
The negative vorticity region 
grows horizontally along the
plate, 
away from the starting vortex, until
it reaches the axis at $t\approx 0.113$, as will be shown later.
At the same time, the negative vorticity region is stretched and 
entrained into the leading vortex.
As the negative vorticity layer thickens, the positive boundary layer vorticity
above it diffuses, until eventually, after $t=1.0$ shown here, 
all downstream boundary layer vorticity is negative.
Around time $t=3$, the positive vorticity in the starting vortex
has reached the axis of symmetry, $x=0$. It diffuses out of the
recirculation region, so that at $t=5$, much vorticity is outside
the enclosing streamline and moves upwards away from the
plate.  The results at the larger times presented here
are in good agreement with results shown by \cite{koushiels96}.

\begin{figure}
\centering
\includegraphics[height=0.35\textwidth]{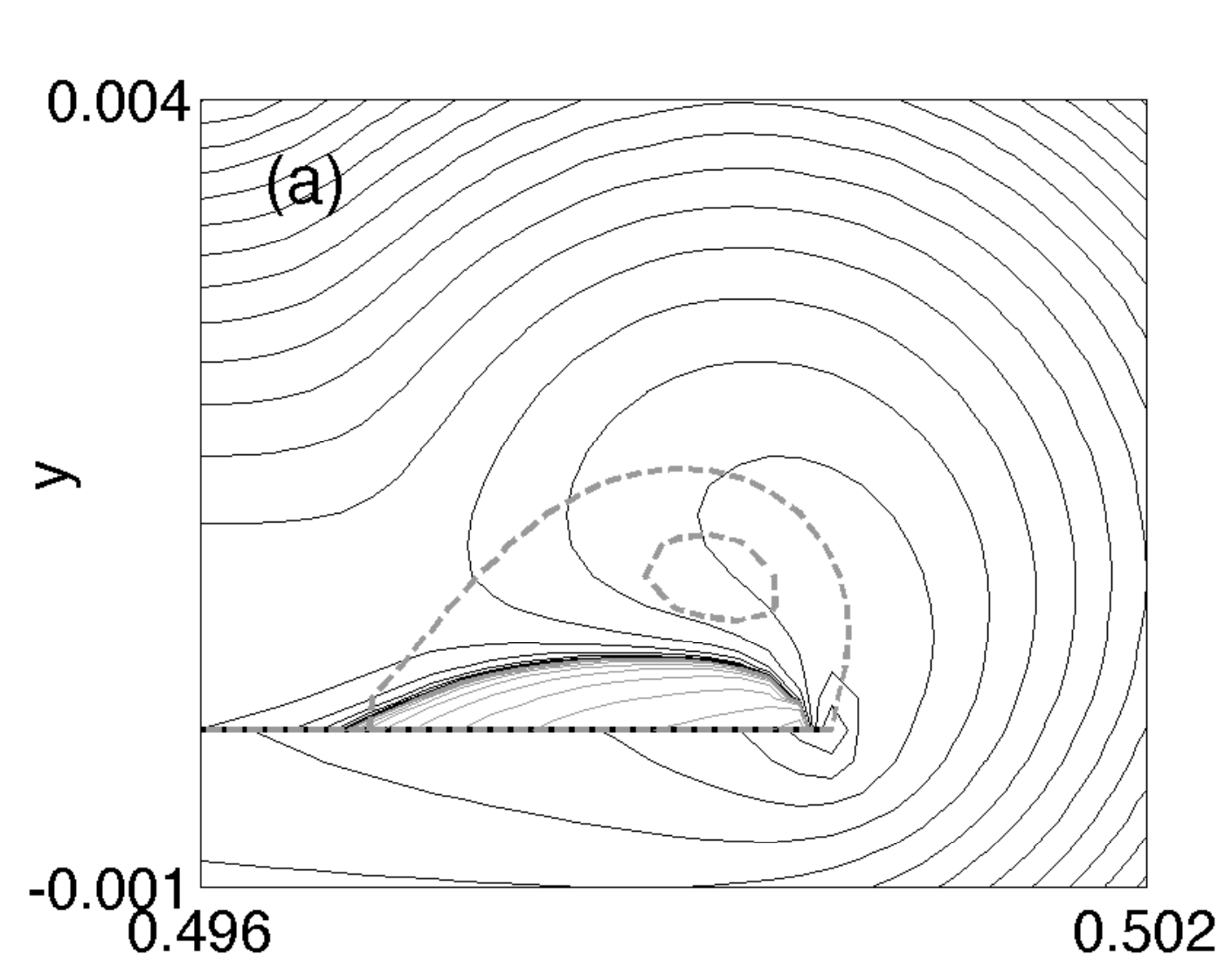}
\includegraphics[height=0.352\textwidth]{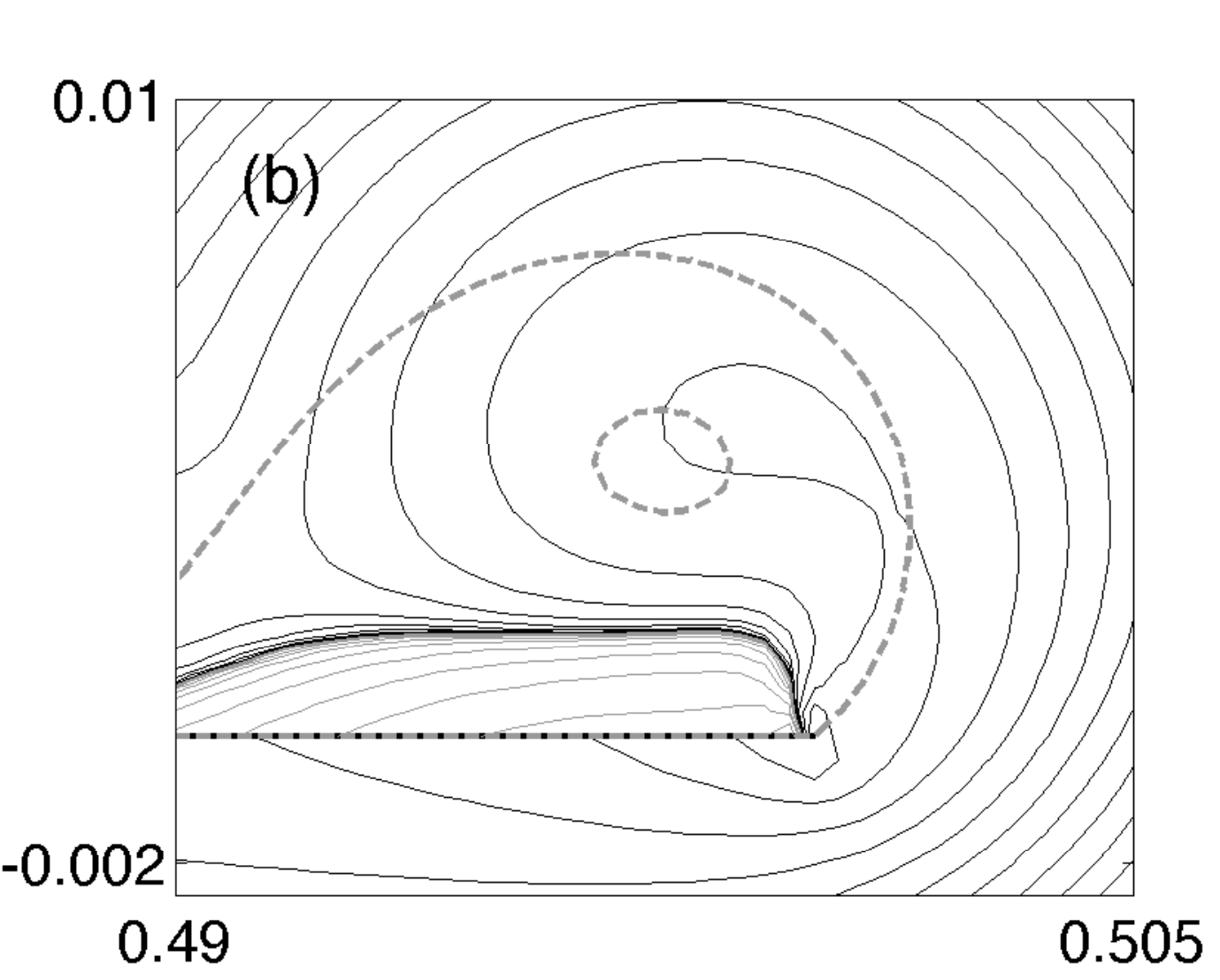}\\
\includegraphics[height=0.369\textwidth]{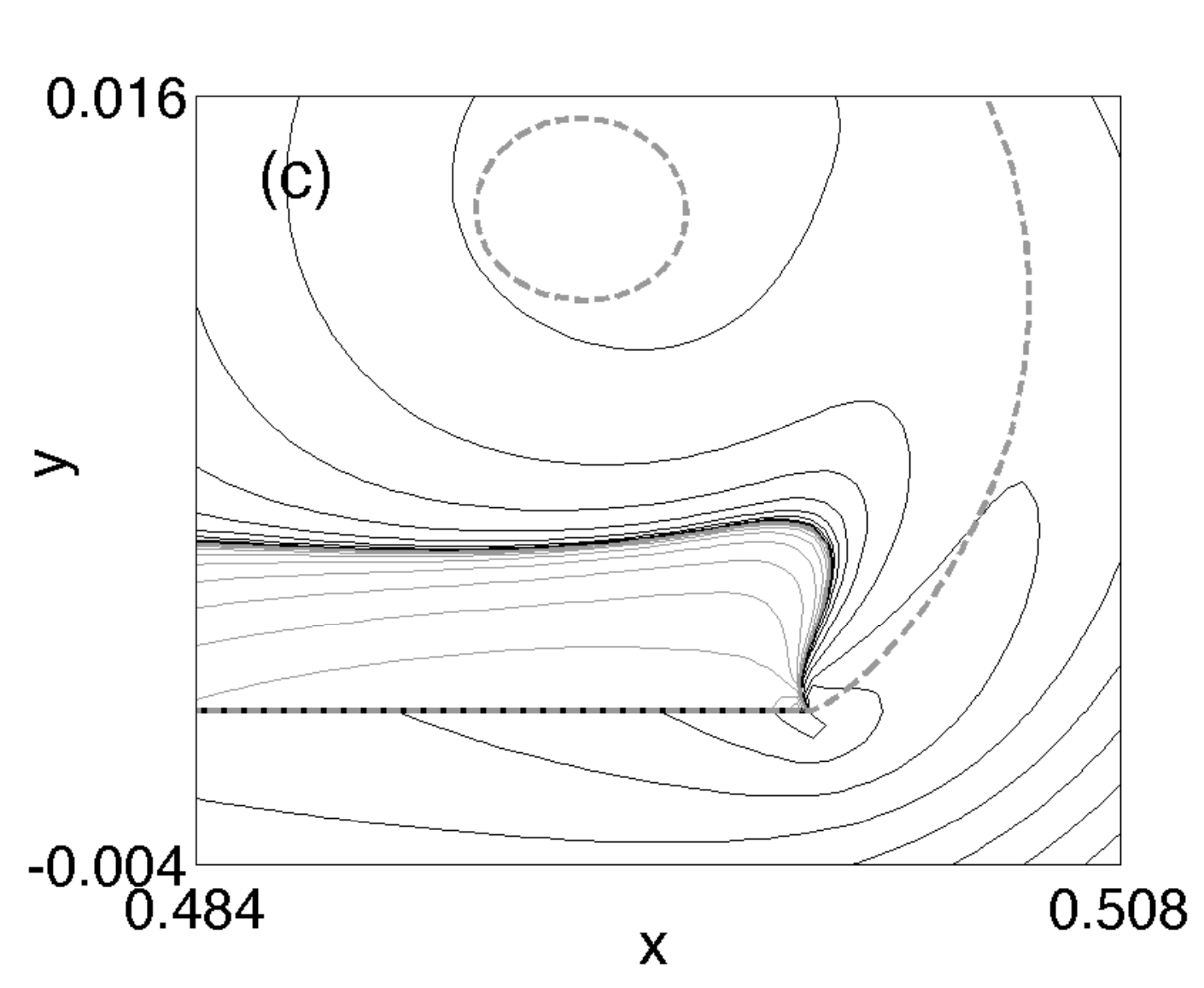}
\includegraphics[height=0.35\textwidth]{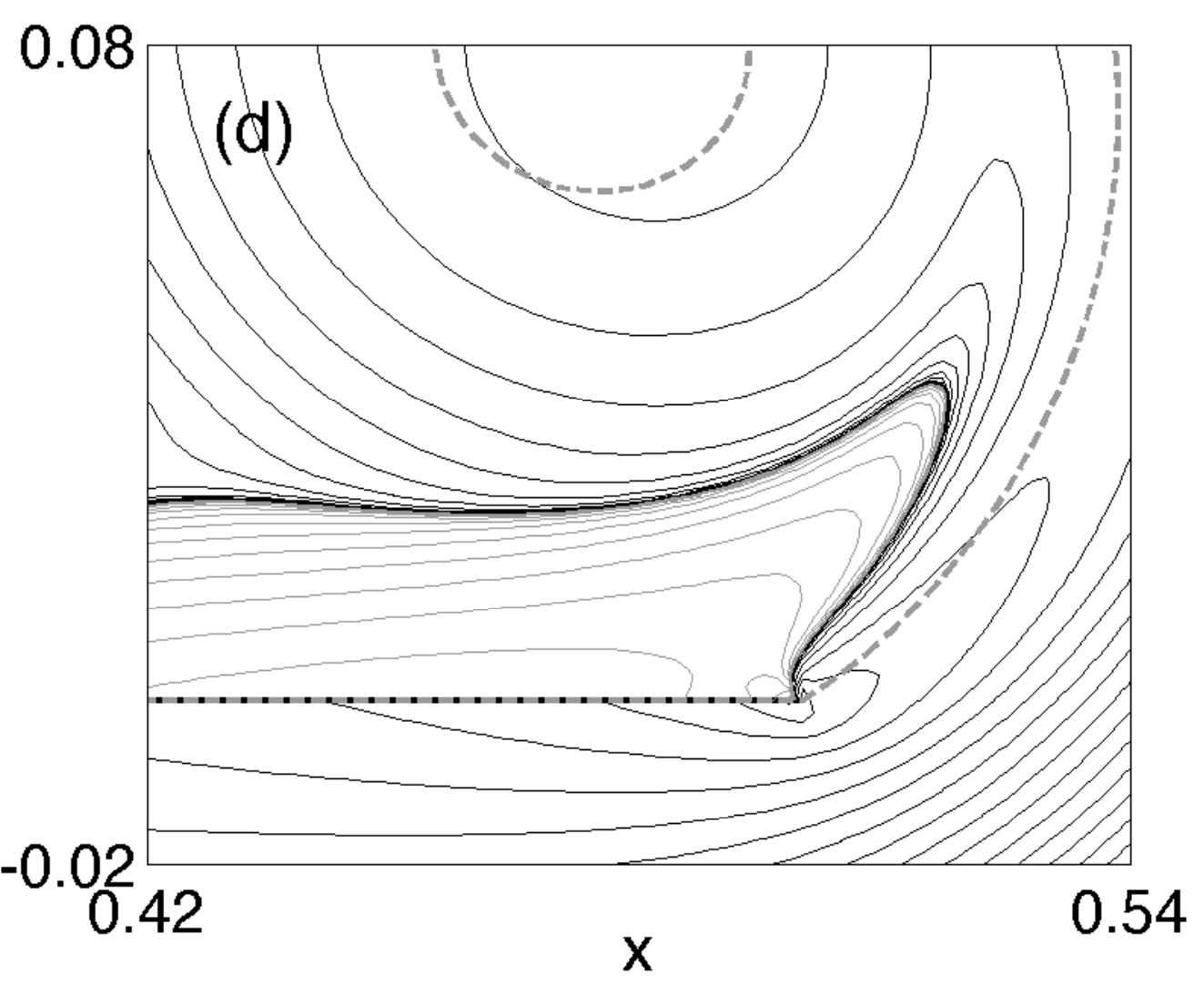}
 \caption{Closeup of vorticity contours near the plate tip,
for $Re=500$, $p=0$, at (a) $t=0.0002$, (b) $t=0.0016$, (c) $t=0.008$ and 
(d) $t=0.1$. Vorticity contour levels are $\pm 2^{[-8:15]}$.
}
\label{F:closeup}
\end{figure}

The streaklines, shown in the middle column of figure \ref{F:evol}, 
are obtained by releasing 
a particle at a point near the tip at each timestep, and computing its
evolution with the fluid velocity. At a given time, 
the figure shows
the position of all particles at that time that were released previously.
The streakline plots thus mimic what one would observe in a laboratory 
experiment if dye were continously released at a point near the tip.
%
Each released particle circulates around the vortex center. Particles that
have been released earlier travel closer to the center and thus,
the resulting streakline has a spiral shape. 
The spiral tightens near the center and the number of spiral turns increase in time.
The maximum vorticity near the tip of the plate is convected with 
the particles along the streakline, and diffuses. Thus
the streakline is a good indicator of the centerline of 
the separated shear layer, but not of the overall vortical
region, or of the recirculation region,
both which extend beyond the region occupied by the spiral.
At the times shown, the spiral center is a good indicator of 
the vorticity maximum in the vortex core,
and of the center of fluid rotation.

The streamlines, in the right column, show the 
region of recirculating flow.
This region is enclosed by the $\psi=0$
streamline, which leaves the tip of the plate and reattaches 
on the downstream side, at a short distance behind the vortex.
As the recirculation region 
grows, the enclosing streamline $\psi=0$ first reaches the axis,
between time $t=0.1$ and $0.2$,
and then continues to 
move up along the centerline, $x=0$. It then forms the familiar
rounded symmetric recirculation bubble downstream of the plate,
as observed experimentally and computationally,
before the flow looses its symmetry at later times 
(see, eg, van Dyke 1982, figure 64, and Koumoutsakos \& Shiels 1996, figure 18).


 \begin{figure}
 \centering
\includegraphics[height=0.36\textwidth]{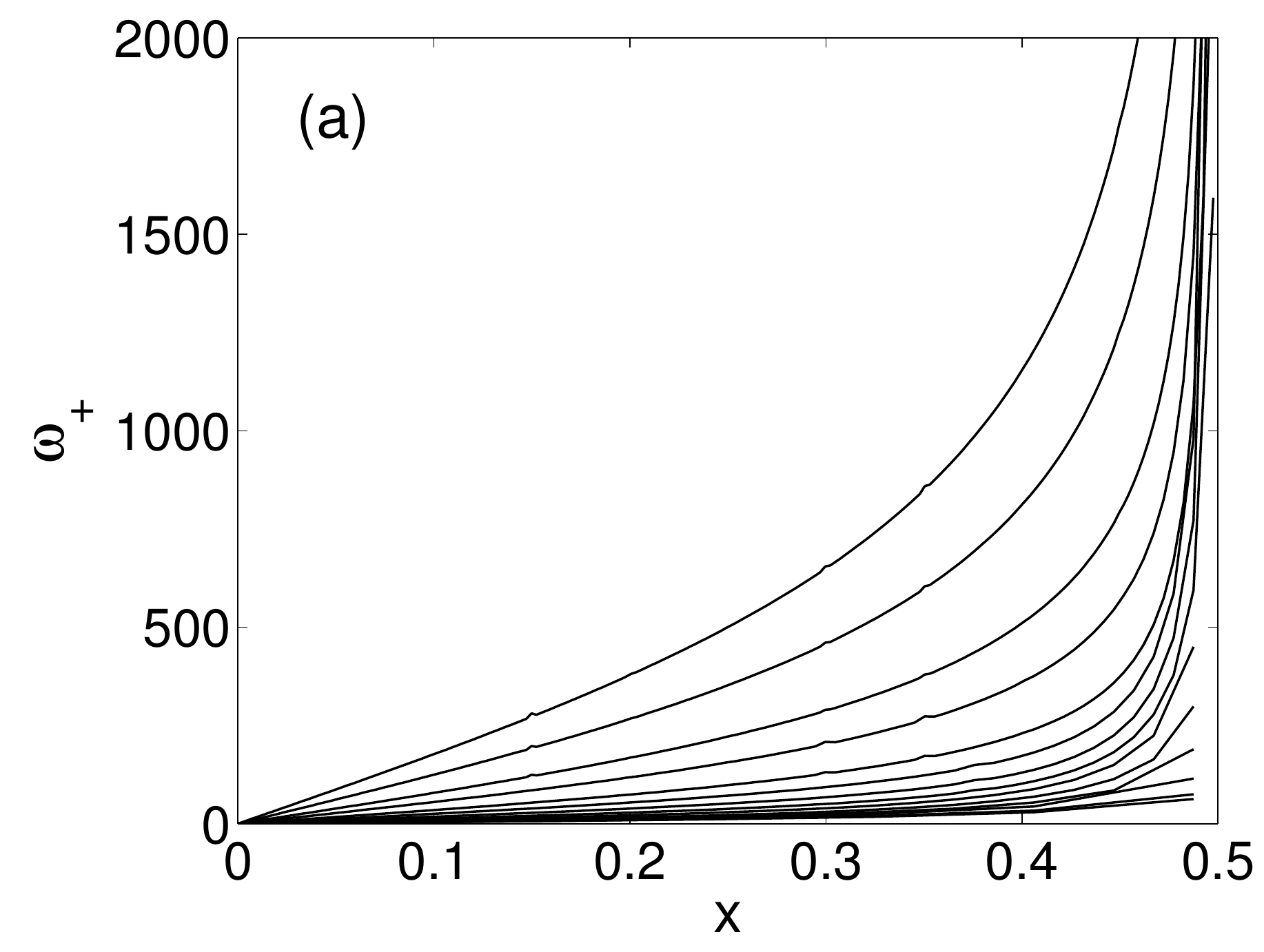}
\includegraphics[height=0.36\textwidth]{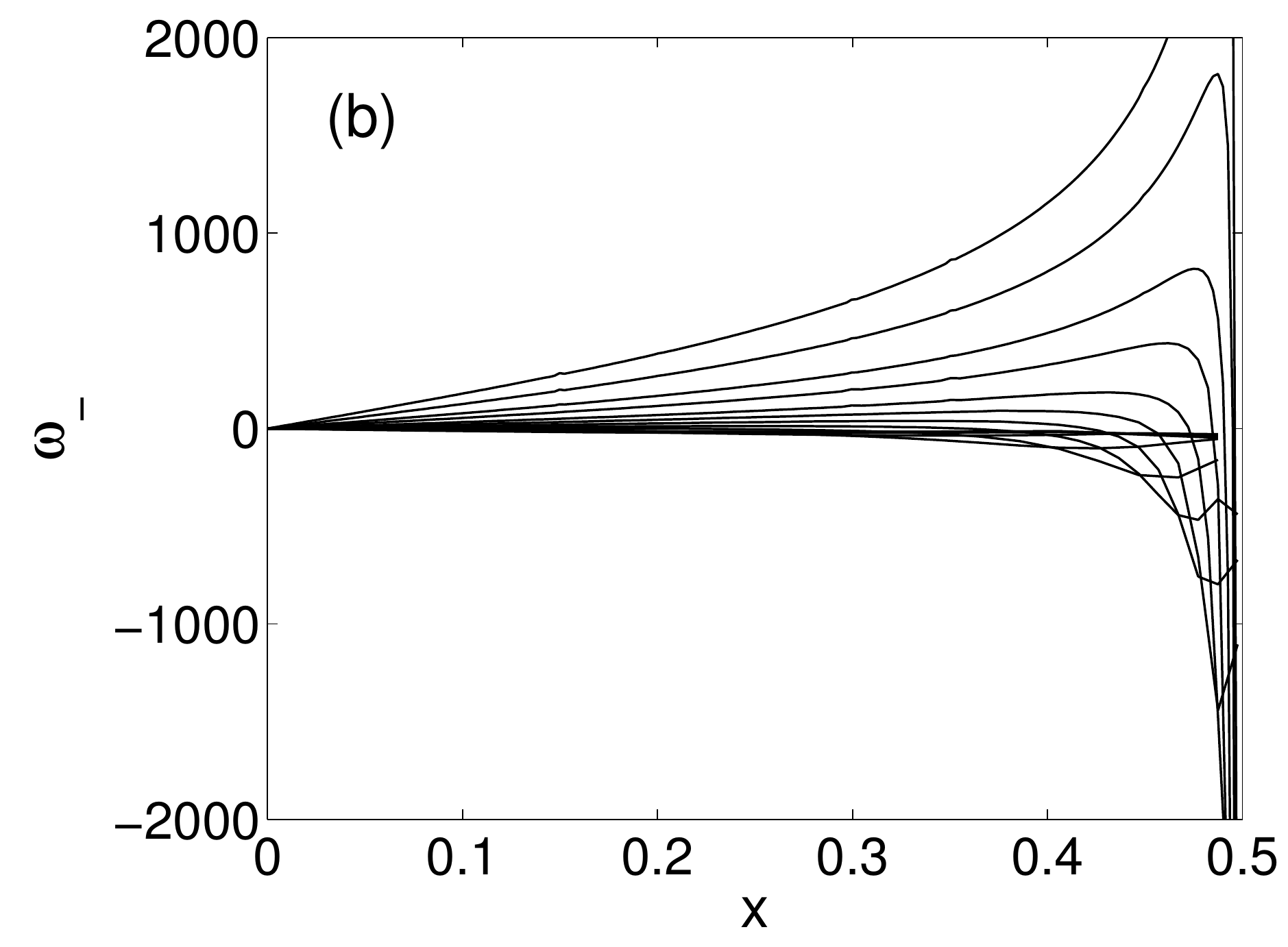}
\caption{
Wall vorticity (a) $\omega_+(x)$, and (b) $\omega_-(x)$,
on the upstream and downstream sides of the plate, respectively,
\vs $x$, at a sequence of times 
$t=$0.0002, 0.0004, 0.001, 0.002, 0.005, 0.01, 0.02, 
0.04, 0.08, 0.2, 0.5, 1, 3, 5, for  $Re=500$.
The vorticity decreases in magnitude as time increases. 
}
\label{F:wallvort}
%

 \centering
\includegraphics[width=0.5\textwidth]{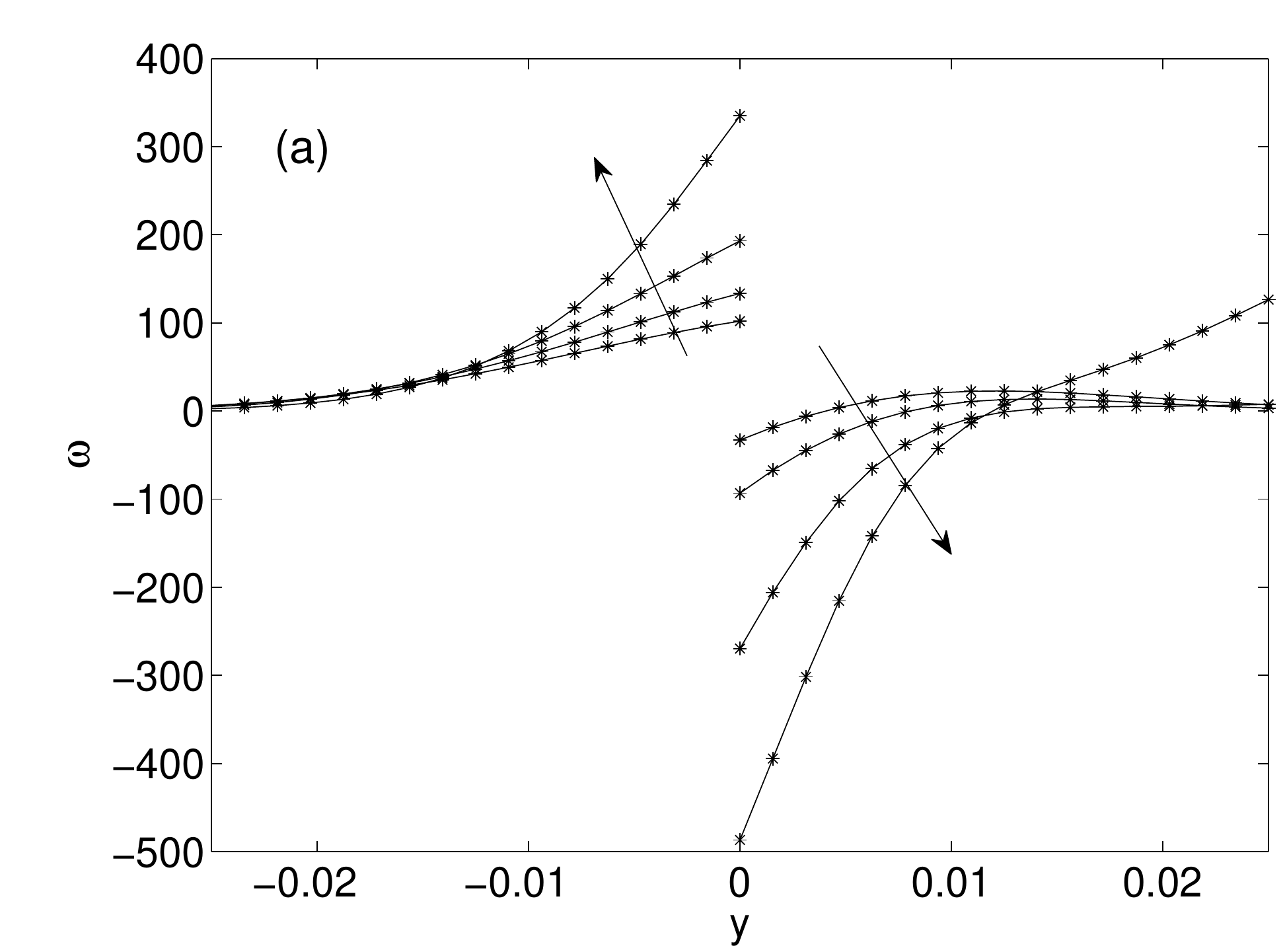}
\caption{Vorticity $t=0.04$ along the lines $x$=0.4, 0.425, 
0.45 and 0.475, as functions of $y$.
The arrows indicate
increasing values of $x$.}
\label{F:vortacross}
\end{figure}

Figure \ref{F:closeup} shows a closeup of the vorticity contours 
near the tip of the plate, plotted as the black and grey 
solid contour lines. It also shows two of the flow streamlines 
as dashed curves. One is the $\psi=0$ streamline enclosing
the region of recirculating flow, the other is a closed streamline 
near the center of rotation.
Figure \ref{F:closeup}(a) shows that very early, at $t=0.0002$,
the negative vorticity region is already well-formed, all along
the wall inside the recirculation region.
Positive vorticity has begun to concentrate downstream, near the tip of the plate,
but it does not yet have a local vorticity maximum that could identify 
a vortex core. On the other hand, the 
recirculation region is well-formed and has a well-defined
center of rotation. Thus, this early on, the center of rotation does
not agree with a maximum in 
the core vorticity.
%
Figure \ref{F:closeup}(b), which plots the solution a little 
later, at $t=0.0016$, shows 
a local vorticity maximum emerging in the center of the leading vortex.
This local maximum 
remains well defined and grows in time, as  
shown in figures \ref{F:closeup}(c,d).
As time increases, the center of rotation and the position of
the vorticity maximum are in better agreement.
Figure \ref{F:closeup}(d) also shows the entrainment
of the negative vorticity by the leading vortex. 

 \begin{figure}
 \centering
\includegraphics[width=0.423\textwidth]{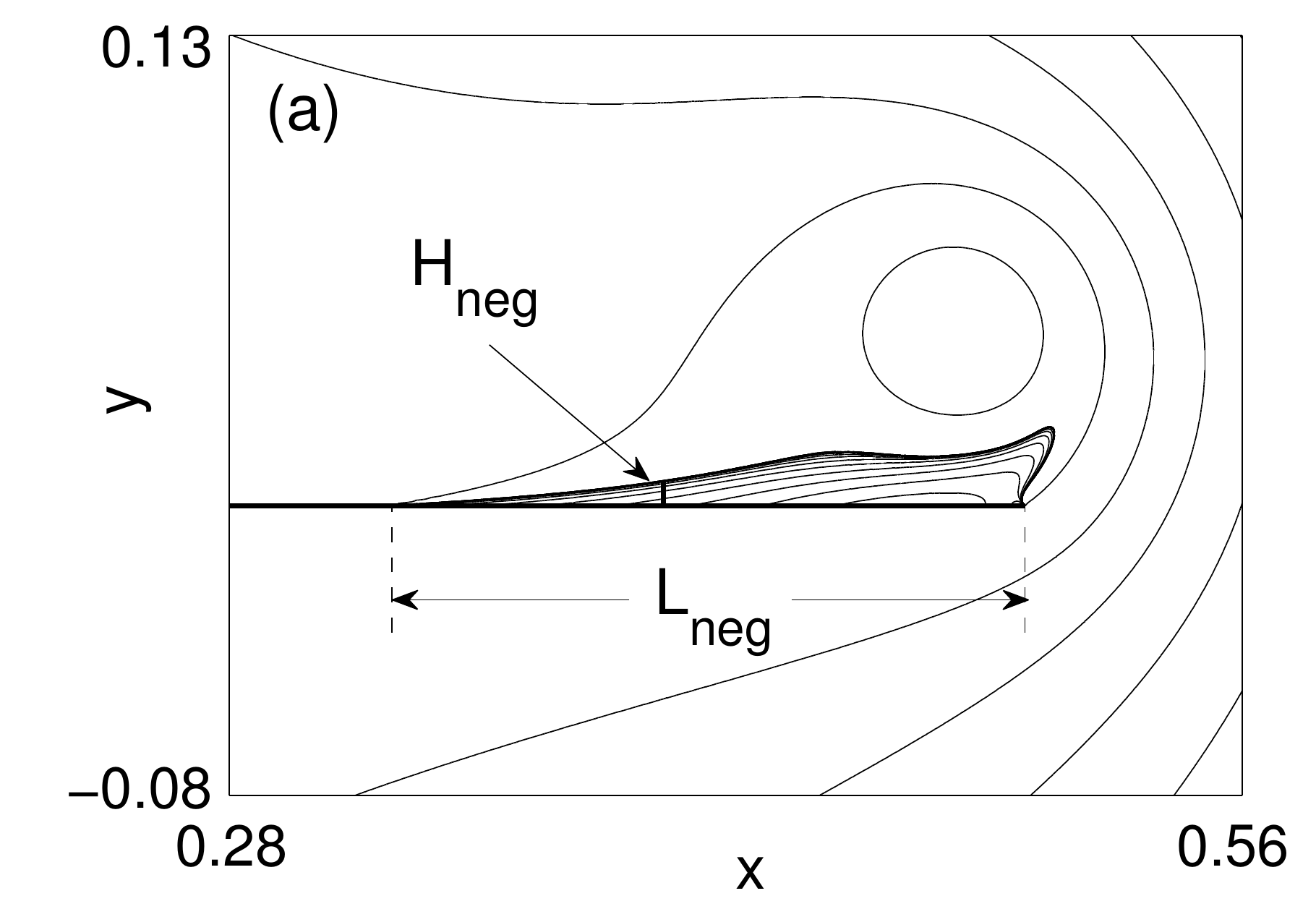}
\includegraphics[width=0.42\textwidth]{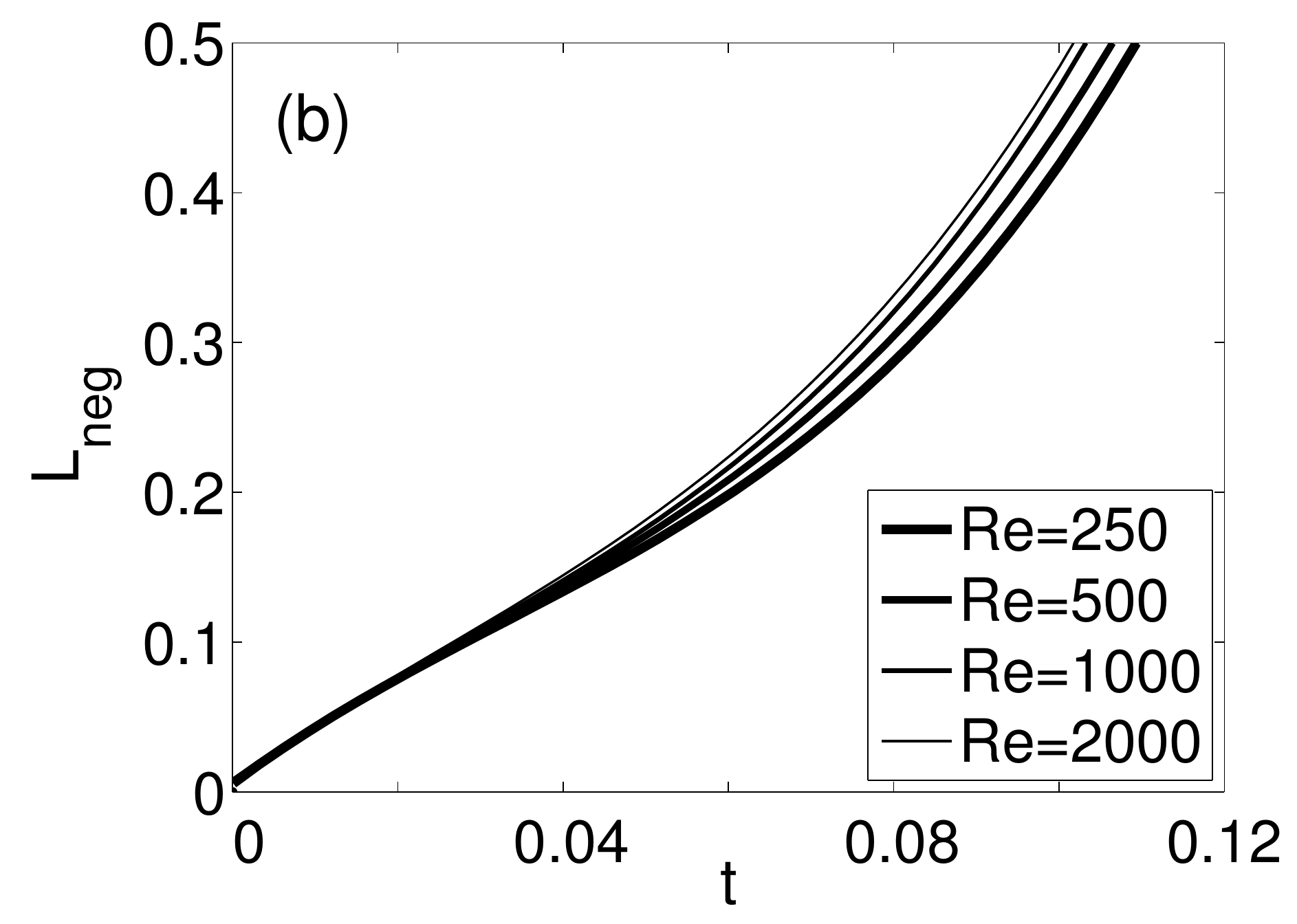}\\
\includegraphics[width=0.42\textwidth]{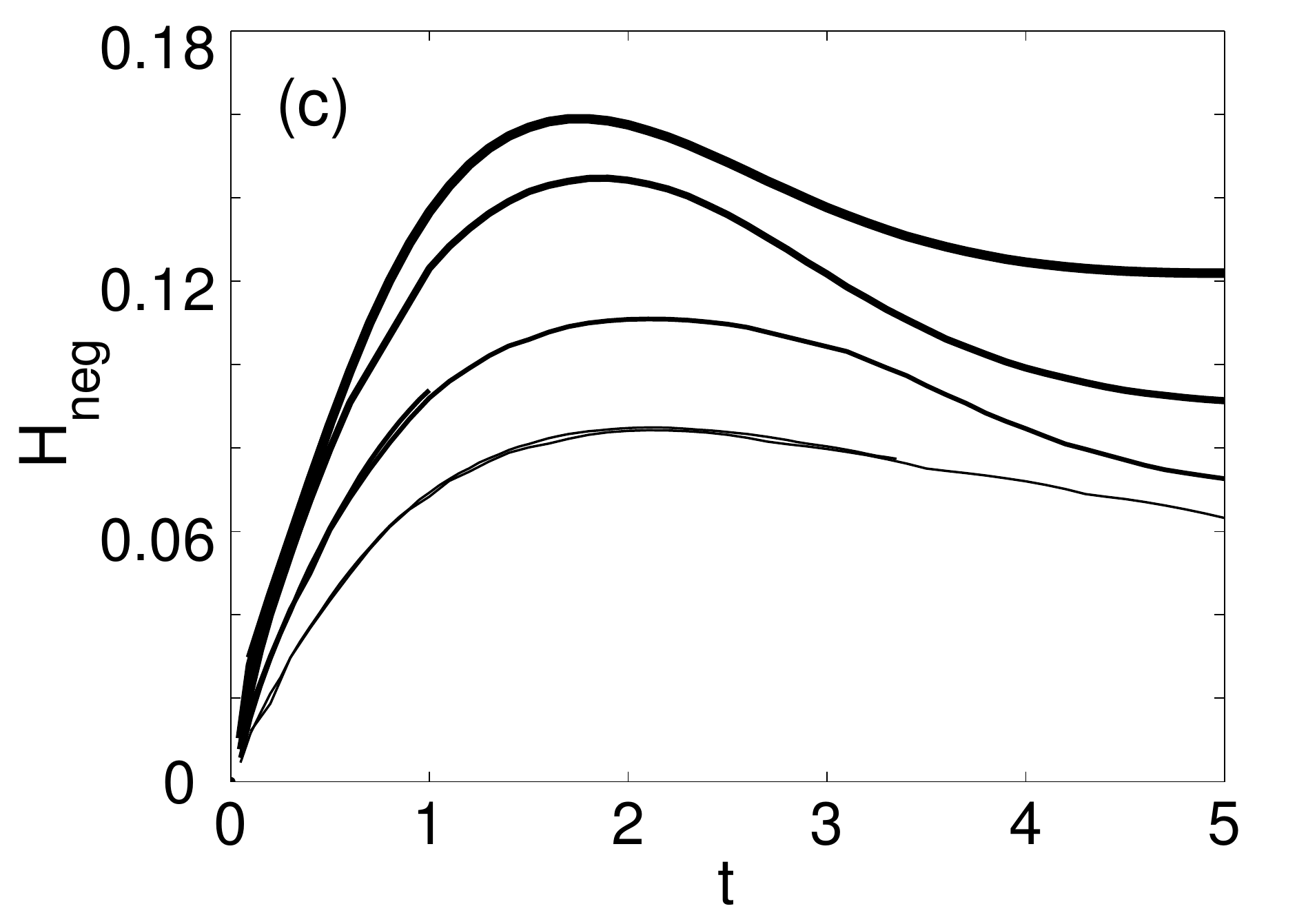}
\includegraphics[width=0.42\textwidth]{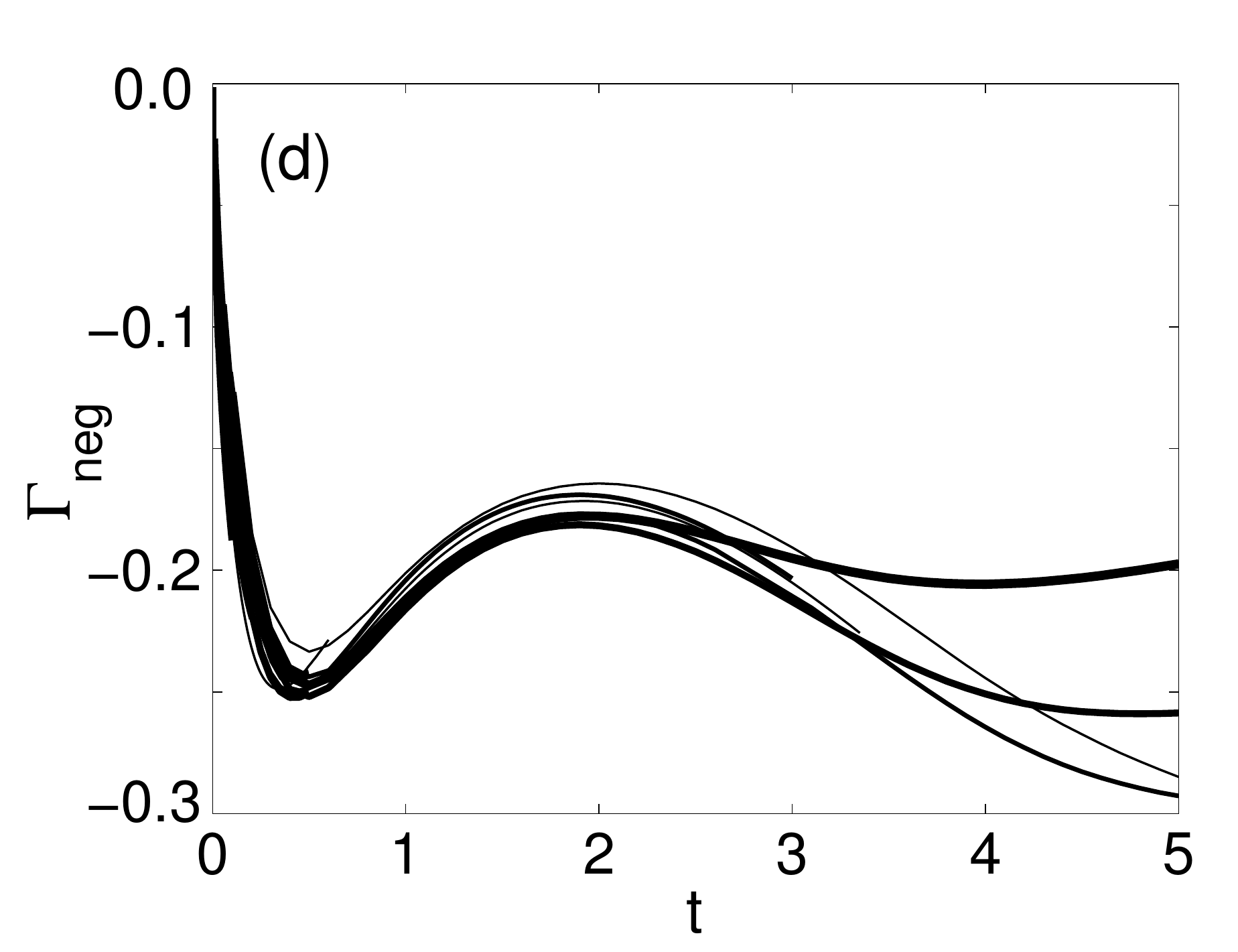}
\caption{Quantifiers of negative vorticity region.
(a) Negative vorticity contours and streamlines at $t=0.04$,
showing definition of length $L_{neg}$ and thickness $H_{neg}$.
(b) Length $L_{neg}$ \vs t.
(c) Thickness $H_{neg}$ \vs $t$.
(d) Integral negative vorticity $\Gamma_{neg}$ \vs $t$.
Figures (b,c,d) show results for $Re=250,500,1000,2000$, as indicated
in the legend in (b).
}
\label{F:negvort}
\end{figure}


Figures \ref{F:wallvort}-\ref{F:negvort} show details of the vorticity profiles on and near the plate.
Figure \ref{F:wallvort}(a) plots the wall vorticity $\omega_+$ on the upstream side of the
plate at a sequence of times.
The vorticity is initially unbounded at the plate tip,
zero at the axis, and positive, increasing, in between. 
As time increases, it remains positive but decreases in magnitude.
Figure \ref{F:wallvort}(b) plots the wall vorticity $\omega_-$ on the downstream side of 
the plate.
At $t=0+$, the vorticity is also positive everywhere, 
but negative unbounded at the tip.
The interval along the plate in which the vorticity is negative grows,
and the maximal magnitude decreases.
The wall vorticity values are responsible for the drag forces
normal to the wall parallel to the background flow, 
presented later in \S 4.5.
Figure \ref{F:vortacross} plots the vorticity at a fixed time t=0.04,
along four vertical lines crossing the region of negative vorticity, 
$x=0.4$, 0.425, 0.45, 0.475, as a function of $y$. 
The arrows indicate the direction of increasing $x$.
The vorticity is positive on the upstream side of the plate ($y<0$), 
and negative downstream 
($y>0$). As $x$ increases towards the tip, the wall vorticity
and their gradients
$\partial \omega/\partial y$ 
increase in magnitude. 
The wall vorticity gradients are responsible for the lift 
forces parallel to the wall, normal to the background flow, presented in \S 4.5.

Figure \ref{F:negvort} gives more information on the region of 
negative vorticity.  It plots its 
length $L_{neg}$ and a characteristic thickness $H_{neg}$,
both of which are illustrated in figure \ref{F:negvort}(a),
and the integral negative vorticity $\Gamma_{neg}$.
For later reference, results are plotted for various Reynolds
numbers, as indicated in the legend in figure \ref{F:negvort}(b).
The length $L_{neg}$, plotted in figure \ref{F:negvort}(b),
is also the length of the recirculation region. 
It increases until it reaches $L_{neg}=0.5$, which is when
the recirculation region reaches the axis.
For $Re=500$, this occurs at $t\approx 0.113$, 
for larger $Re$, it occurs sooner. 
After this time, negative vorticity covers all of the downstream plate wall.
%
The thickness $H_{neg}$,
plotted in figure \ref{F:negvort}(c),
is chosen to be the thickness of the negative vorticity region at $x=0.4$.
It increases initially, reaches a maximum, and
then decreases again as the negative vorticity is 
entrained by the leading vortex.
Larger Reynolds number flows have thicker regions of negative vorticity.
The behaviour at places other than $x=0.4$ is similar. 
The integral negative vorticity in the right hand plane, 
\begin{equation}
\Gamma_{neg} =\int_{\omega<0\atop x\ge 0} \omega \, dA~,
\end{equation}
is plotted in figure \ref{F:negvort}(d). 
Remarkably, even though the layer thickness $H_{neg}$ depends heavily 
on $Re$, its integral circulation is practically independent of $Re$, 
at least until about $t=3$. 
After that, diffusion causes the integral 
negative vorticity to decrease in magnitude, with larger decrease for 
lower $Re$.
%


\subsection{Dependence on $Re$}

\begin{figure}
\centering
   \includegraphics[height=1.701truein]{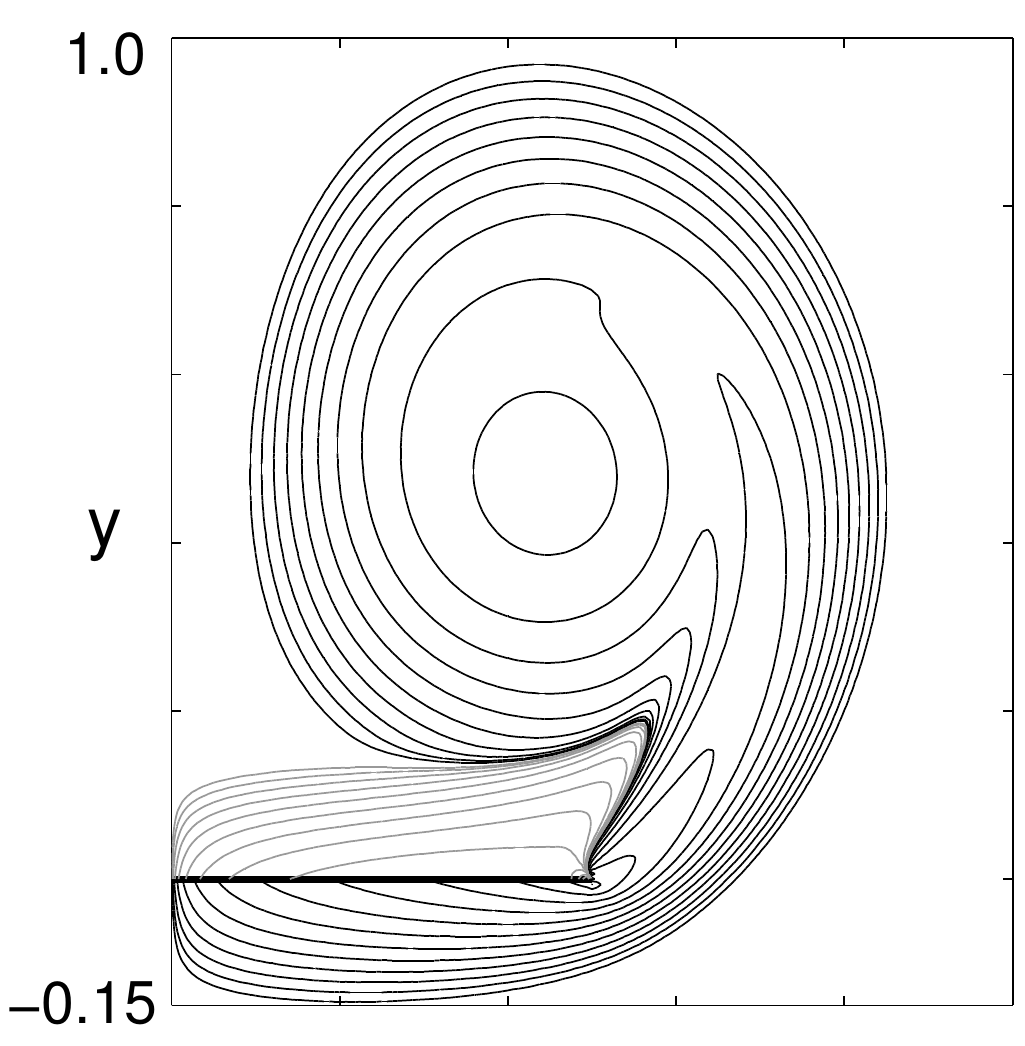}
\hskip4.6pt
   \includegraphics[height=1.701truein]{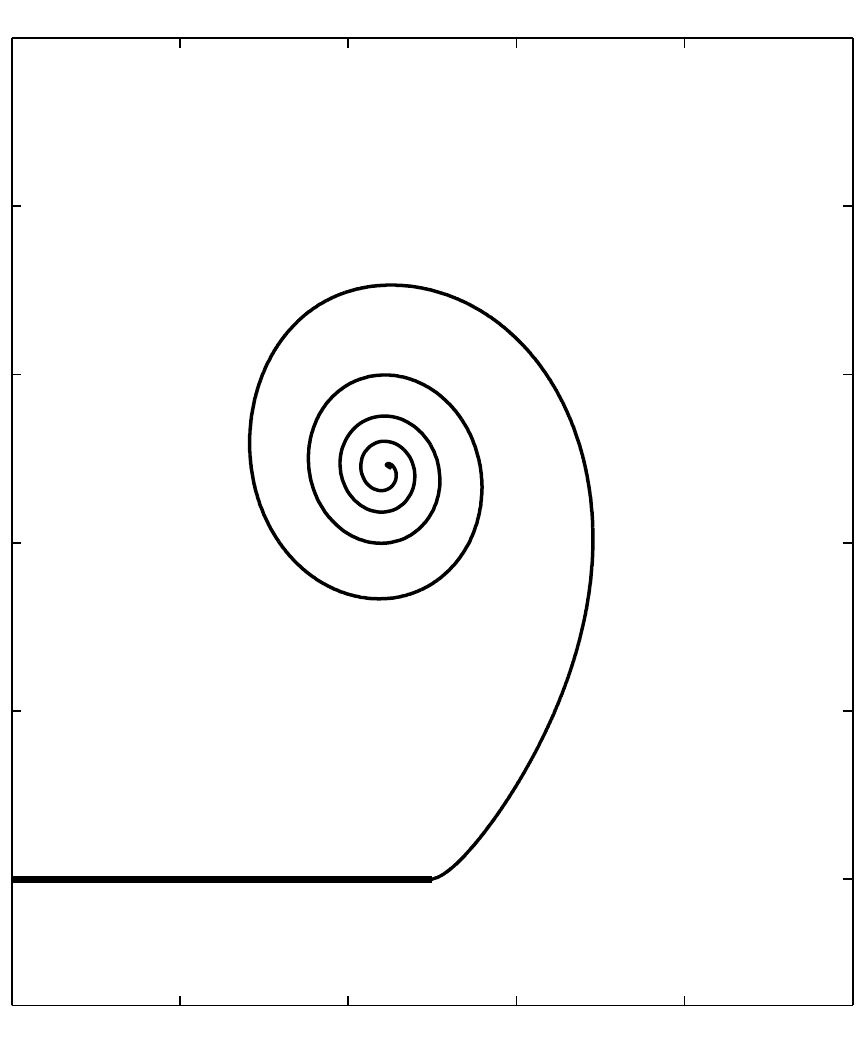}
\hskip4.6pt
   \includegraphics[height=1.701truein]{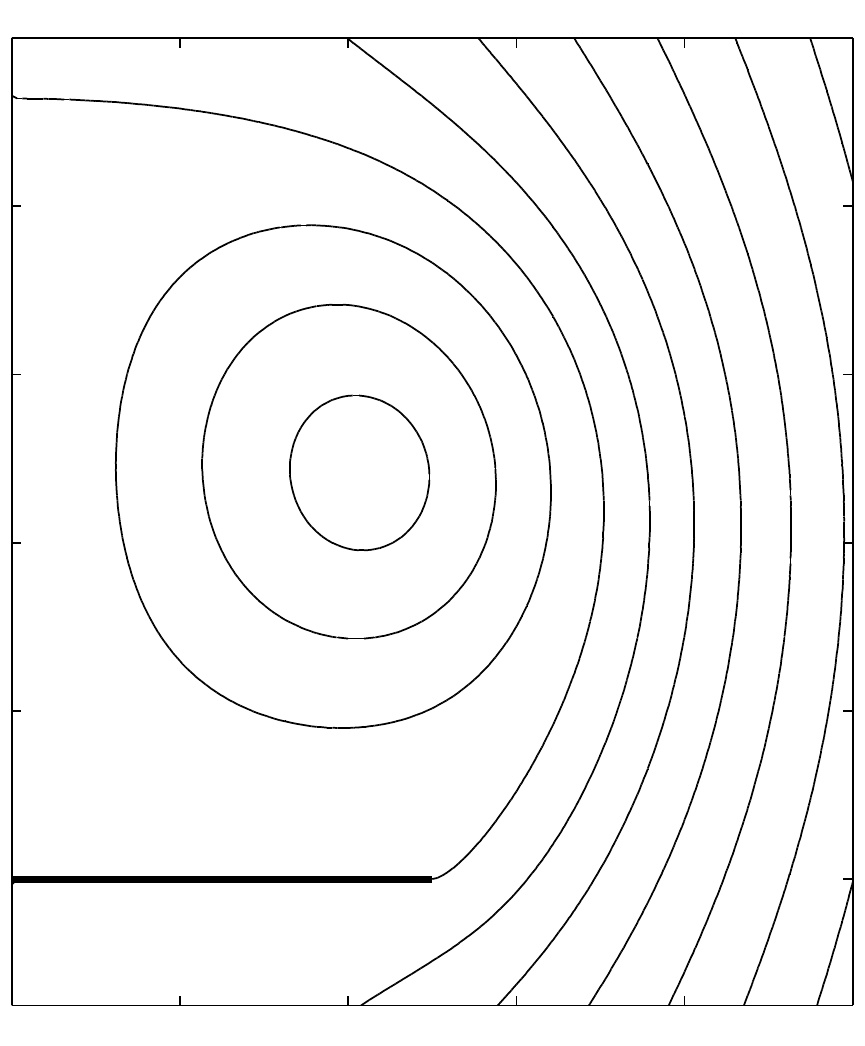}  
\hskip4.6pt

   \includegraphics[height=1.701truein]{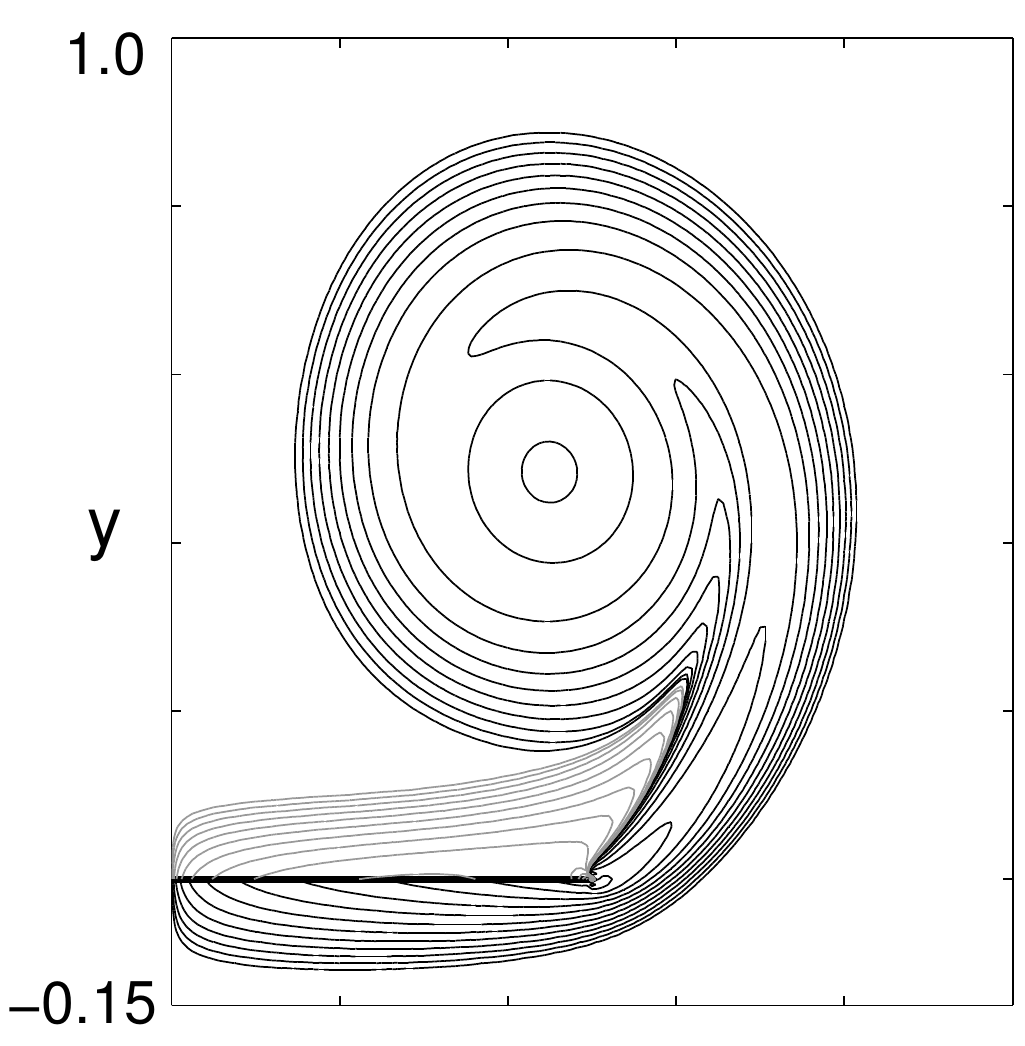}
\hskip4.6pt
   \includegraphics[height=1.701truein]{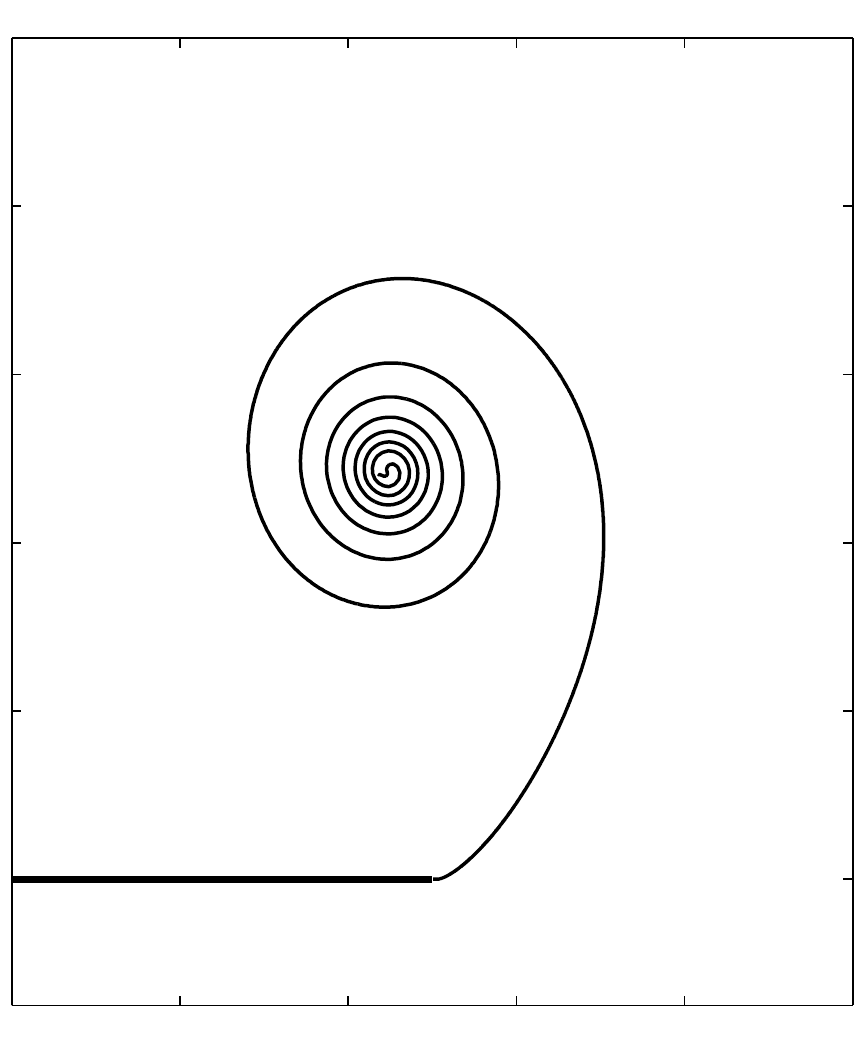}
\hskip4.6pt
   \includegraphics[height=1.701truein]{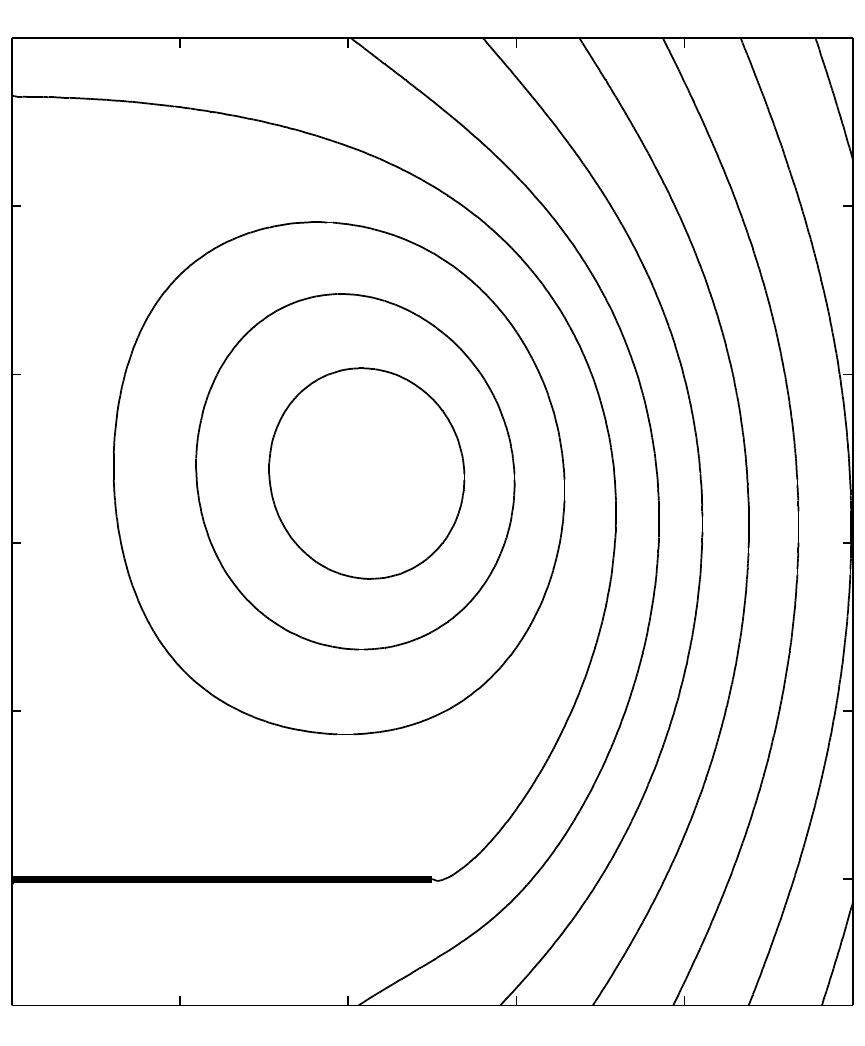}  
\hskip4.6pt

   \includegraphics[height=1.701truein]{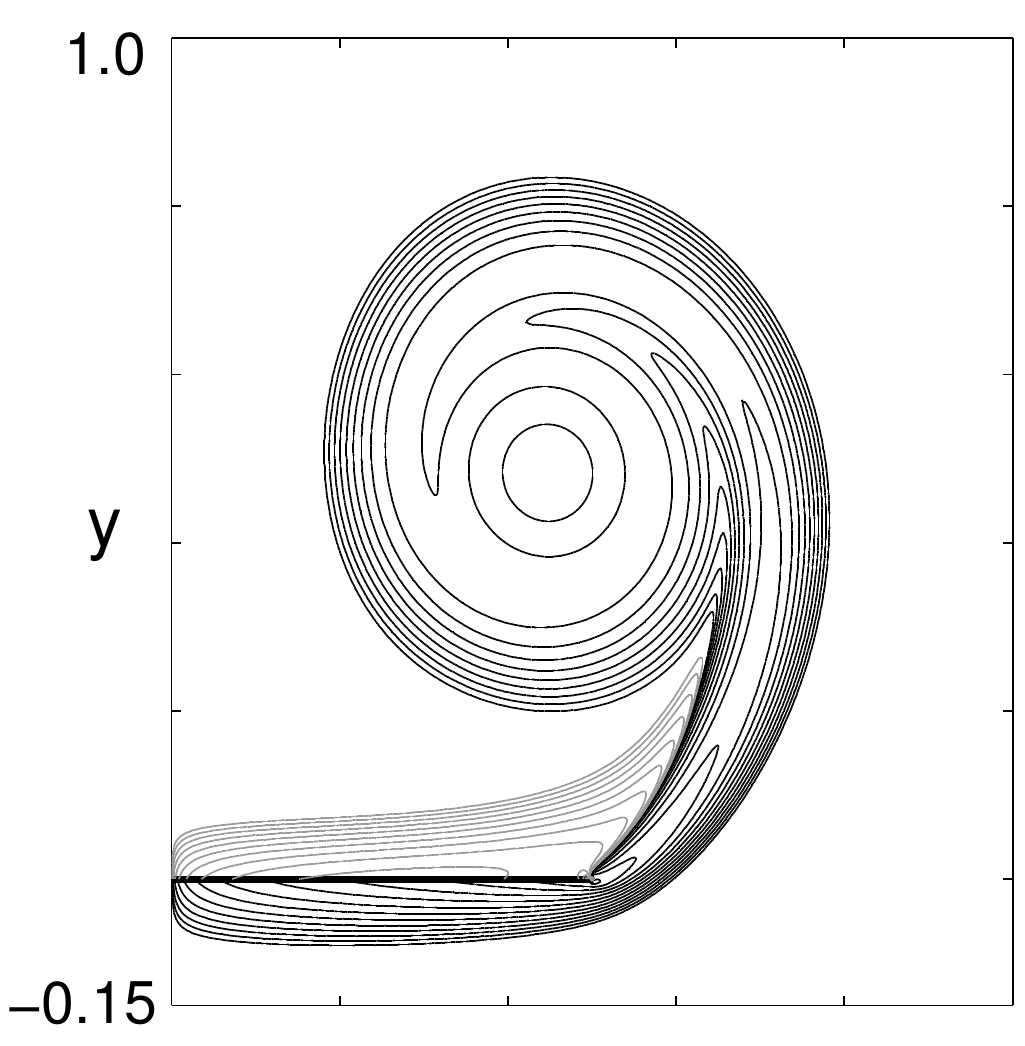}
\hskip4.6pt
   \includegraphics[height=1.701truein]{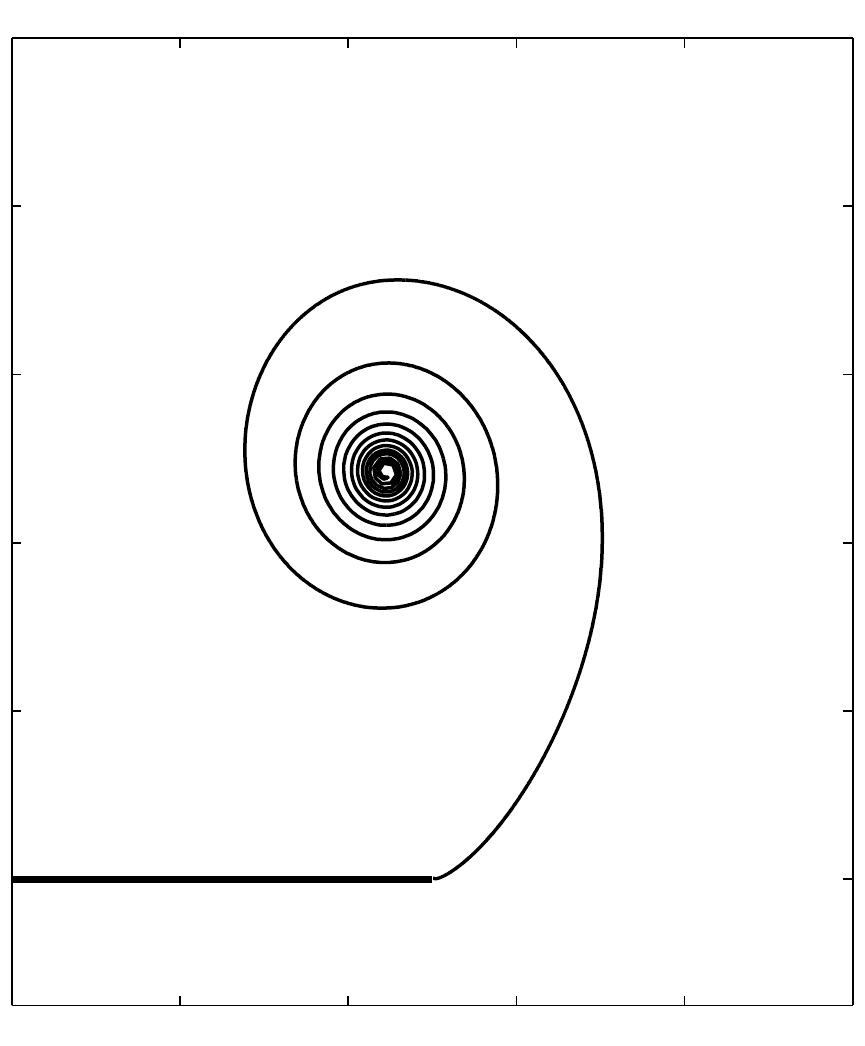}
\hskip4.6pt
   \includegraphics[height=1.701truein]{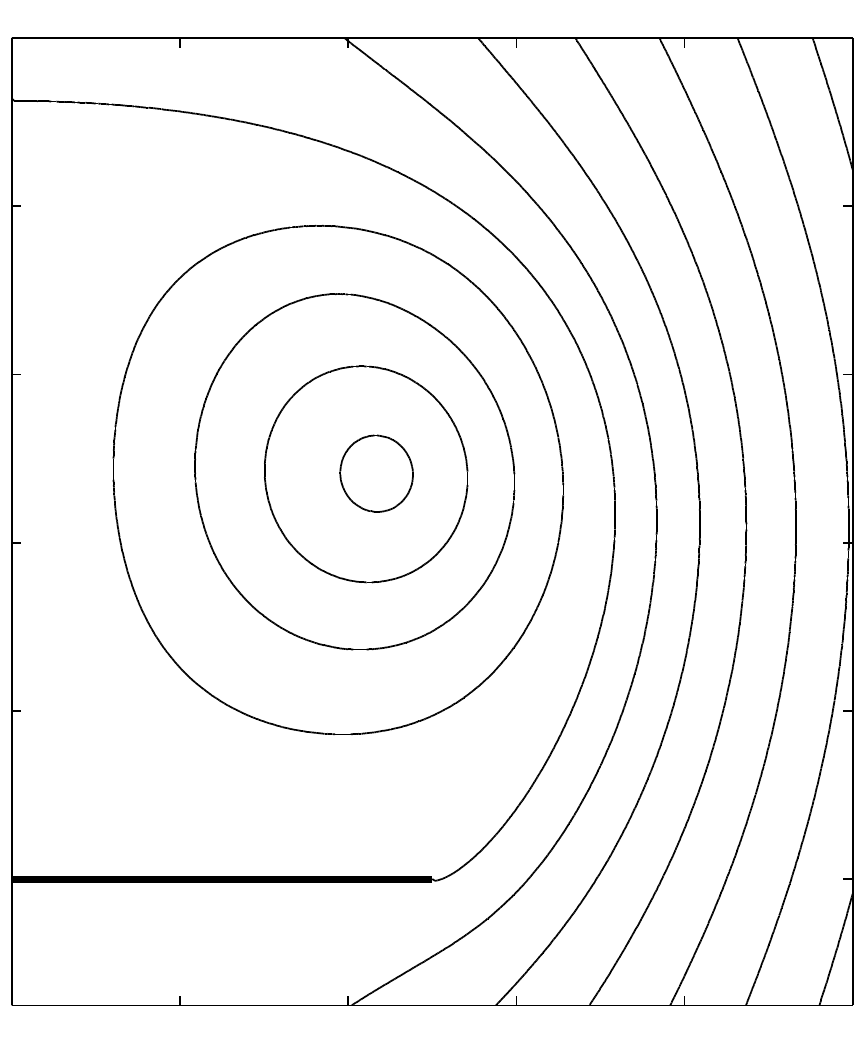}  
\hskip4.6pt

   \includegraphics[height=1.84truein]{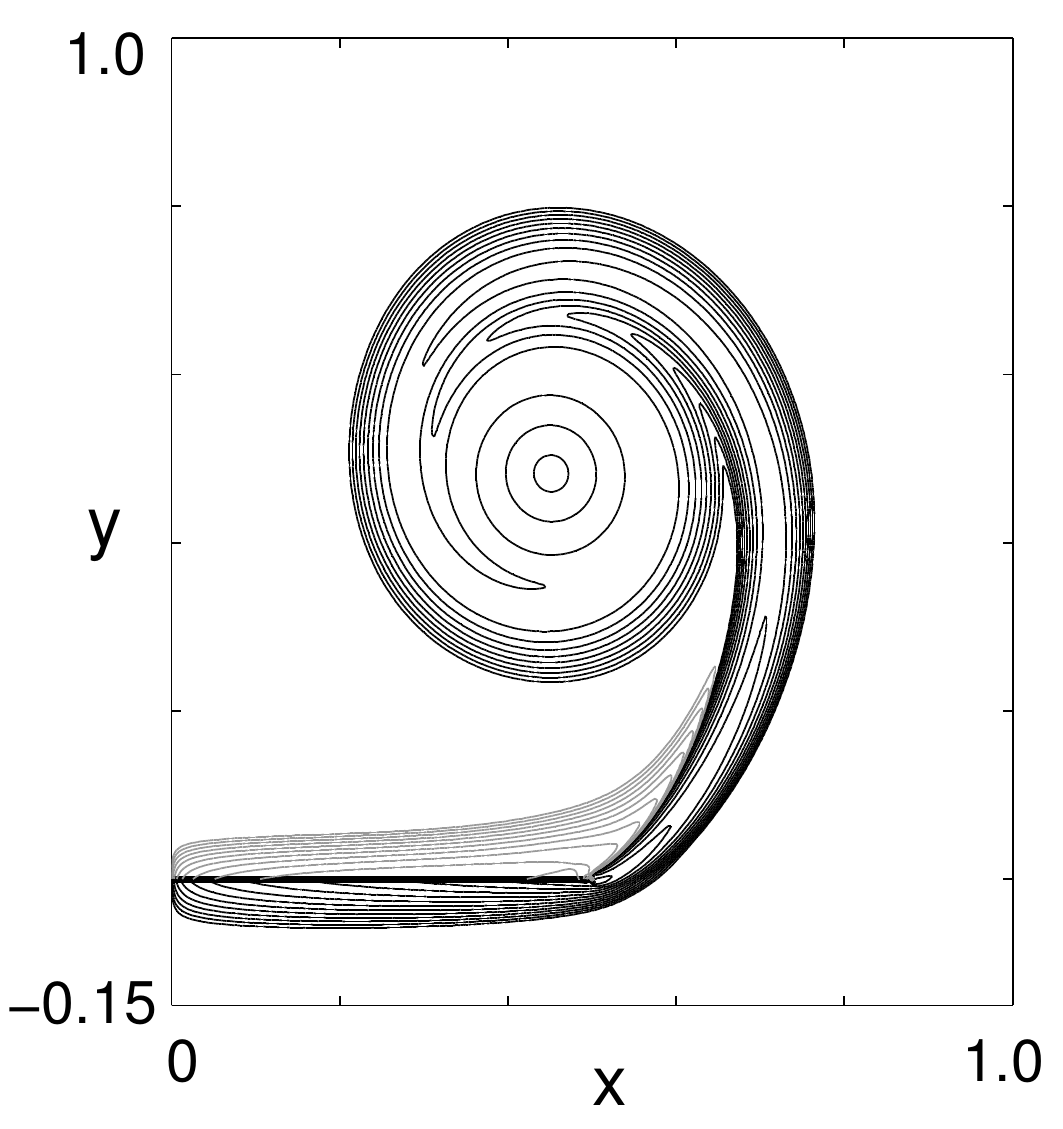}
   \includegraphics[height=1.80truein]{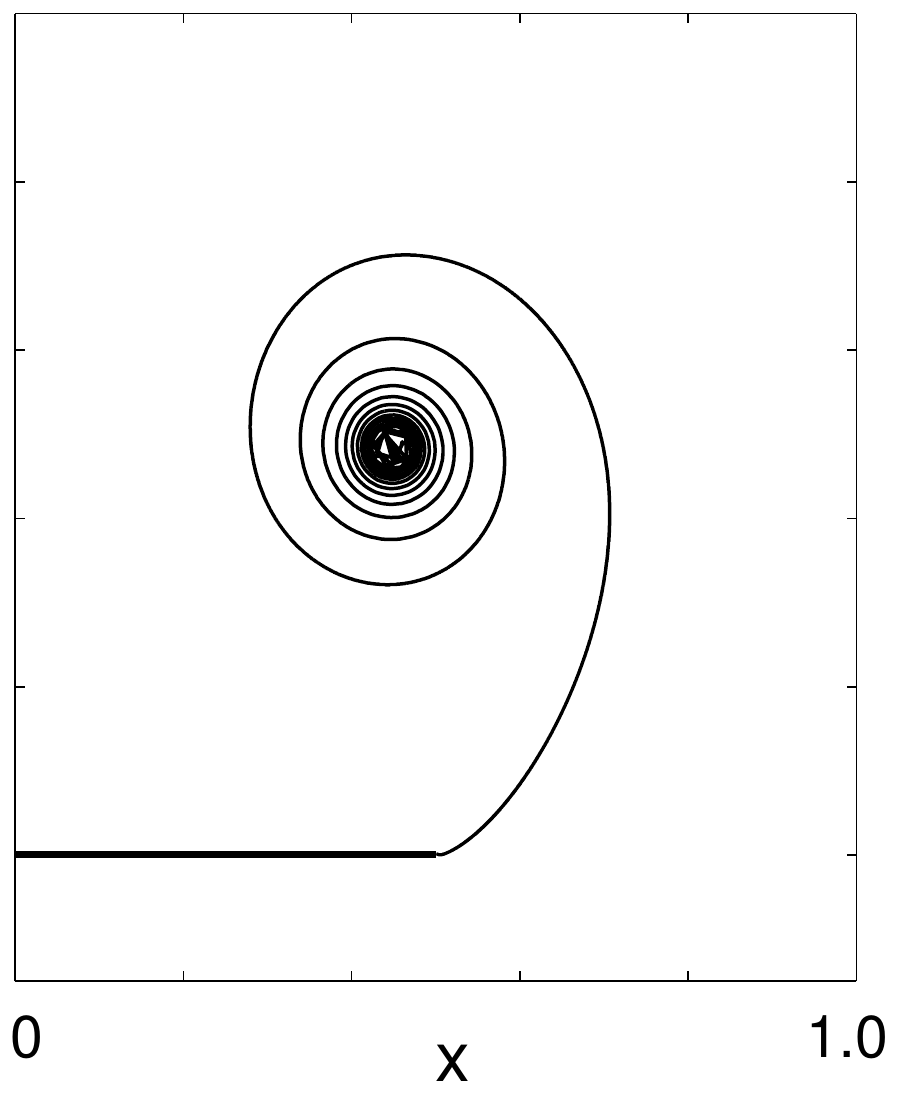}
   \includegraphics[height=1.80truein]{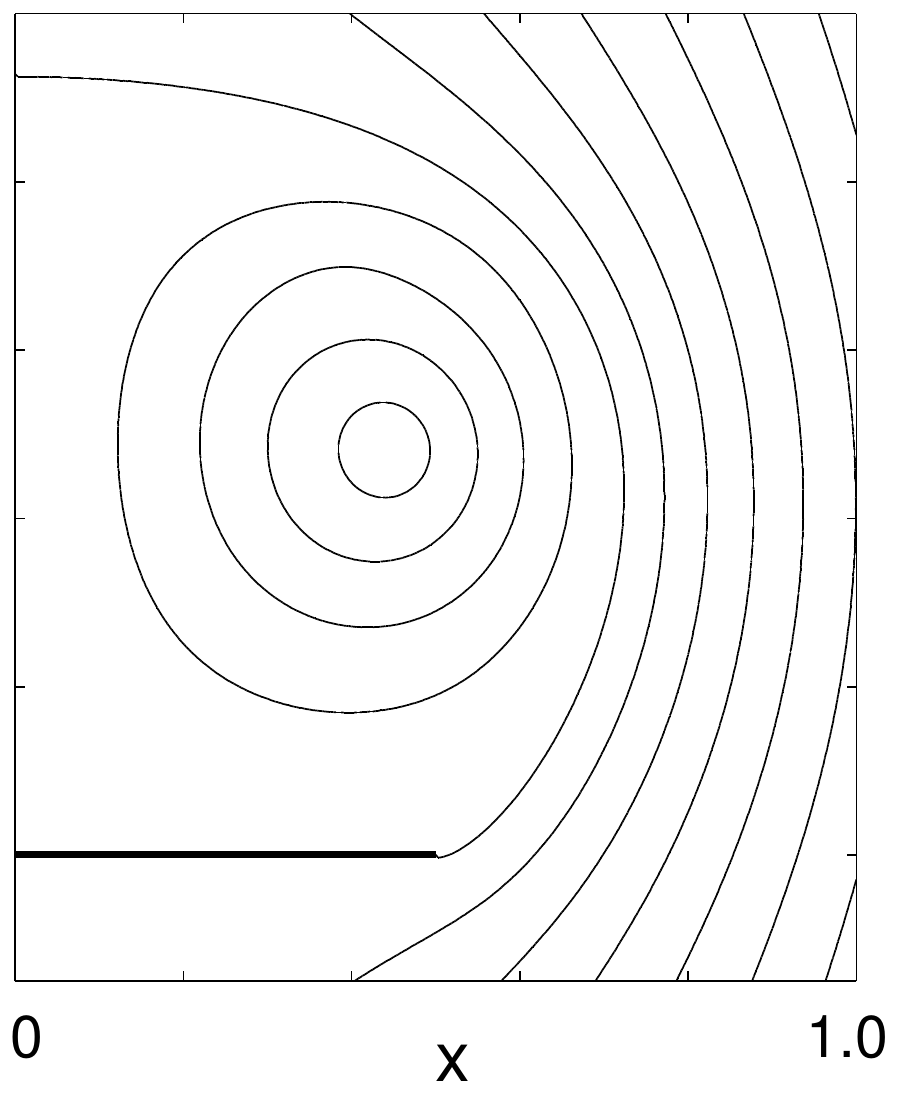}  

\caption{Vorticity, streak lines and streamlines at $t$=1
for $Re$=250, 500, 1000 and 2000.
Vorticity contour levels are $\pm 2^{[-5:12]}$, 
and streamlines contour levels are  $[-1:0.1:1]$.}
\label{FigRe_vort}
\end{figure}

Figure 10 shows the dependence of vorticity contours, streamlines and streaklines
at a fixed time $t=1$ on Reynolds number, for $Re\in [250,2000]$.
%
%
As $Re$ increases, the vorticity contours
show well-known features:
the wall boundary layer thickness 
decreases; the separated shear layer thickness decreases,
and its spiral roll-up becomes more evident;
the thickness of the negative vorticity region decreases,
as already seen in figure 9(c).  
For larger $Re$, the separated vorticity 
is supported in a smaller, more compact region. 


Some dependence on $Re$ is also observed in the spiral streaklines. 
Most noticeably, the spiral roll-up near the center is tighter
for larger $Re$, and there are more spiral turns. 
The spiral size does not depend much on $Re$,
and is, in particular, not a good indicator of the size of the
vortex structure. 
%
%
%
The size of the recirculation region 
does not depend much on $Re$ either.  The streamline 
density within the vortex increases with $Re$, indicating larger
gradients, that is, larger fluid velocities. 

A measure of the boundary layer thickness is given by 
the thickness $\delta$ of the positive vorticity region
on the upstream side of the plate, at $x=0.2$, plotted in figure \ref{F:blayer}. 
Here, $\delta$ is the thickness of the region 
with $\omega\ge 2^{-5}$. The figure shows that 
when plotted against $t/Re$,
the results for all $Re$ collapse onto a line of slope $1/2$, and thus
\begin{equation}
 \delta \sim (t/Re)^{1/2}.
\end{equation} 
asymptotically, as $t\to 0$. This is in agreement with results for 
self-similar flow past infinite plates.
The thickness at other values of $x$ is qualitatively similar.

\begin{figure}
\centering
\includegraphics[height=0.40\textwidth]{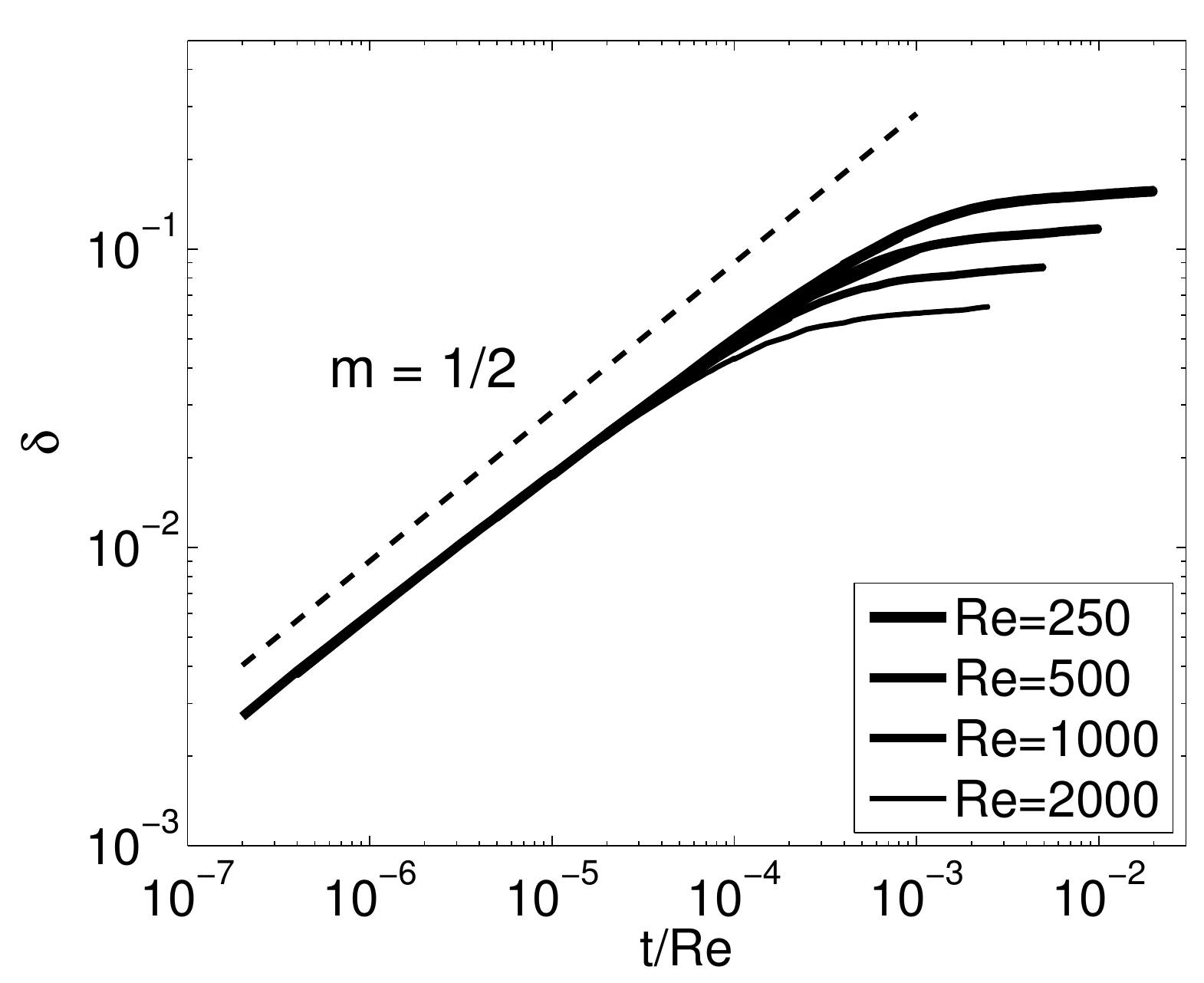}
\caption{
Boundary layer thickness $\delta$ \vs $t/Re$.
Results are shown for $Re$=250, 500, 1000 and 2000, as indicated.}
\label{F:blayer}
\end{figure}

\begin{figure}
\centering
\includegraphics[height=0.336\textwidth]{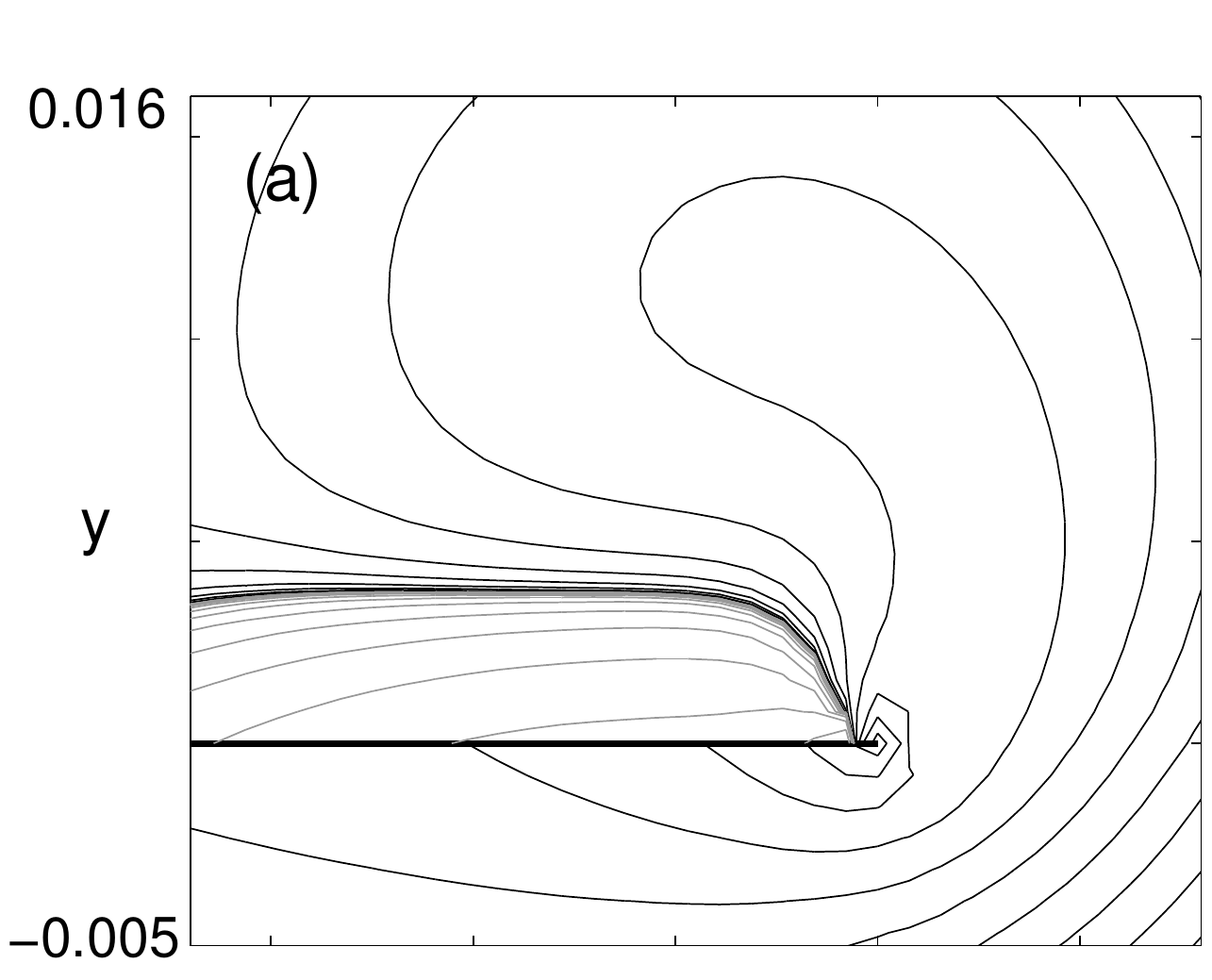}
\hskip15pt
\includegraphics[height=0.336\textwidth]{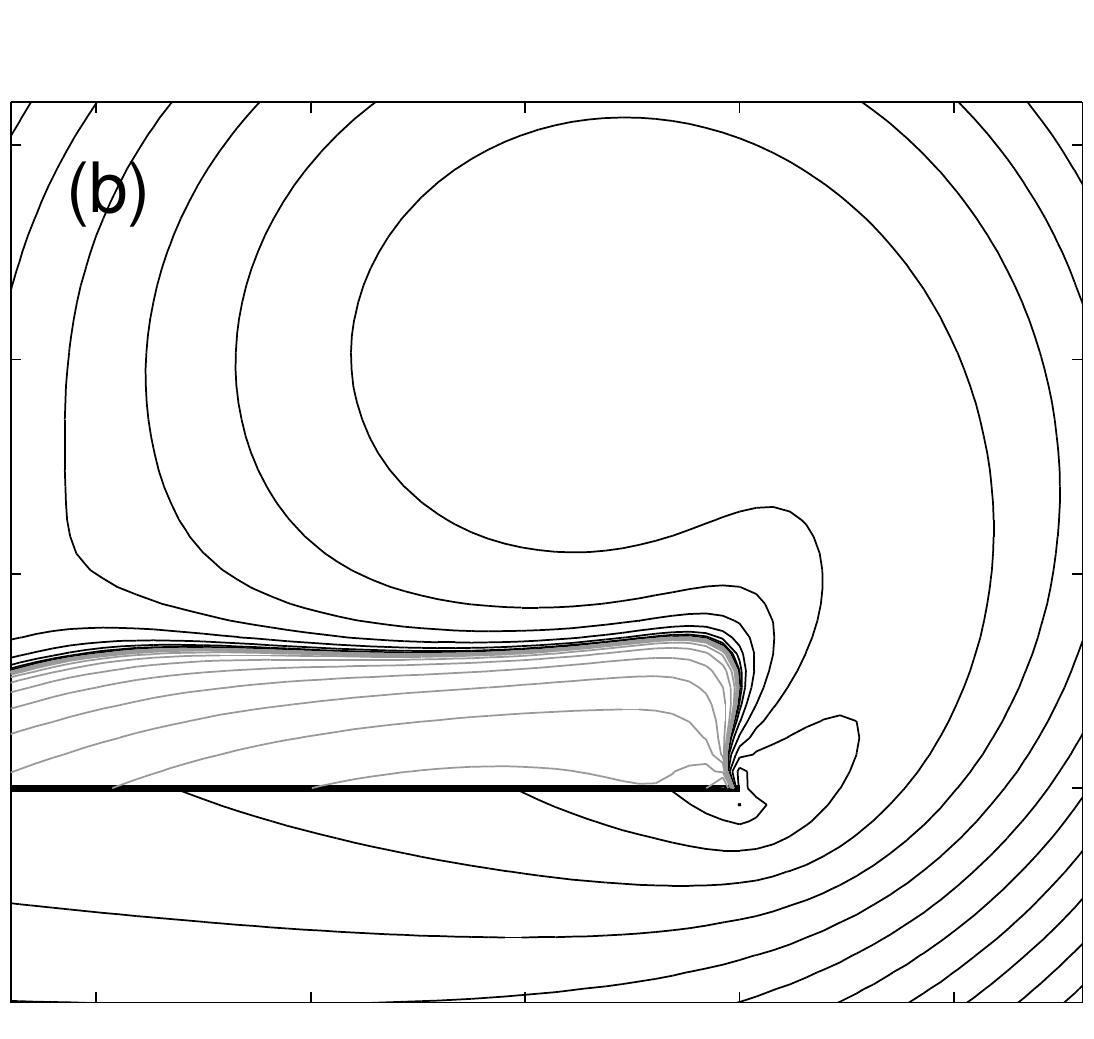}
\hskip15pt 

\includegraphics[height=0.367\textwidth]{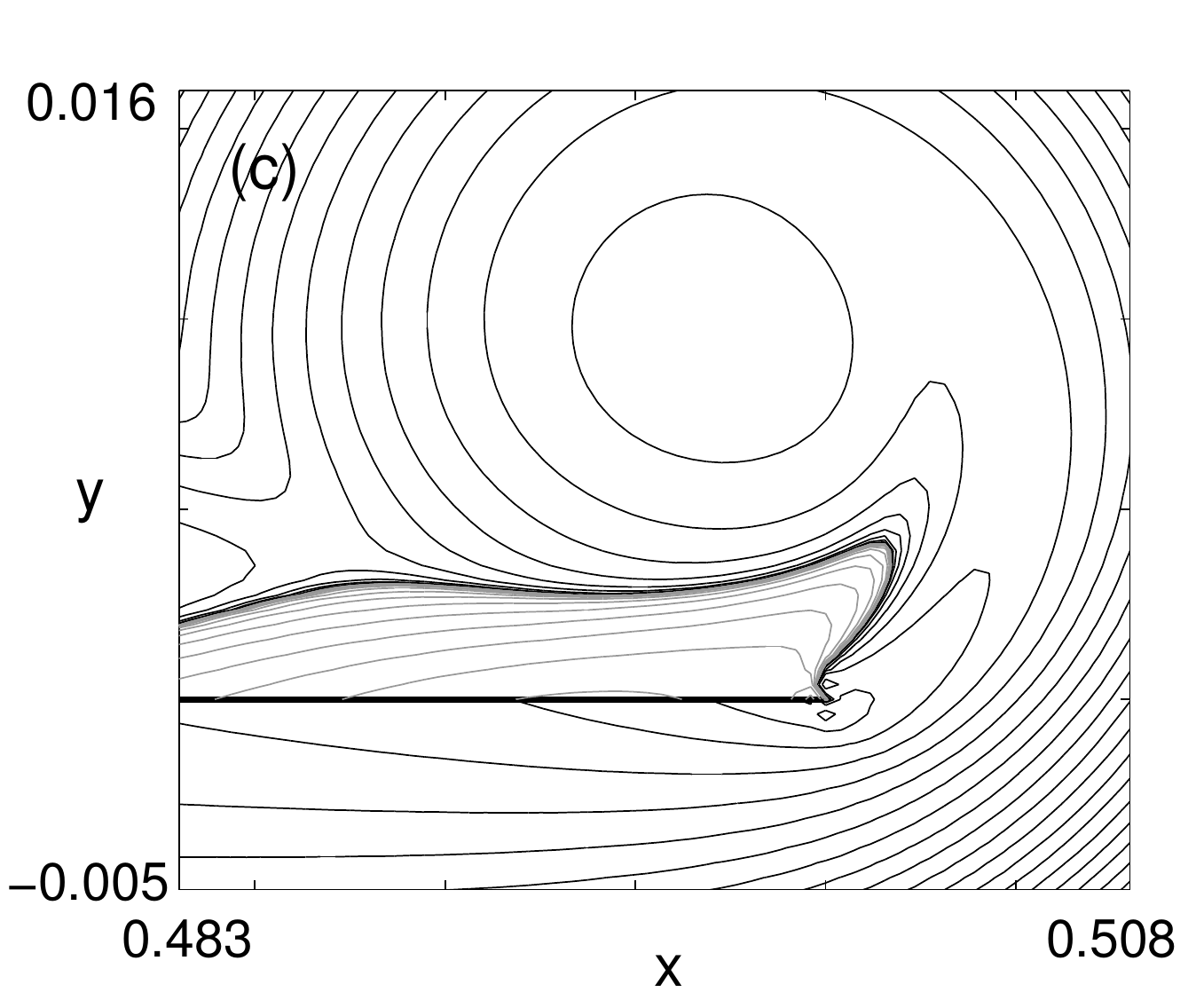}
\includegraphics[height=0.367\textwidth]{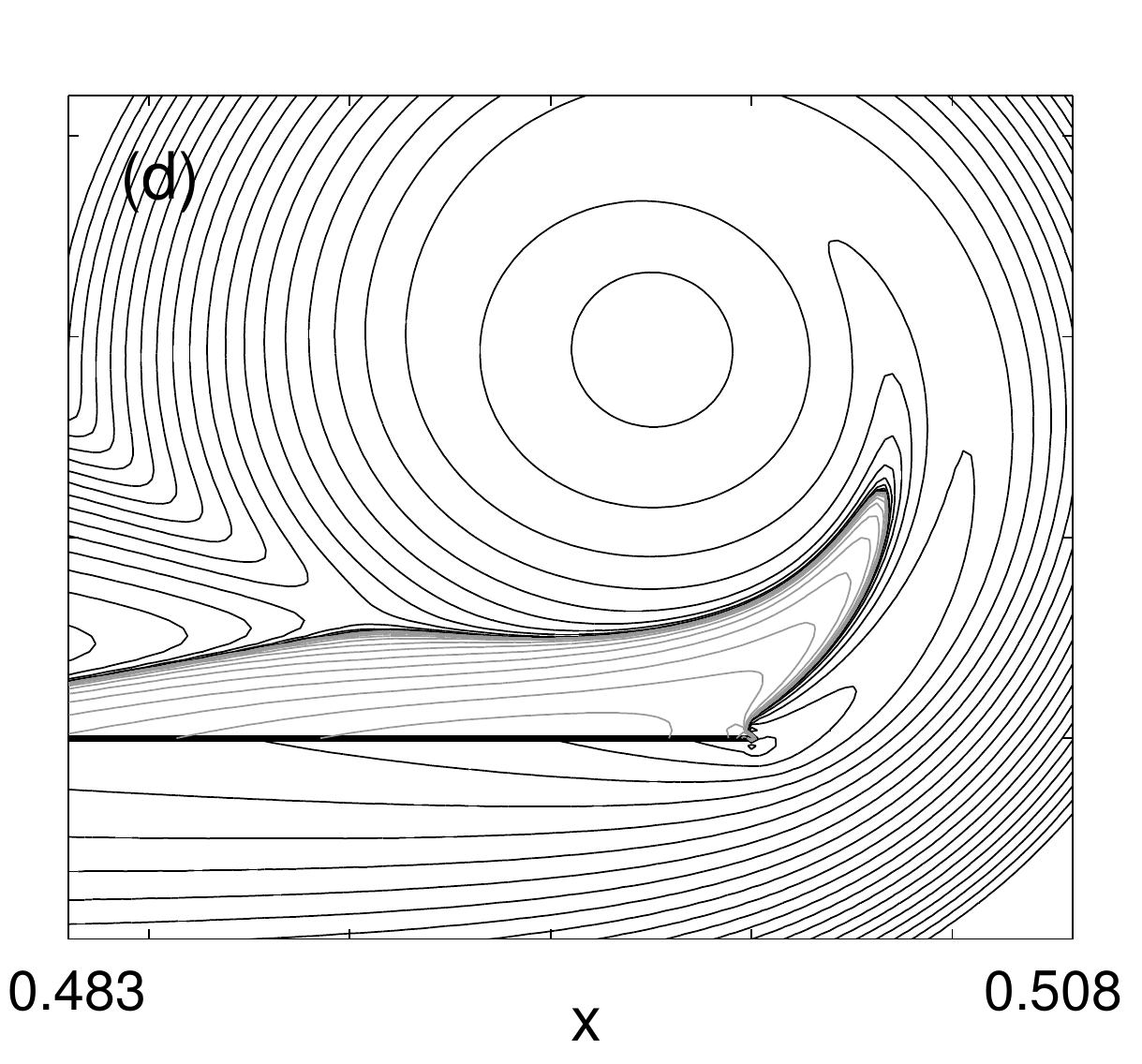}
 \caption{Closeup of vorticity contours near the plate tip,
at $t=0.005$, with 
(a) $Re=250$, (b) $Re=500$, (c) $Re=1000$, (d) $Re=2000$.
Vorticity contour levels are $\pm 2^{[-8:15]}$.
}
\label{F:closeupre}
\end{figure}

Figure \ref{F:closeupre} shows a closeup of the 
vorticity at a fixed early time, for 
different Reynolds numbers. 
With low $Re$, as in figure \ref{F:closeupre}(a), the vorticity has not 
yet formed a local maximum away from the tip and the negative 
vorticity is not yet entrained past the tip. For larger $Re$, 
the core vorticity increases, the negative vorticity region 
lengthens and rolls up around the spiral.
These features at a fixed time, as $Re$ 
increases, are similar to the features observed with
fixed $Re$, as time increases (see figure 6), up to scale,
which is as expected at early times in which the presence of a length
scale is not yet noticeable.


\subsection{Core vorticity and trajectory}
\begin{figure}
\begin{center}
\includegraphics[height=2.0in]{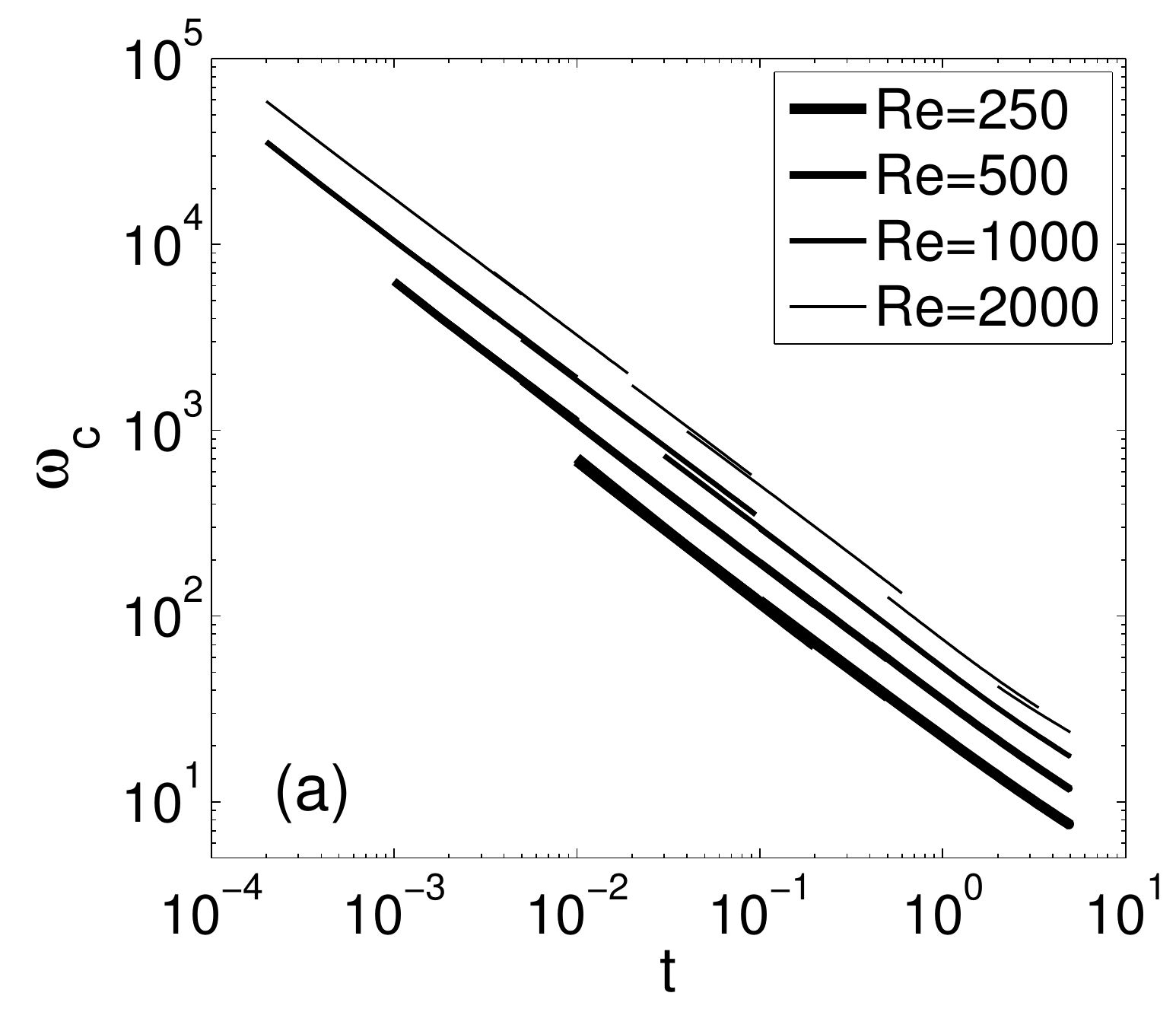}
\hskip4pt
\includegraphics[height=2.01in]{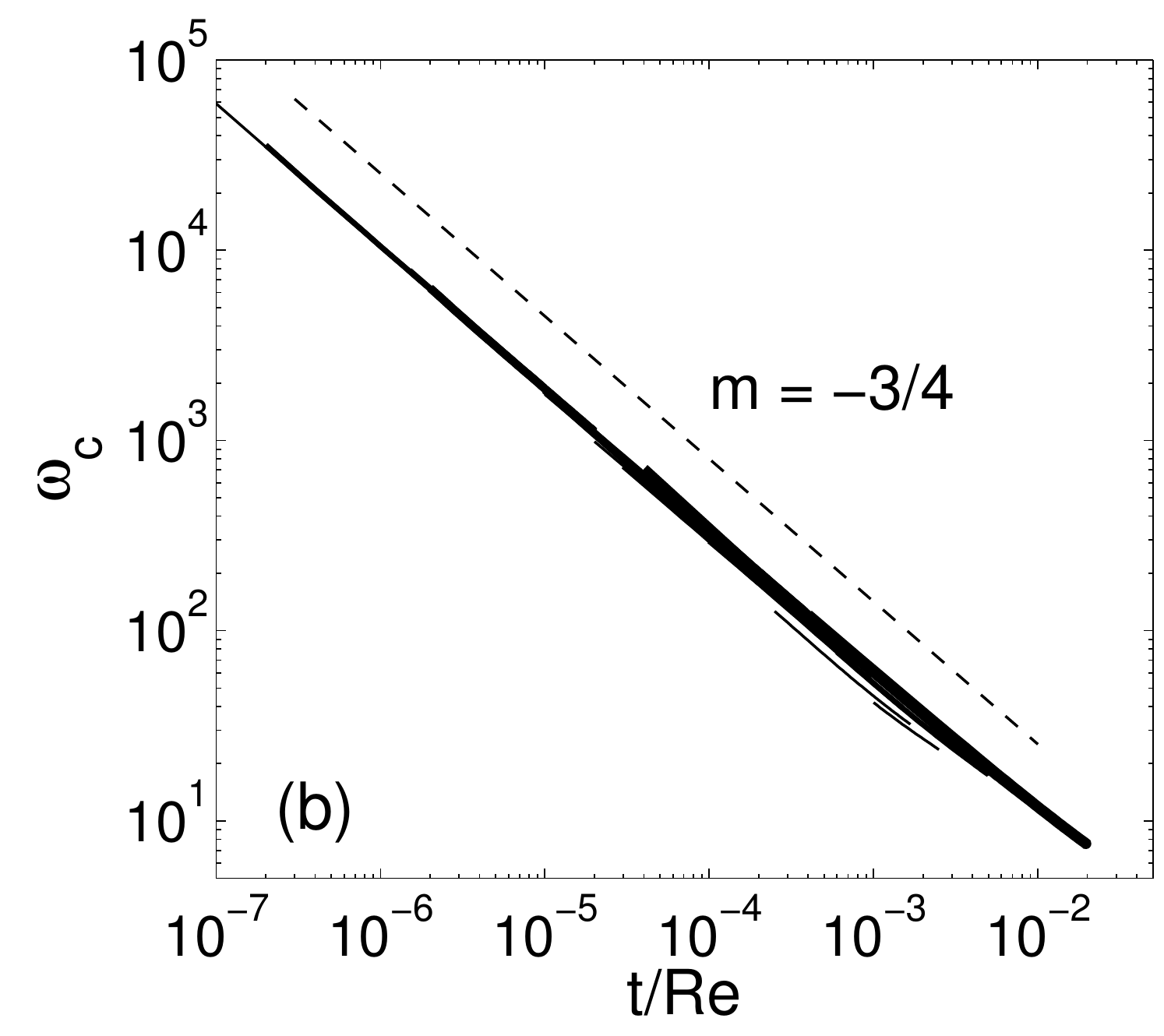}

\includegraphics[height=2.0in]{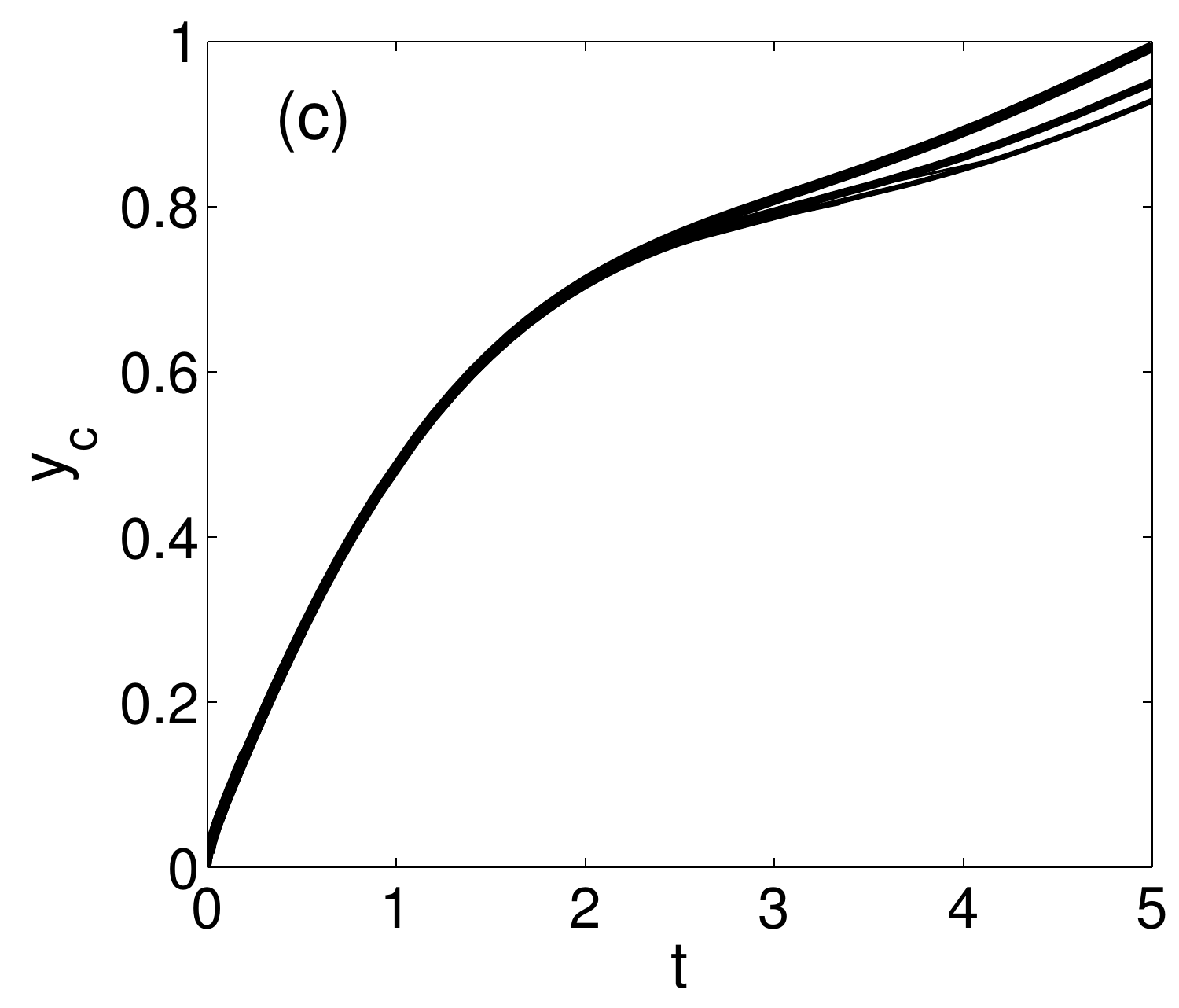}
\hskip4pt
\includegraphics[height=2.03in]{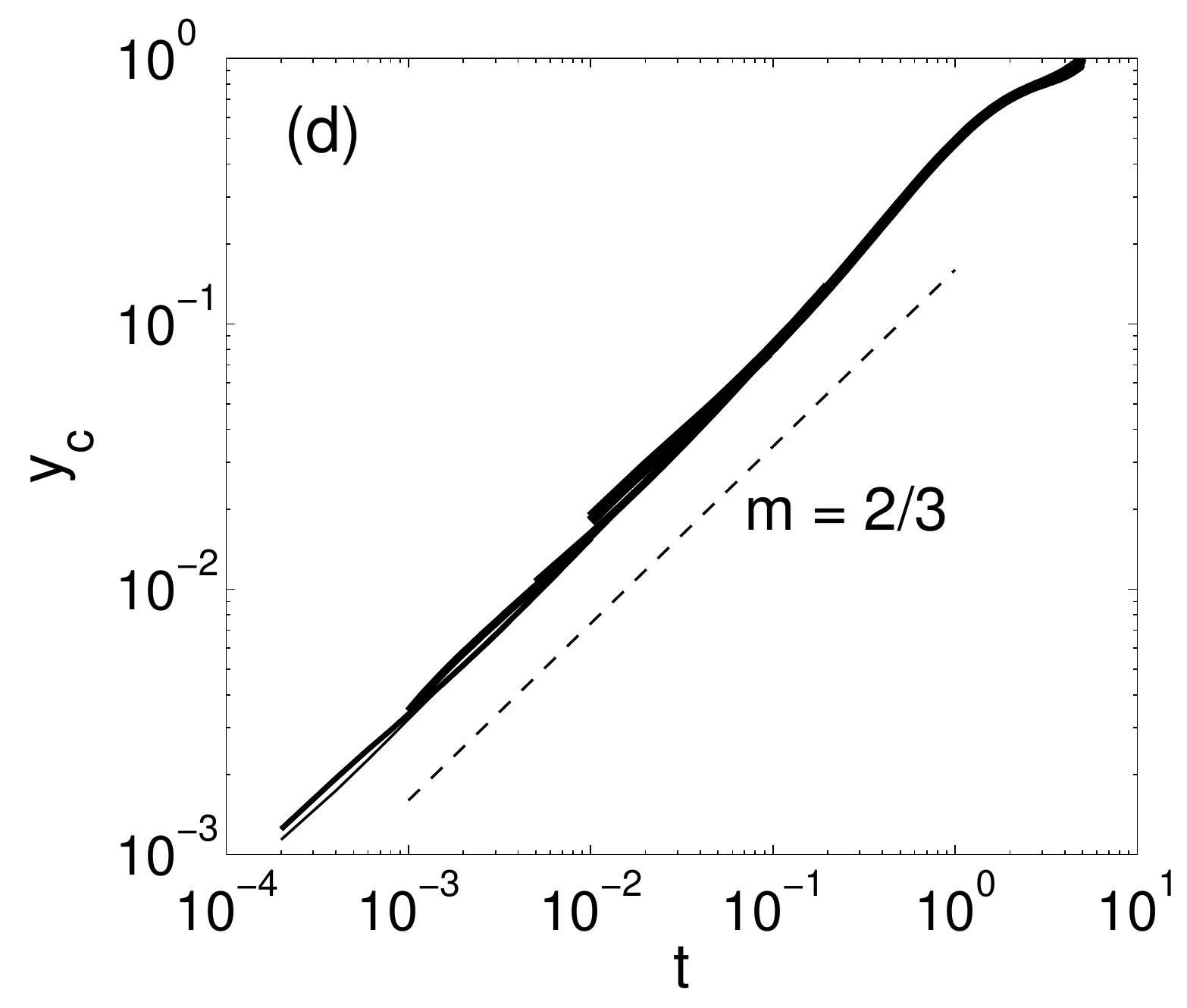}

\includegraphics[height=2.03in]{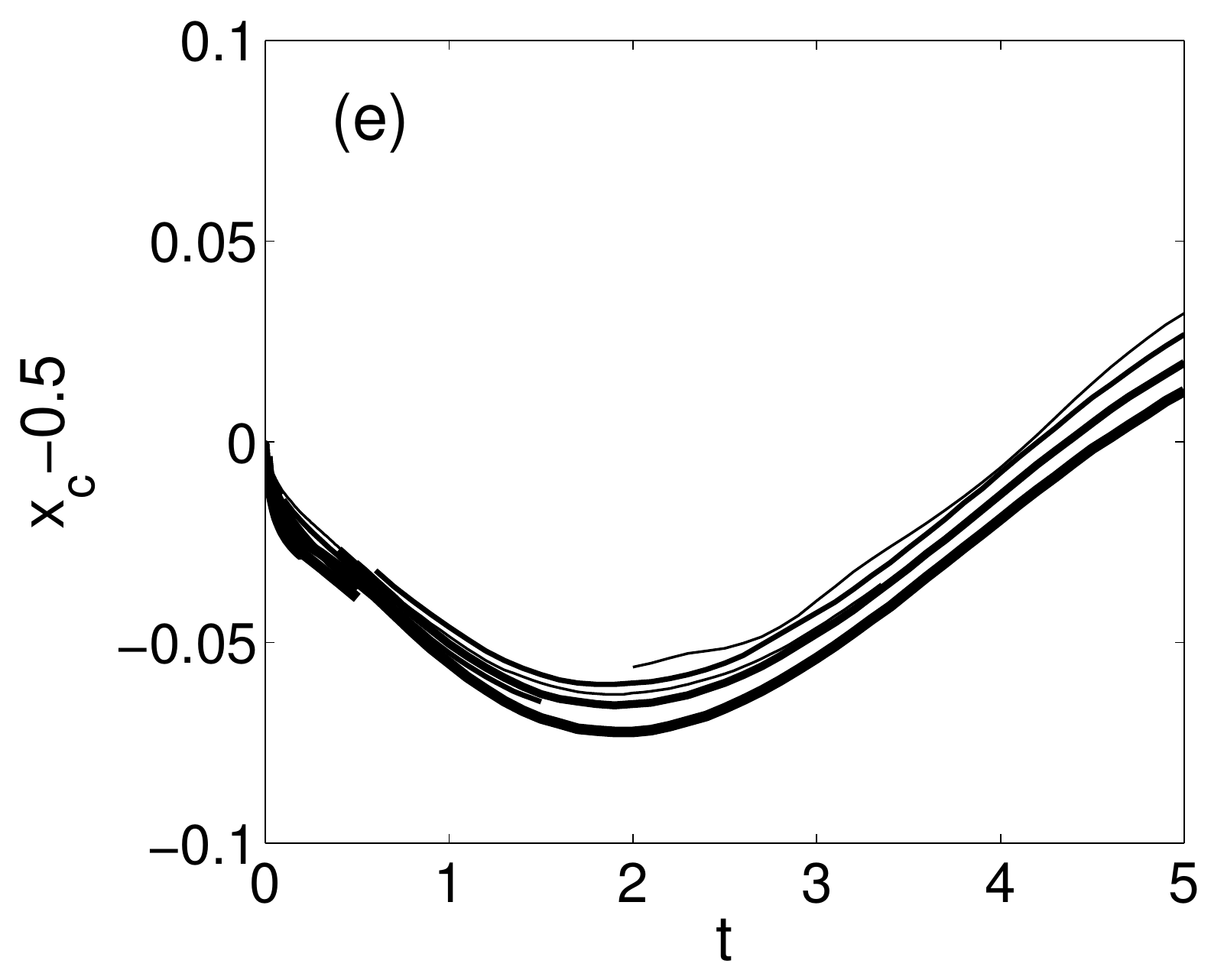}
\includegraphics[height=2.06in]{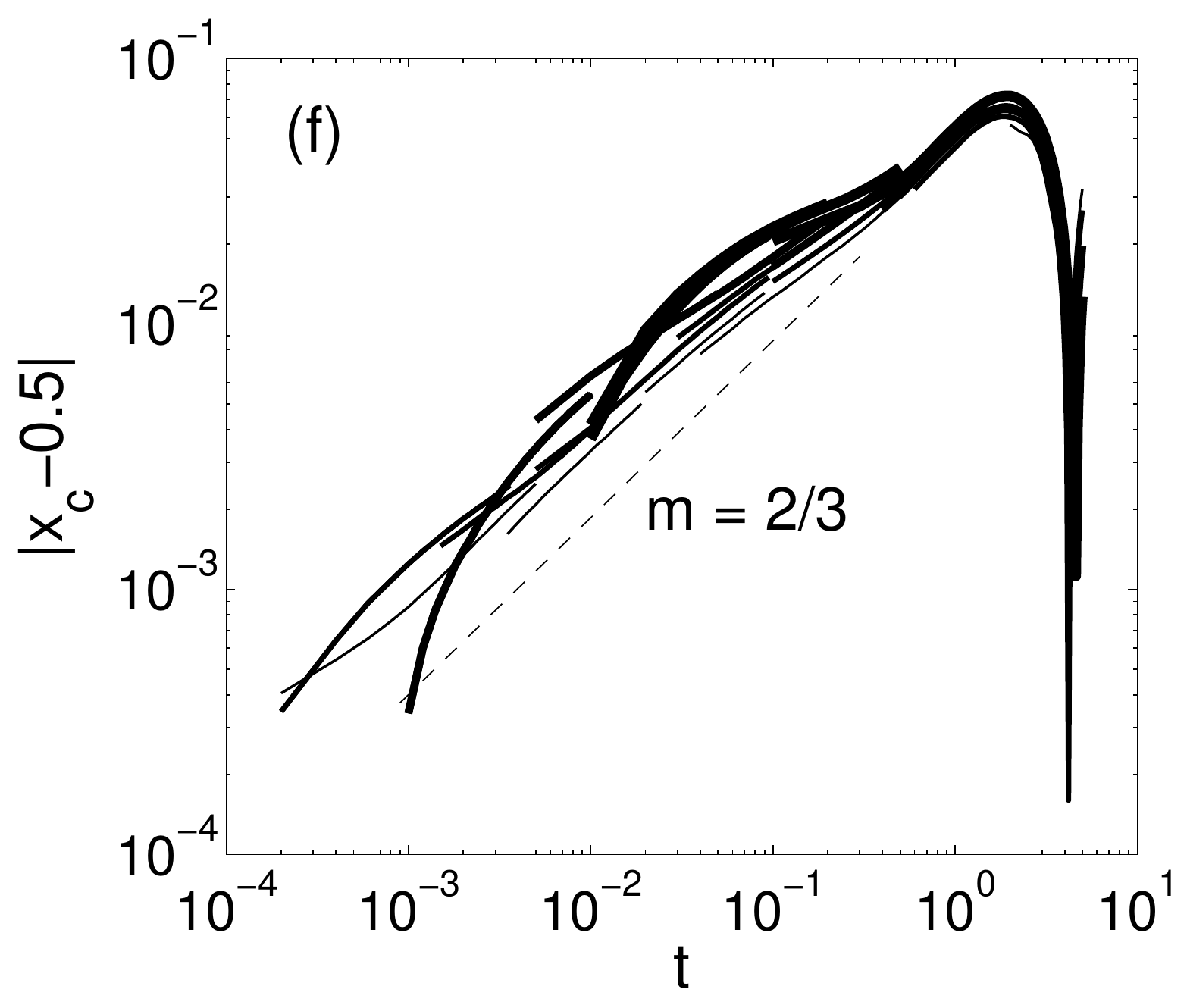} \hskip6pt\\
\end{center}
\caption{Core vorticity and trajectory, for $Re=250,500,1000,2000$ as indicated.
(a) Core vorticity $\omega_c$ \vs $t$.
(b) Core vorticity $\omega_c$ \vs scaled time $t/Re$.
(c,d) Vertical core position $y_c$ \vs $t$. 
(e,f) Horizontal core displacement $x_c-0.5$ from edge.
The dashed lines have the indicated slopes.}
\label{F:core}
\end{figure}

The core position and vorticity are defined as the coordinates $(x_c,y_c)$ 
and vorticity $\omega_c$ at the local vorticity maximum in the leading vortex 
(see figure \ref{F:conv}d). They are defined only after this local 
maximum away from the tip of the plate has formed. 
As seen in figure \ref{F:closeupre}, this occurs earlier for higher $Re$.
This section investigates their scaling behaviour and dependence on $Re$.
We note that alternatively, the core position can be defined as
the center of rotation, which exists at all times, but this 
option is not explored here.

Figure \ref{F:core} plots the core coordinates and vorticity computed
for all $Re$, as indicated in the legend in figure \ref{F:core}(a), and all
values of $h$ used, as given in table 2.
Thus, each subplot in figure \ref{F:core} shows results for 
about 20 different time series, computed with different $Re$ and resolutions.
Figure \ref{F:core}(a) plots 
the values of $\omega_c$ \vs t. The values 
show remarkably little dependence on the 
resolution: only for the largest Reynolds numbers do we see small jumps in the  
values of $\omega_c$ as $h$ is doubled. 
The values $\omega_c$ appear linear in the logarithmic scale, with a vertical
shift between different $Re$. By plotting the results versus $t/Re$, in figure \ref{F:core}(b),
all the data collapse onto one curve, 
and agree for over more than 5 decades in time with the approximation, 
\begin{equation}
\omega_c\approx 0.335{\left(Re\over t\right)}^{3/4}.
\end{equation}
Thus, at any fixed time, the core vorticity increases as $Re^{3/4}$.
For fixed $Re$, it decreases in time as $t^{-3/4}$.
We do not know of an analytical result that explains this observation.

Figures \ref{F:core}(c,d) plot the vertical displacement $y_c$ of the 
vortex core from the plate, on a linear and a 
logarithmic scale respectively, versus t. 
Here again, the values for all 
$Re$ and all resolutions computed collapse onto one curve, 
with no apparent dependence on $h$, and
only a small dependence on $Re$ visible at later times.
The data for all $Re$ scales for about 4 decades as 
\begin{equation}
y_c\approx 0.37t^{2/3}
\end{equation}
with deviations from this line visible after approximately $t=1$.
This scaling agrees with the self-similar inviscid spiral roll-up
of semi-infinite free vortex sheets, or of separated vortex sheets at the
edge of a semi-infinite plate (Kaden 1931, Pullin 1978).

Figure \ref{F:core}(e) plots the horizontal displacement $x_c-0.5$
of the vortex core from the plate, on a linear scale.
Figure \ref{F:core}(f) plots the absolute value on a logarithmic scale.
These values are an order of magnitude smaller than those of $y_c$,
and less well resolved, with the dependence on $h$ more visible.
To within the available resolution, the results depend little on $Re$.
The linear scale shows that $x_c-0.5$ is initially negative.
Up to about $t=2$, the symmetric vortices at each end of the plate
more slightly inward, as they would in the self-similar inviscid case.
After that time they begin moving outwards again, with 
$x_c>0.5$ at the final time computed, $t=5$.
The logarithmic scale shows that until about $t=1$, $x_c$ also
satisfies the self-similar inviscid scaling, with
\begin{equation}
x_c-0.5\approx -0.1t^{2/3}~.
\end{equation}

\begin{figure}
\begin{center}
\includegraphics[height=2.03in]{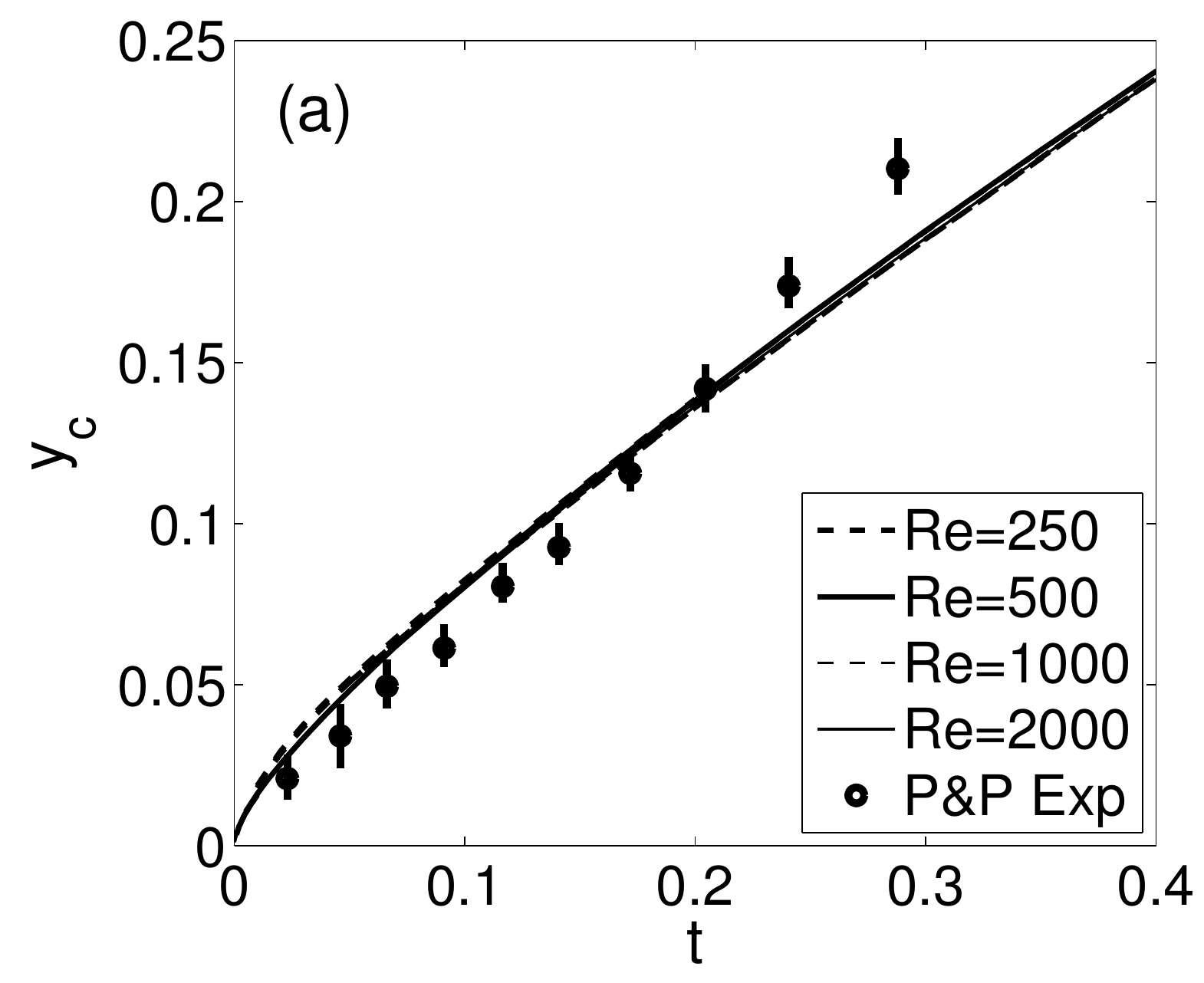}
\includegraphics[height=2.03in]{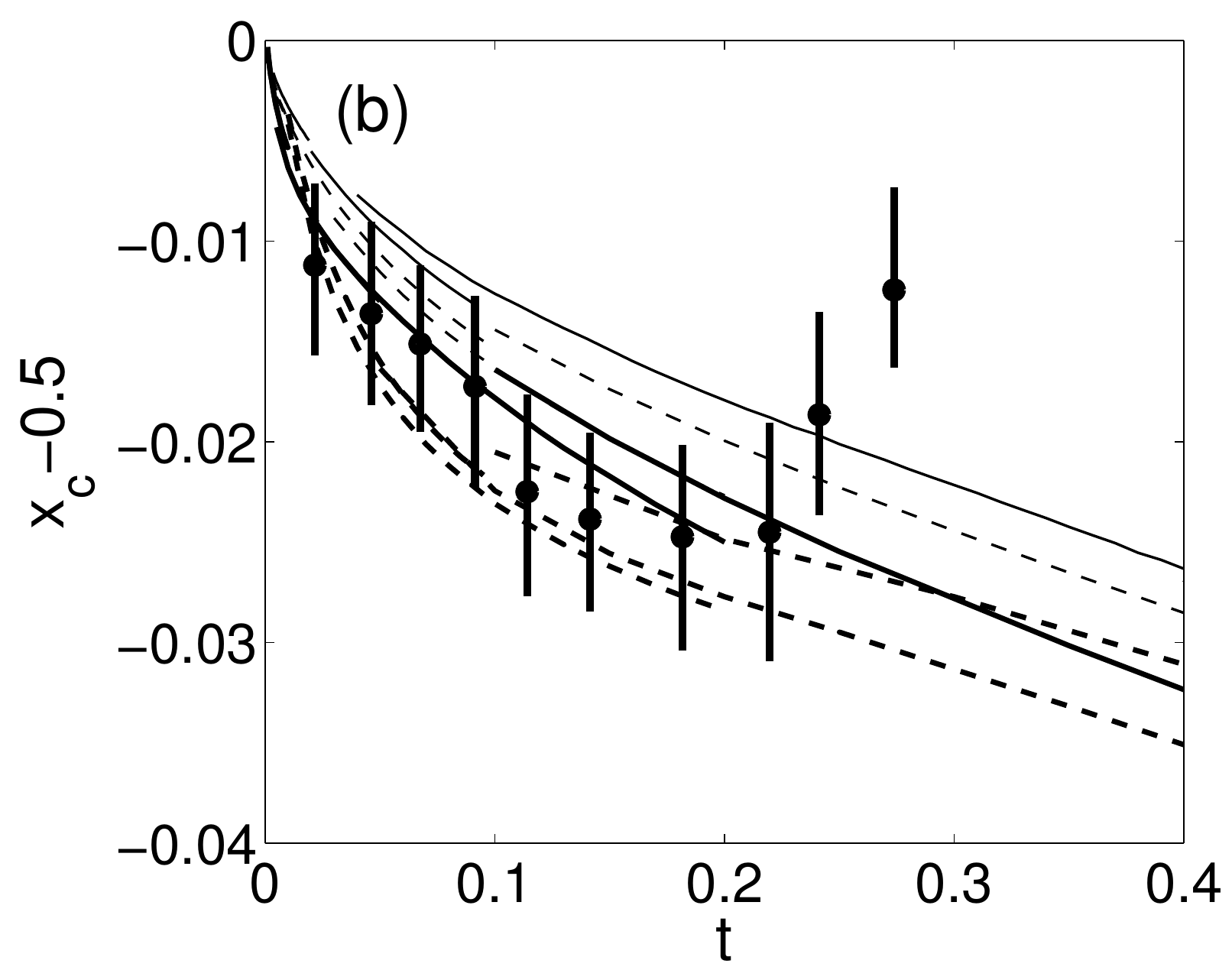}
\end{center}
\caption{Comparison of vortex core coordinate with 
experimental data of Pullin \& Perry (1980). 
(a) Vertical displacement $y_c$.
(b) Horizontal displacement $x_c$.
The computed data for all Reynolds number and the experimental data is shown,
as indicated in the legend. The vertical bars through the experimental
data are the error bars indicated in P\&P.
}
\label{F:pullin}
\end{figure}

The closest data on viscous flow at early times available for comparison in the 
literature are the experimental measurements of Pullin \& 
Perry (1980), who measured vortex core positions in 
flow past wedges and compared them to similarity theory. 
Figure \ref{F:pullin} reproduces their results for the smallest wedge considered,
of wedge angle $\beta=5^o$, together with our computed results.
The experimental data span an early time interval $t\in[0,0.3]$.
The vertical displacement $y_c$ is in quite good quantitative agreement
with the computed values. The horizontal displacement $x_c$ overlaps with
the present better resolved values at the lower Reynolds numbers, except for 
the last two data points. 
The experimental data was obtained at larger Reynolds number of $Re\approx 6000$,
but the data is expected to be practically independent of $Re$, as
is clearly the case for the values of $y_c$. 
We cannot explain the deviation of the computation
from the last two experimental data points in figure \ref{F:pullin}(b).


\begin{figure}
\centering
   \includegraphics[width=2.5truein]{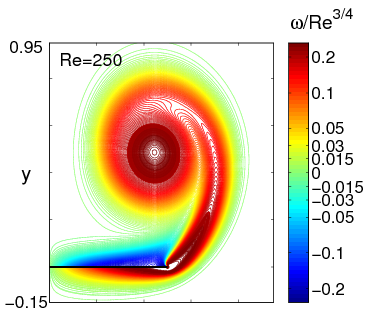}
   \includegraphics[width=2.5truein]{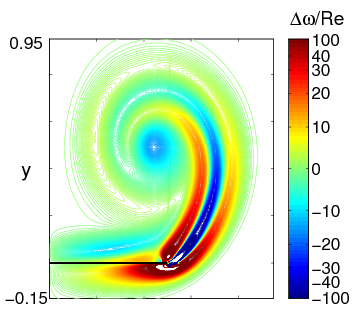}\\

\vskip-4pt
   \includegraphics[width=2.5truein]{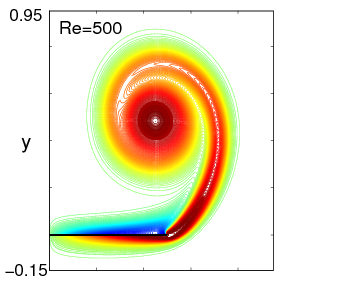}
   \includegraphics[width=2.5truein]{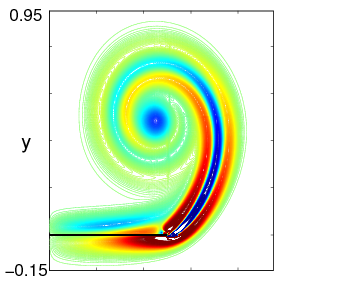}\\

\vskip-4pt
   \includegraphics[width=2.5truein]{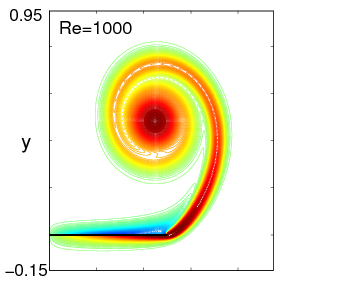}
   \includegraphics[width=2.5truein]{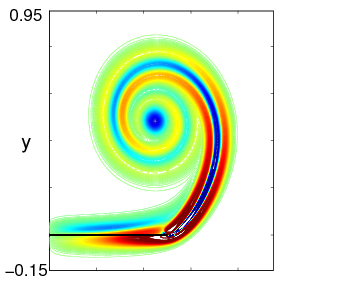}\\

\vskip-4pt
   \includegraphics[width=2.5truein]{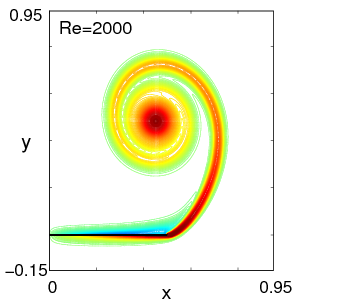}
   \includegraphics[width=2.5truein]{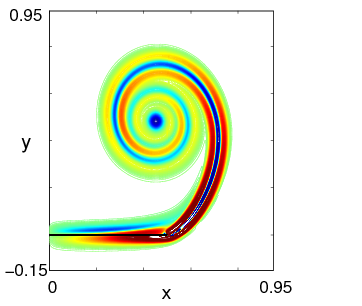}

\caption{Scaled vorticity $\omega/Re^{0.75}$ (left column) 
and dissipation $\Delta \omega/Re$ (right column) at $t=1$, 
for $Re$=250, 500, 1000 and 2000, from top to bottom. 
The colorbar is the same for all values of $Re$, and shown for 
highest $Re$ only.}
\label{F:dissip}
\end{figure} 

With knowledge of the core vorticity scaling, we now plot scaled vorticity, 
as well as dissipation.  Figure \ref{F:dissip} shows color coded contours of the scaled 
fluid vorticity, $\omega/Re^{3/4}$, at $t=1$, for the indicated values of $Re$
(left column), and of the corresponding dissipation 
$\Delta\omega/Re$ (right column).  
For clarity, we note that the plots
consist of equally spaced contour curves of $A\tanh(q/A)$, for a chosen value
of $A$ below the true maximum, where $q$ is either the scaled vorticity or dissipation. 
As a result,
contours of arbitrarily large levels can be shown. Also as a result, the values 
of the quantity $q$ shown in the attached colorbars are not equally spaced in
the color level.

At this time, the vorticity attached to the upper plate wall is 
all negative, and all remaining vorticity is positive.
The vorticity contours 
show the decrease of boundary and shear layer thicknesses, 
and the increasingly visible spiral shear layer roll-up as $Re$ increases.
In a crossection at any point through the shear layer,
the vorticity is largest in the middle of the layer, and decreases to local
minima in between different spiral turns. 
We note that at this rather late time, the vorticity maximum 
only approximately satisfies the scaling 
$\omega_c\sim Re^{3/4}$,
and the scaled values 
plotted in the figure decrease from 0.359 to 0.250.

The dissipation plot indicates the magnitude of viscous diffusion 
in the flow.
The dissipation at the vortex core, where the vorticity has a maximum, 
is negative, which causes the maximum vorticity to decrease.
The largest absolute values at the 
core increase as $Re$ increases, from 
16.16 to 54.96, in a manner consistent with the scaling
\begin{equation}
{\Delta\omega\over Re}\sim Re^{3/4}~.
\end{equation}
Across each of the spiral shear layer turns the dissipation changes sign
twice. It is positive at the local vorticity minima in between spiral
turns, causing these minimum values to increase,
and it is negative at the local maxima in the middle of the layer, 
causing these maxima to decrease. 
As a result, the spiral turns are more clearly visible in the dissipation
plots than in the vorticity contours. 
The dissipation is largest in magnitude near the tip of the plate, where
it reaches values well above 200.

\subsection{Shed vortex circulation}

\begin{figure}
 \centering
\includegraphics[height=0.382\textwidth]{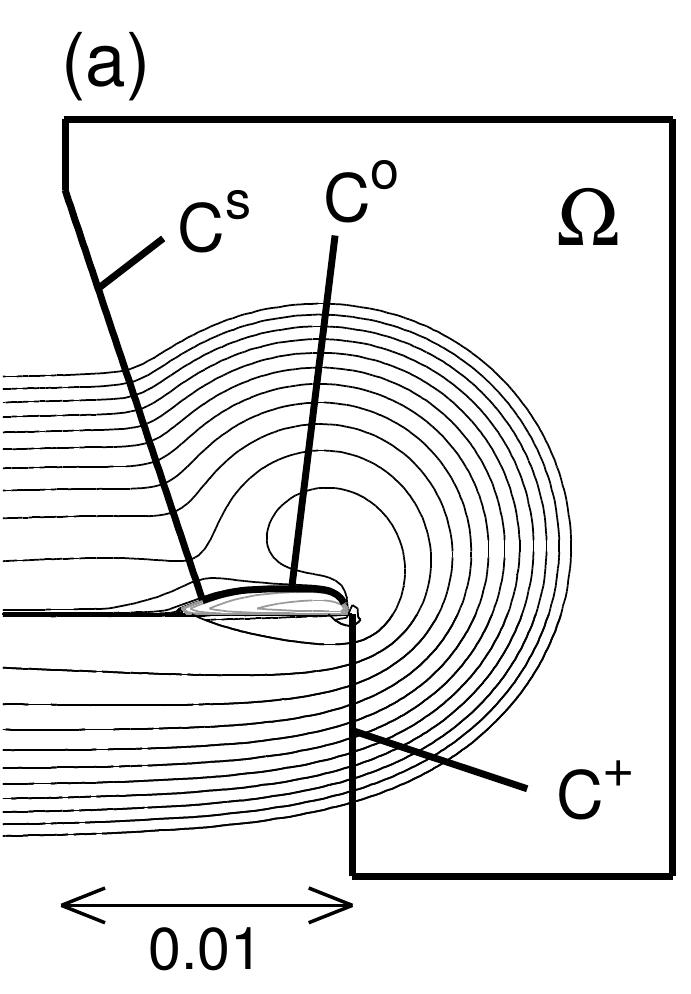}
\hskip10pt
\includegraphics[height=0.382\textwidth]{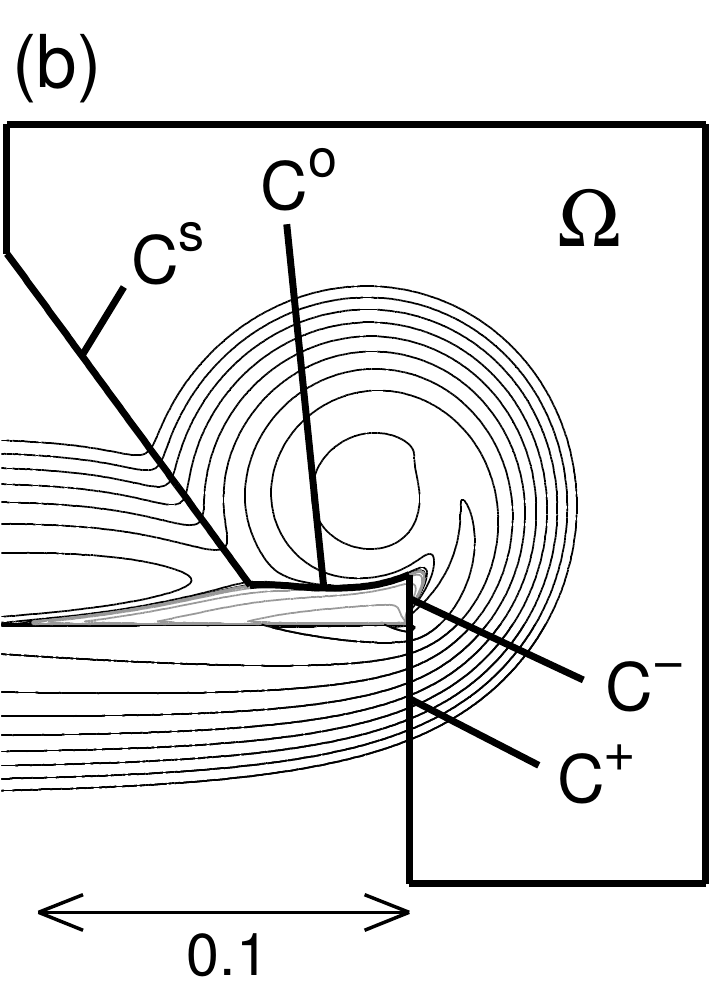}
\hskip10pt
\includegraphics[height=0.382\textwidth]{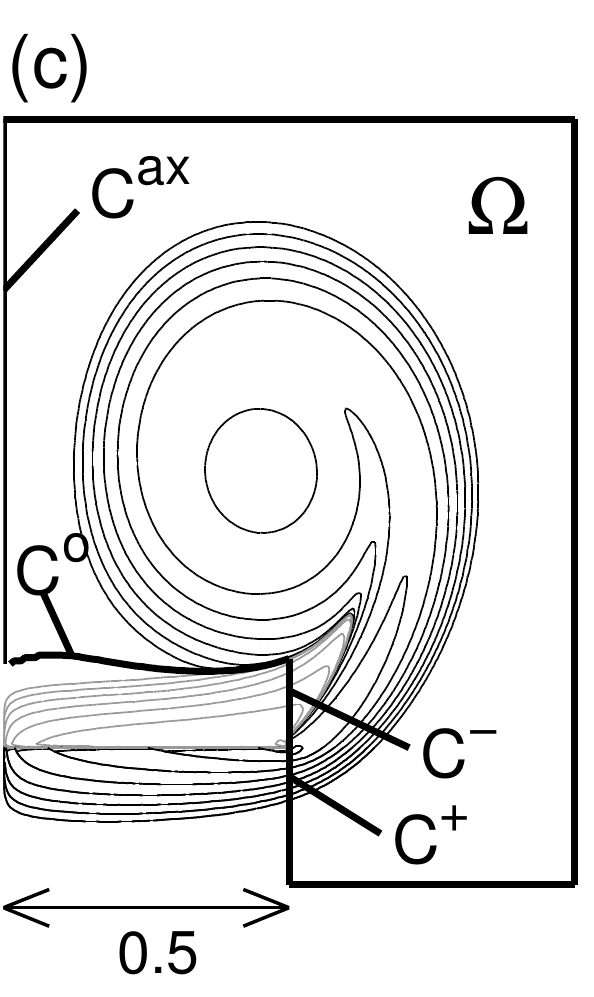}
 \caption{Sketch defining the domain $\Omega$ and portions 
$C^+,C^-,C^s,C^o,C^{ax}$ of its boundary with nonzero circulation 
flux.}
\label{F:gamdef}
\end{figure}
 
\subsubsection{Definitions}
One of our main interest in performing the present simulations was
to obtain circulation shedding rates for the viscous flow, for
which little data is available in the literature.
We are interested in resolving the shed circulation over
large time scales, including early times. However, at
early times the vorticity in the leading vortex is not
clearly distinguished from the boundary layer vorticity.
Here, we define the shed circulation to be
\begin{equation}
\Gamma=\int_{\Omega(t)} \omega \,dA~,
\end{equation}
where the region $\Omega(t)$ is defined in figure \ref{F:gamdef}.
The circulation is normalized by $UL$.

The definition of $\Omega$ is defined slightly differently in different
time-regimes of the flow, with continuous transitions between them.
At early times, 
when the region of negative vorticity has not yet been
entrained past the vertical line $x=0.5$, we follow the
sketch in 
figure \ref{F:gamdef}(a). 
On the upstream side ($y<0$), 
the region $\Omega$ is bounded by the vertical
line through the tip, $C^+$. On the downstream side ($y>0$), 
it is bounded by the
zero vorticity contour $C^o$ that separates negative from positive vorticity, 
and by a slant line $C^s$
through points of high curvature visible in the vorticity
contours. 
That is, the region $\Omega$ is defined to 
include all vorticity to the right of the tip, to 
exclude the negative boundary layer vorticity,
and is limited on the left by the slant line.
Figure \ref{F:gamdef}(b) shows the vorticity at intermediate times,
when the negative vorticity on the downstream side
has been entrained past the vertical line $x=0.5$.
Here we include all vorticity, positive or negative, to the
right of the vertical line through the tip, which introduces 
a vertical piece of boundary $C^-$ above the plate.
Figure \ref{F:gamdef}(c) shows the vorticity at later times,
when all the positive boundary
layer vorticity on the downstream wall has diffused and effectively
vanished. At this time the vortex is bounded on the left not
by the slant line, but by the axis $C^{ax}$.
Each subplot in figure \ref{F:gamdef} shows a typical length scale, indicating that
the three regimes in time span three decades of length scales
in space.

In order to determine the effect of viscous diffusion relative
to inviscid convection of vorticity into $\Omega$, and also
to determine the suitability of the definition above, 
we consider shedding rates of vorticity through the various
components of the boundary of $\Omega$. 
By applying the Transport Theorem, the Navier Stokes Equations,
and the Divergence Theorem, one finds that
\begin{equation}
\begin{split}
\frac{d\Gamma}{dt}
&=\frac{d}{dt}\int_{\Omega(t)}\omega(\xb,t)\,dA\\
&=
\int_{\Omega(t)}\frac{\partial \omega}{\partial t}\,dA 
+ 
\int_{\partial\Omega}\omega(\ub_{bd}\cdot\nb)\,ds\\
&=
\int_{\Omega(t)}\left[-(\ub\cdot\grad)\omega +{1\over Re}\lap\omega \right]\,dA 
+ 
\int_{\partial\Omega}\omega(\ub_{bd}\cdot\nb)\,ds\\
&=
\int_{\partial\Omega(t)}\left[-\omega\ub\cdot\nb +{1\over Re}\grad\omega\cdot \nb 
+ 
\omega(\ub_{bd}\cdot\nb)\right]\,ds
=
\frac{d\Gamma_c}{dt}
+\frac{d\Gamma_{d}}{dt}
+\frac{d\Gamma_m}{dt}
\end{split}
\end{equation}
where $\nb$ is the outward normal,
and $\ub_{bd}$ is the velocity of the boundary.
That is, across each piece of the boundary there is 
a contribution to the vorticity due to convection, diffusion, and 
the moving boundary.
We denote these components by subscripts $c$, $d$, and $m$ respectively.
Notice that the only moving boundary portions are $C^o$ and $C^s$, 
and the latter has no nonzero vorticity moving with it or convecting through it. 
Similarly, there is no convection of vorticity through $C^{ax}$.
Thus the nonzero contributions to the circulation shedding rate are
the 9 components
\begin{equation*}
\frac{d\Gamma^+_c}{dt}~,\quad
\frac{d\Gamma^-_c}{dt}~,\quad
\frac{d\Gamma^s_c}{dt}~,\quad
\frac{d\Gamma^+_{d}}{dt}~,\quad
\frac{d\Gamma^-_{d}}{dt}~,\quad
\frac{d\Gamma^o_{d}}{dt}~,\quad
\frac{d\Gamma^s_{d}}{dt}~,\quad
\frac{d\Gamma^{ax}_{d}}{dt}~,\quad
\frac{d\Gamma^s_{m}}{dt}
\end{equation*}
where the superscript refers to the portion of the boundary, and
the subscript refers to the component of the shedding rate.
For conciseness, we combine two of the viscous components 
into one: 
\begin{equation*}
\frac{d\Gamma^-_{d}}{dt} +\frac{d\Gamma^o_{d}}{dt}
\rightarrow \frac{d\Gamma^-_{d}}{dt}~.
\end{equation*}
Below, we first investigate the 8 circulation shedding rates
for $Re=500$, and then determine dependence on $Re$.

\subsubsection{$Re=500$}

\begin{figure}
 \centering
\hbox{ }\hskip9pt
\includegraphics[height=0.367\textwidth]{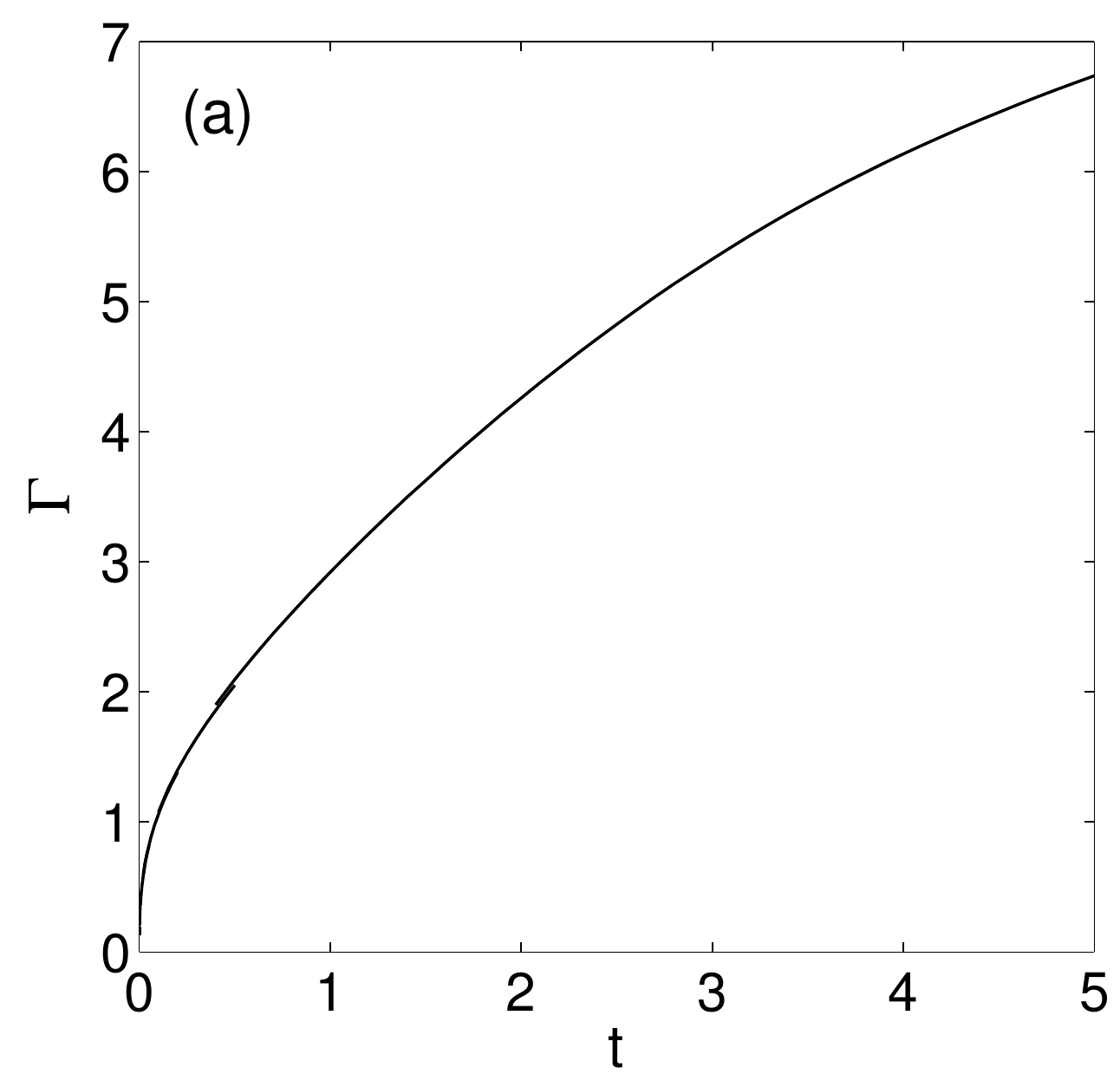}
\hskip5pt
\includegraphics[height=0.380\textwidth]{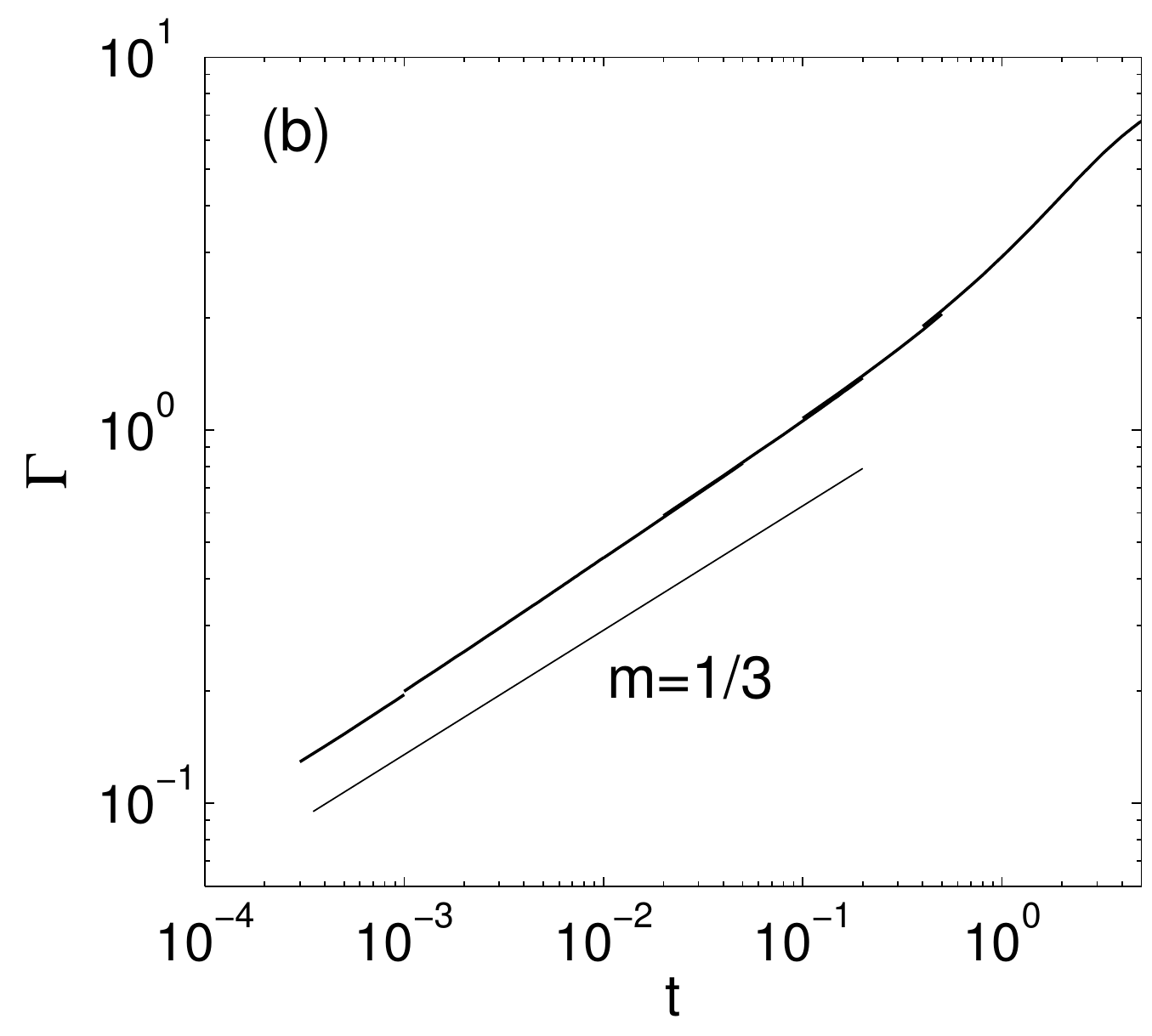}
\hbox{ }\hskip10pt
\includegraphics[height=0.362\textwidth]{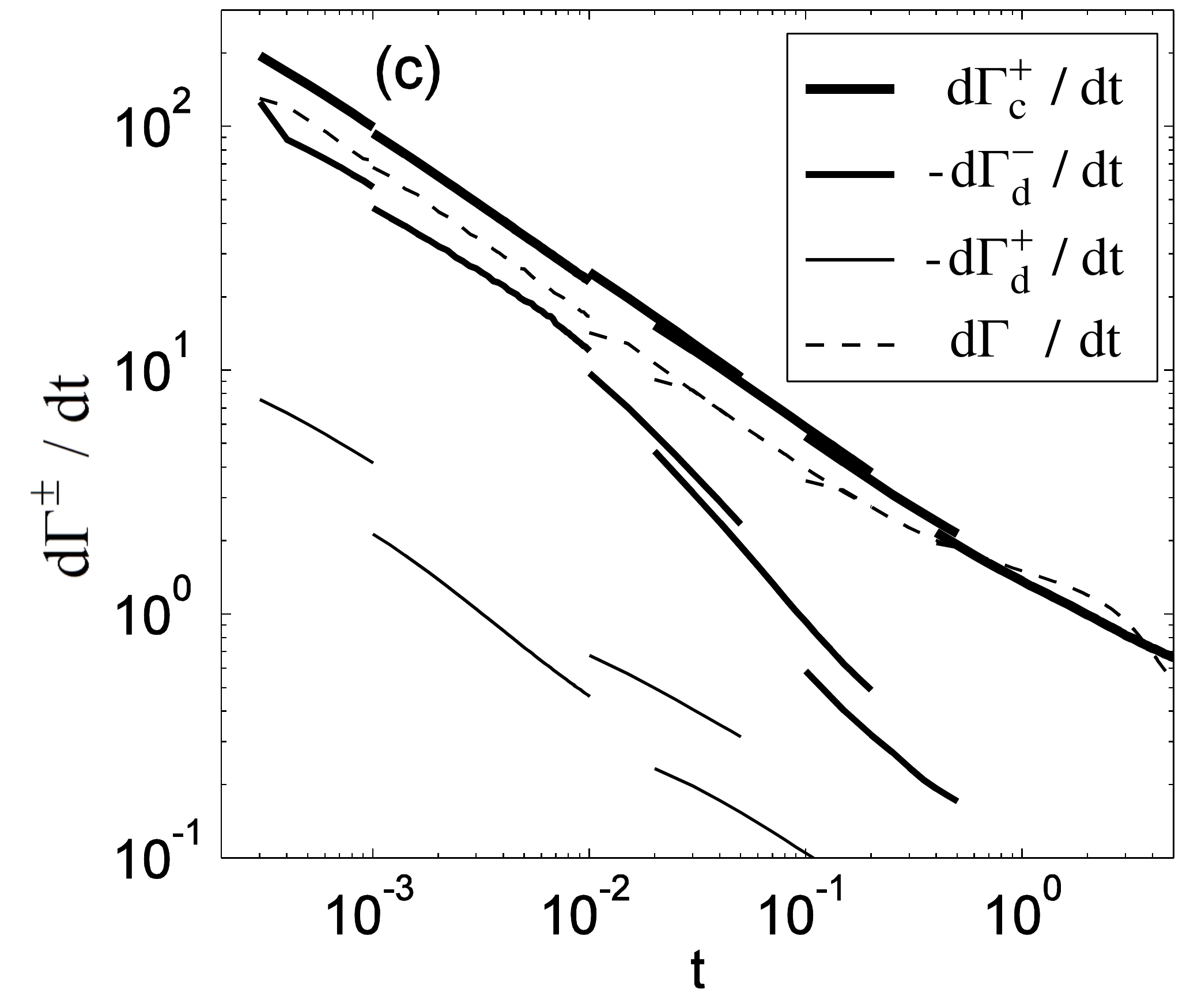}
\hskip18pt
\includegraphics[height=0.367\textwidth]{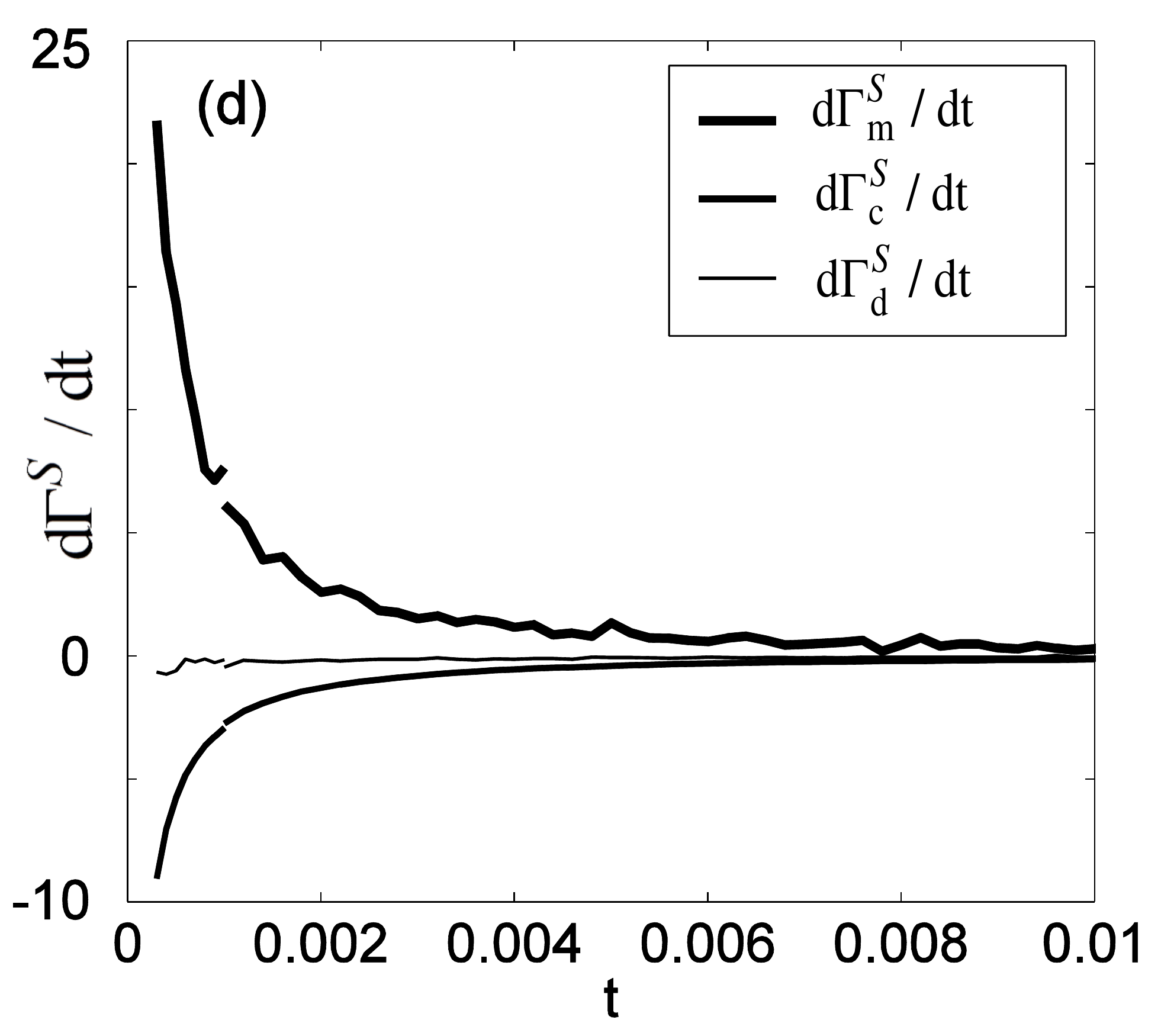}

\caption{
Circulation and shedding rates for $Re=500$.
(a) Circulation $\Gamma(t)$, linear scale.
(b) Circulation $\Gamma(t)$, logarithmic scale.
(c) Components 
$d\Gamma^{+}_c(t)/dt$,
$d\Gamma^{+}_{d}(t)/dt$,
$d\Gamma^{-}_{d}(t)/dt$,
$d\Gamma/dt$, on a logarithmic scale
(d) Components 
$d\Gamma^{s}_c(t)/dt$,
$d\Gamma^{s}_{d}(t)/dt$,
$d\Gamma^{s}_{m}(t)/dt$.}
\label{F:gam500}
\end{figure}

Figures \ref{F:gam500}(a,b) plot the shed circulation computed using
the above definition for Re=500. The logarithmic scale in figure 17(b)
shows that the circulation satisfies the scaling behaviour
predicted by inviscid similarity theory (Pullin 1978)
surprisingly well,
$\Gamma(t)\sim t^{1/3}~.$
Moreover, it shows that the circulation is
resolved over more than 4 decades in time.
Finally, we note that just as the data for $\omega_c$ and $y_c$,
the circulation data shows remarkable
independence of the meshsize used in the computation.
Even though the figure plots the results for all meshsizes
and time intervals given in table 2, 
the data is an almost continuous function of the meshsize.

Figures \ref{F:gam500}(c,d) show some of the circulation flux components. The
largest flux into the region is
the convective component through the
vertical $C^+$ on the upstream side of the plate,
$d\Gamma_c^+/dt$,
shown as the thickest curve in figure \ref{F:gam500}(c). It is
larger than $d\Gamma/dt$, shown as the dashed curve.
The viscous flux components are negative and reduce the
total circulation. Of these, largest in magnitude is the
viscous flux through $C^-\cup C^o$, $d\Gamma_{d}^-/dt$,
shown as the curve of medium thickness.
We conclude that $d\Gamma_c^+/dt$ is most significant,
but the contribution to the total flux due
to viscous diffusion is nonnegligible.

Figure \ref{F:gam500}(d) shows the three flux components through the slant line.
To note is, first, that these are much smaller than
the the largest components shown in figure \ref{F:gam500}(c) and do not contribute
significantly to the circulation, and second,
that they vanish quickly
and are negligibly small after about $t=0.005$.
Thus, not much vorticity leaves or enters through the slant line,
making it a reasonable left boundary to the leading vortex.

\subsubsection{Dependence on $Re$}

\begin{figure}
 \centering
\includegraphics[width=0.44\textwidth]{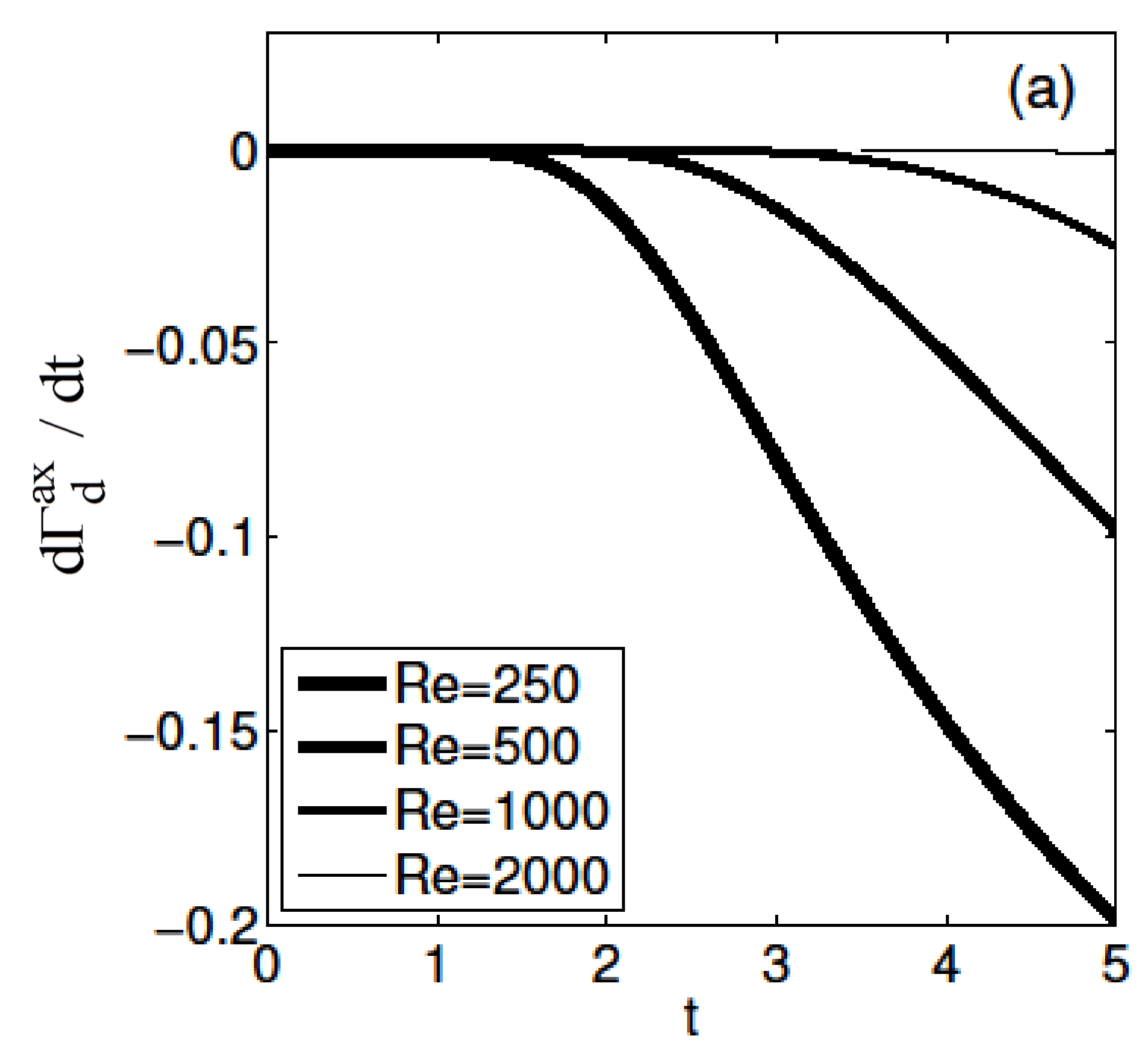}
\includegraphics[width=0.42\textwidth]{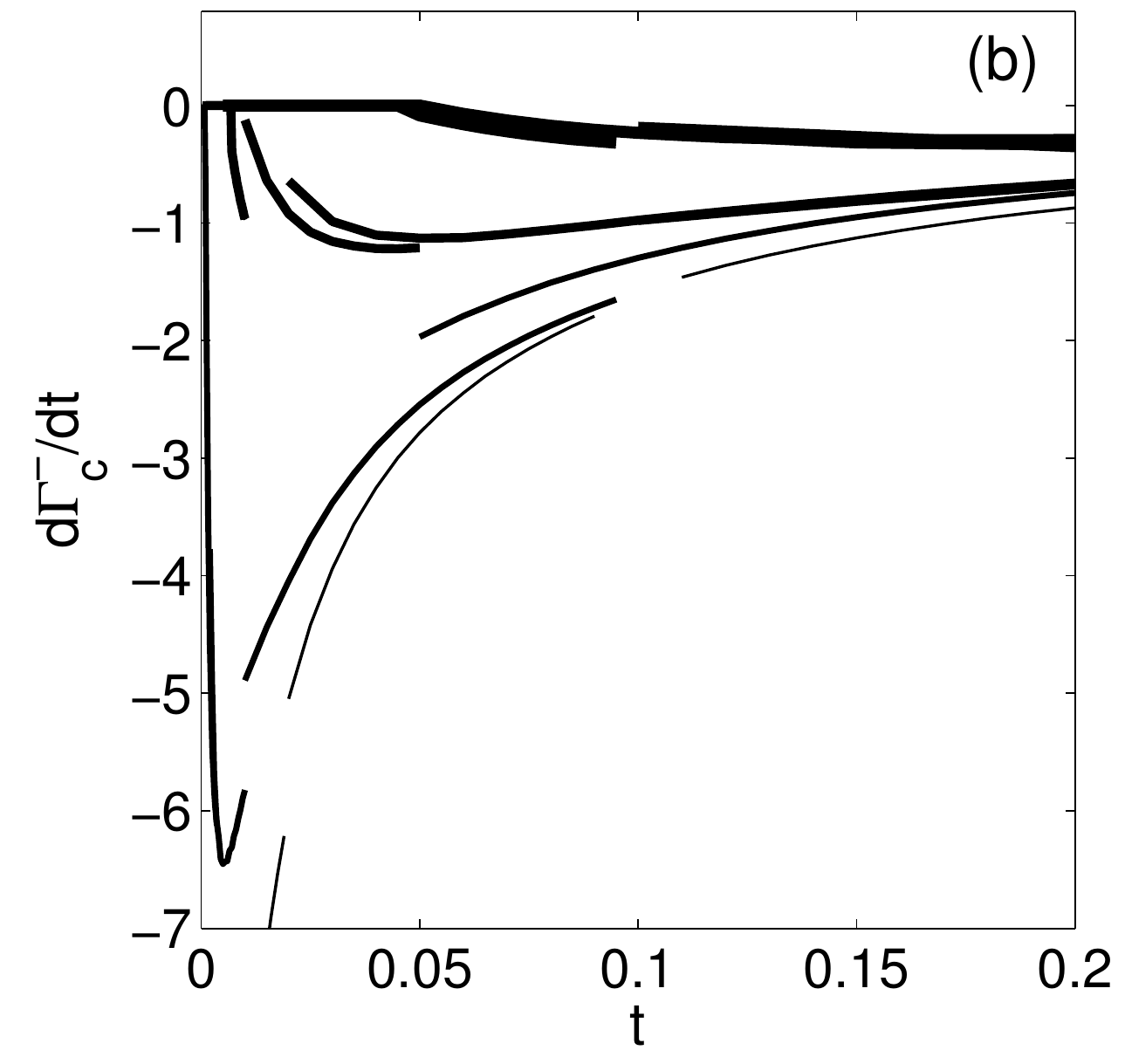}

\includegraphics[height=0.42\textwidth]{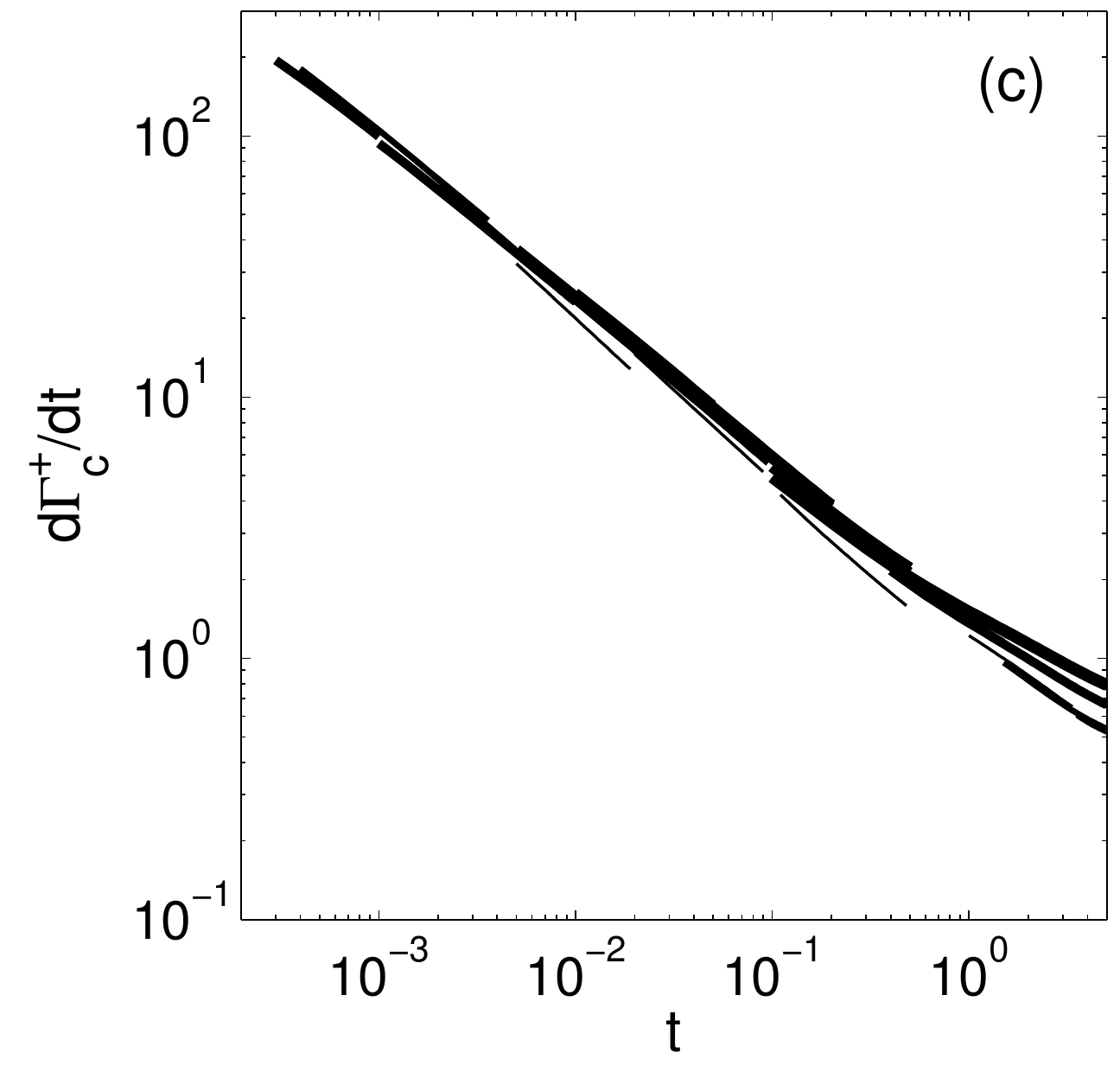}
\includegraphics[height=0.42\textwidth]{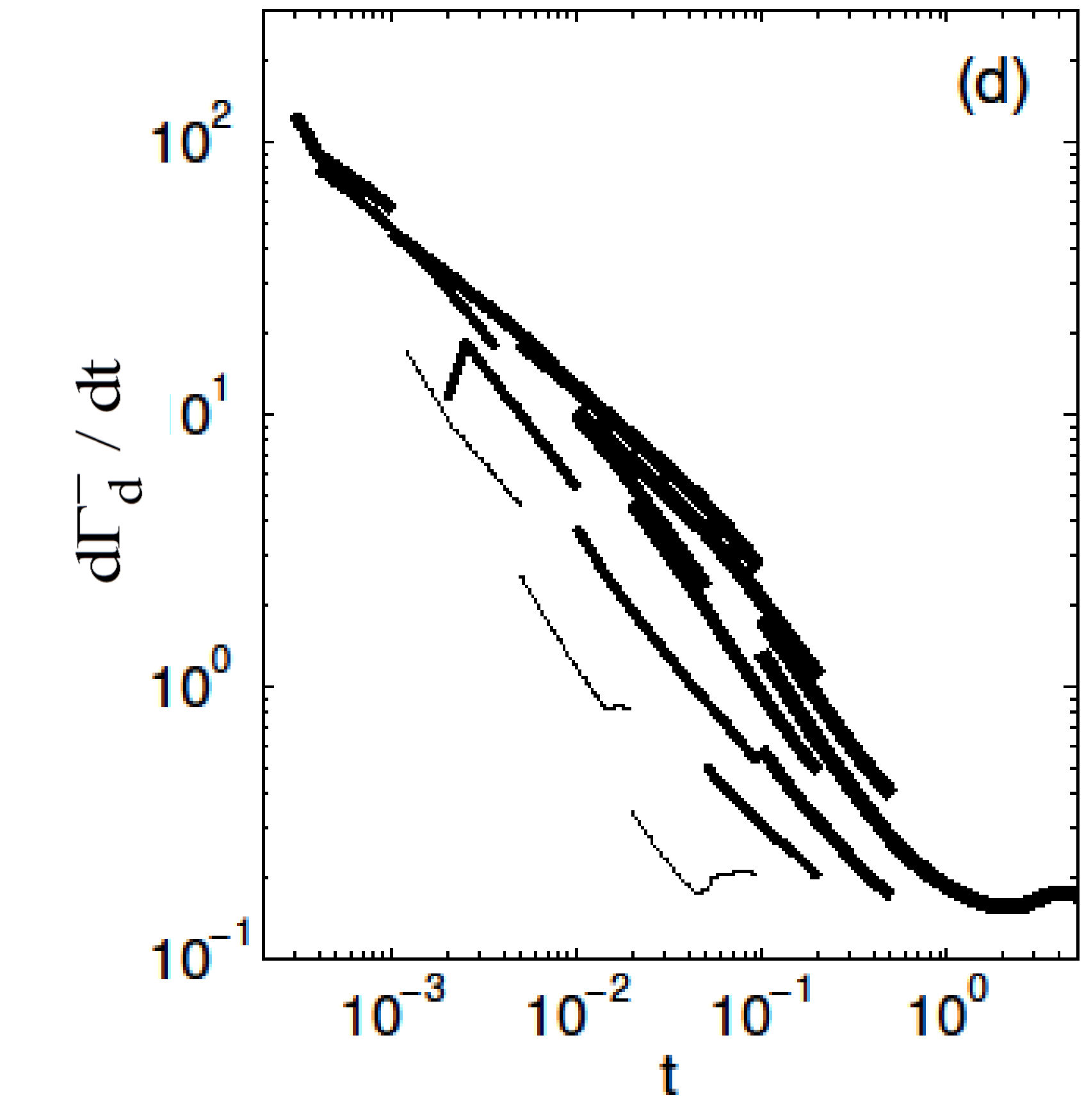}
 \caption{Dependence of circulation shedding rate components 
on Re, for $Re=250,500,1000,2000$.
(a) $d\Gamma_{d}^{ax}/dt$
(b) $d\Gamma_c^-/dt$, 
(c) $d\Gamma_c^+/dt$, 
(d) $d\Gamma^-_{d}/dt$, \vs $t$.}
\label{F:dgamre}
\end{figure}

In order to determine the dependence of the shed circulation on $Re$,
figure \ref{F:dgamre} plots several circulation flux components for 
$Re=250,500,1000,2000$.
Figure \ref{F:dgamre}(a) shows that the viscous loss
of vorticity through the axis depends clearly on $Re$, but
is comparatively very small. Similarly, the convective component
through the vertical $C^-$ on the downstream side of the plate, 
shown in figure \ref{F:dgamre}(b),
depends on $Re$ but is small.
The largest component, the convective component through
the vertical $C^+$ on the upstream side of the plate, shown
in figure \ref{F:dgamre}(c), is completely independent of $Re$.
The next largest component, the viscous component of vorticity diffusion
through
$C^-\cup C^o$, shown in figure \ref{F:dgamre}(d), appears to be quite
independent of $Re$ at early times, based on the results for
$Re=250, 500, 1000$. For $Re=2000$, the results are not fully resolved,
due to the difficulty in resolving
the large vorticity gradients present near the tip of the plate.

We conclude from figure \ref{F:dgamre} that the flux is essentially independent
of $Re$. The convective flux is clearly so,
and the dominant diffusive flux, even though it is large, 
is largely independent on $Re$.
We attribute the latter to the fact that as $Re$ increases, vorticity
gradients increase, but are offset by the $1/Re$ factor in
the diffusive term of equation (4.8),
\begin{equation}
{1\over Re}\int \grad\omega\cdot\nb\,ds\,.
\end{equation}

Figure \ref{F:gamre} plots the circulation $\Gamma(t)$ for $Re=250,500,1000,2000$.
Consistent with the conclusions based on figure \ref{F:dgamre}, the circulation
is basically independent on $Re$ at early times. As time increases,
differences between $Re$
increase slightly. The largest difference
over the range of $Re$ considered here occurs
at the last time computed, $t=5$, and is less than 5\% of the circulation at that time.
The logarithmic plot in figure \ref{F:gamre}(b) shows that all 20 time series computed 
with different values of $Re$ and meshsizes collapse
onto one curve, which is well approximated by
\begin{equation}
\Gamma(t)\approx 2 t^{1/3} ~
\text{ as }~ t\to0~.
\end{equation}

\begin{figure}
 \centering
\includegraphics[width=0.426\textwidth]{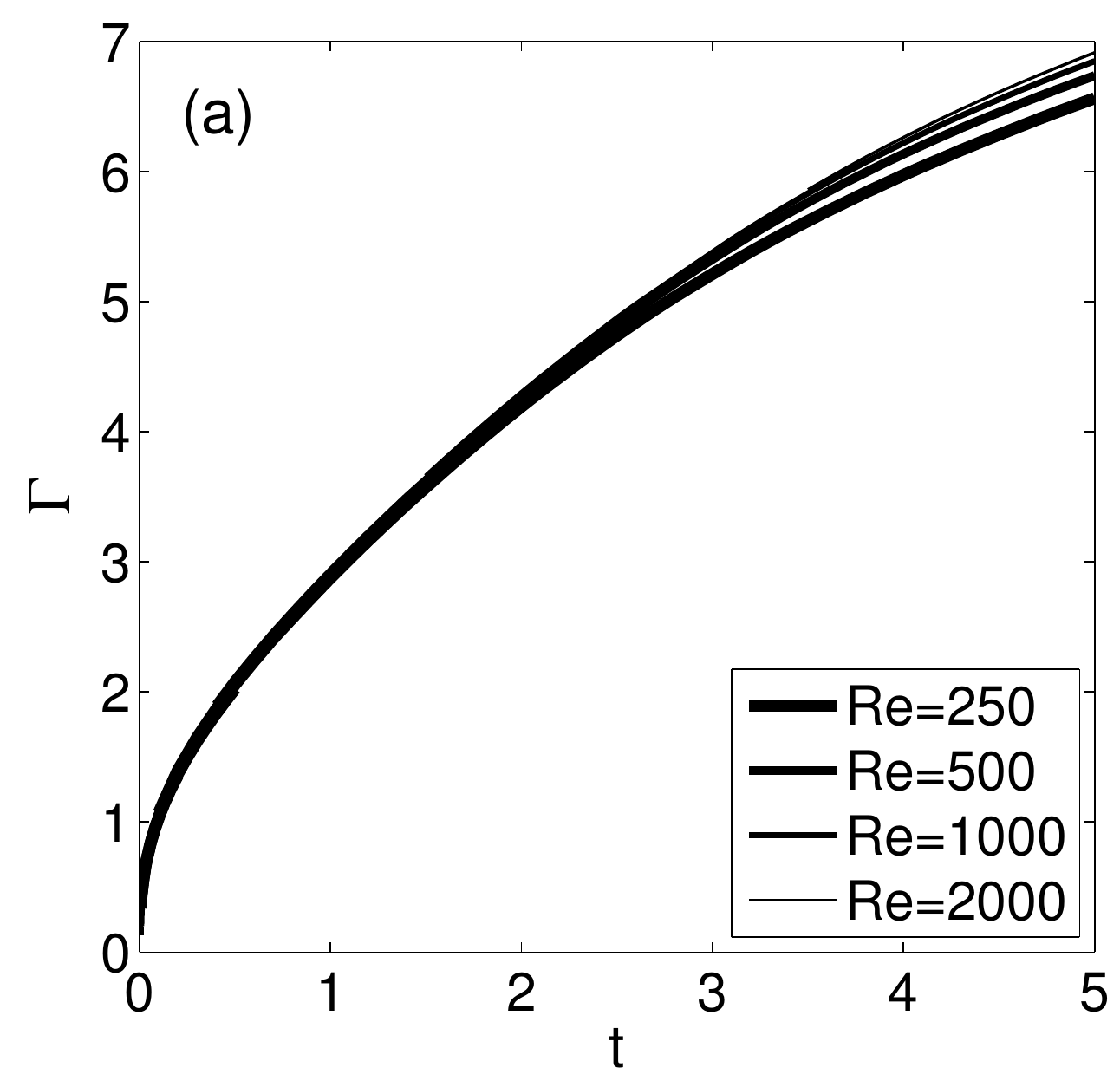}
\includegraphics[width=0.44\textwidth]{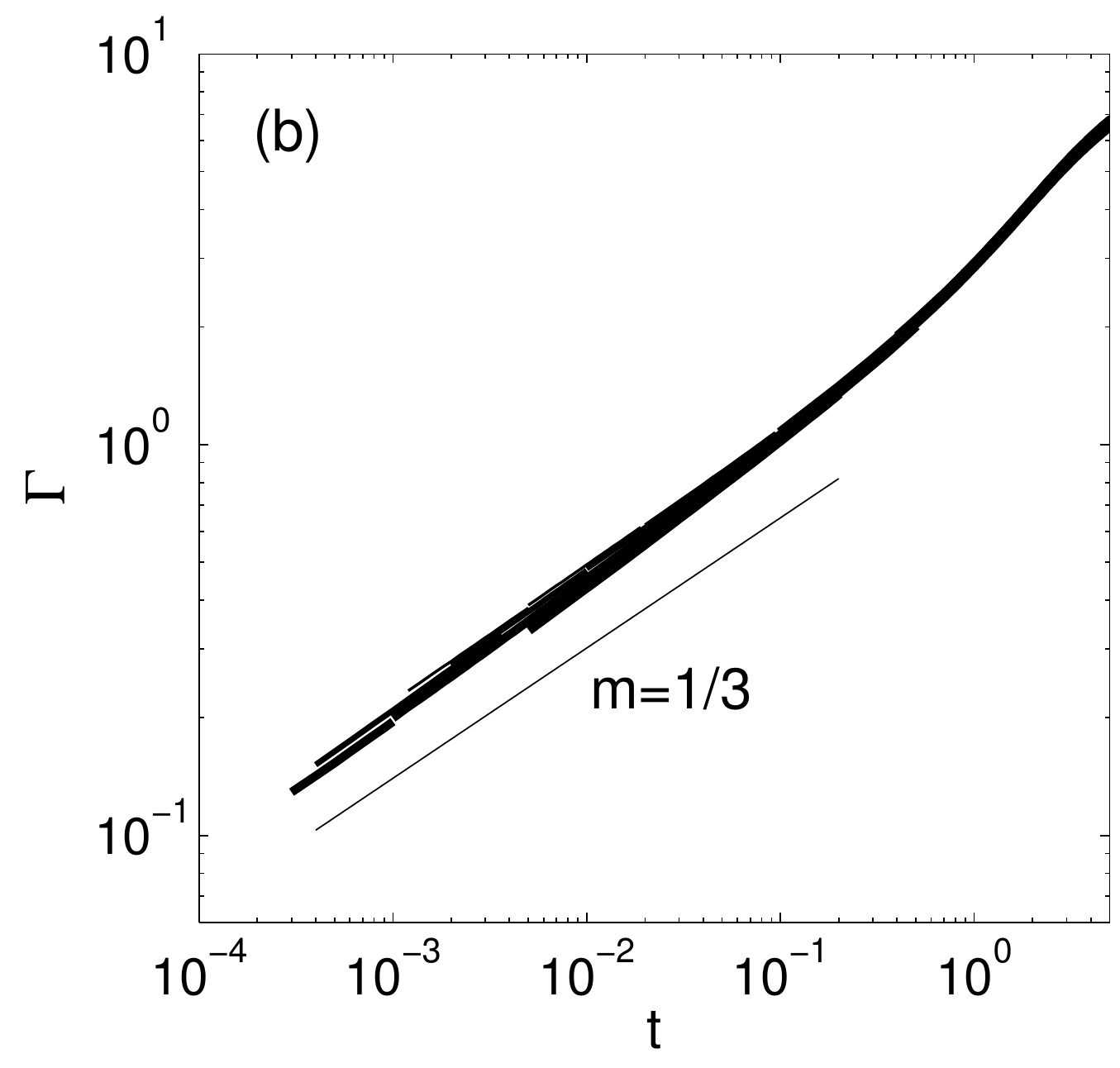}
 \caption{Dependence of total shed circulation $\Gamma(t)$ on $Re$, 
for $Re=250,500,1000,2000$, as indicated. (a) linear scale,
(b) logarithmic scale.}
\label{F:gamre}
\end{figure}

\subsection{Vortex forces}

The vorticity profiles on the plate wall induce drag and lift
forces. Here we compute the drag and lift, $F_D$ and $F_L$,
defined to be total force parallel and normal to the background flow, in
the right half of the plate only. They are given by (Eldredge 2007)  
\begin{subequations}
\begin{equation}
F_D(t)={2\over Re} 
\int_0^{1/2} \left[\omega_+(x,0,t)+\omega_-(x,0,t)\right]\,dx~,
\end{equation}
\begin{equation}
F_L(t)={2\over Re} 
\int_0^{1/2} x\left[{\partial\omega_+\over\partial y}(x,0,t)
+{\partial\omega_-\over\partial y}(x,0,t)\right]dx~,
\end{equation}
\label{E:forces}
\end{subequations}
where all forces are normalized by ${1\over2}\rho U^2L$.
By symmetry, the overall lift force acting on the whole plate is zero,
and the total drag is $2F_D$.

\begin{figure}
 \centering
\includegraphics[height=0.37\textwidth]{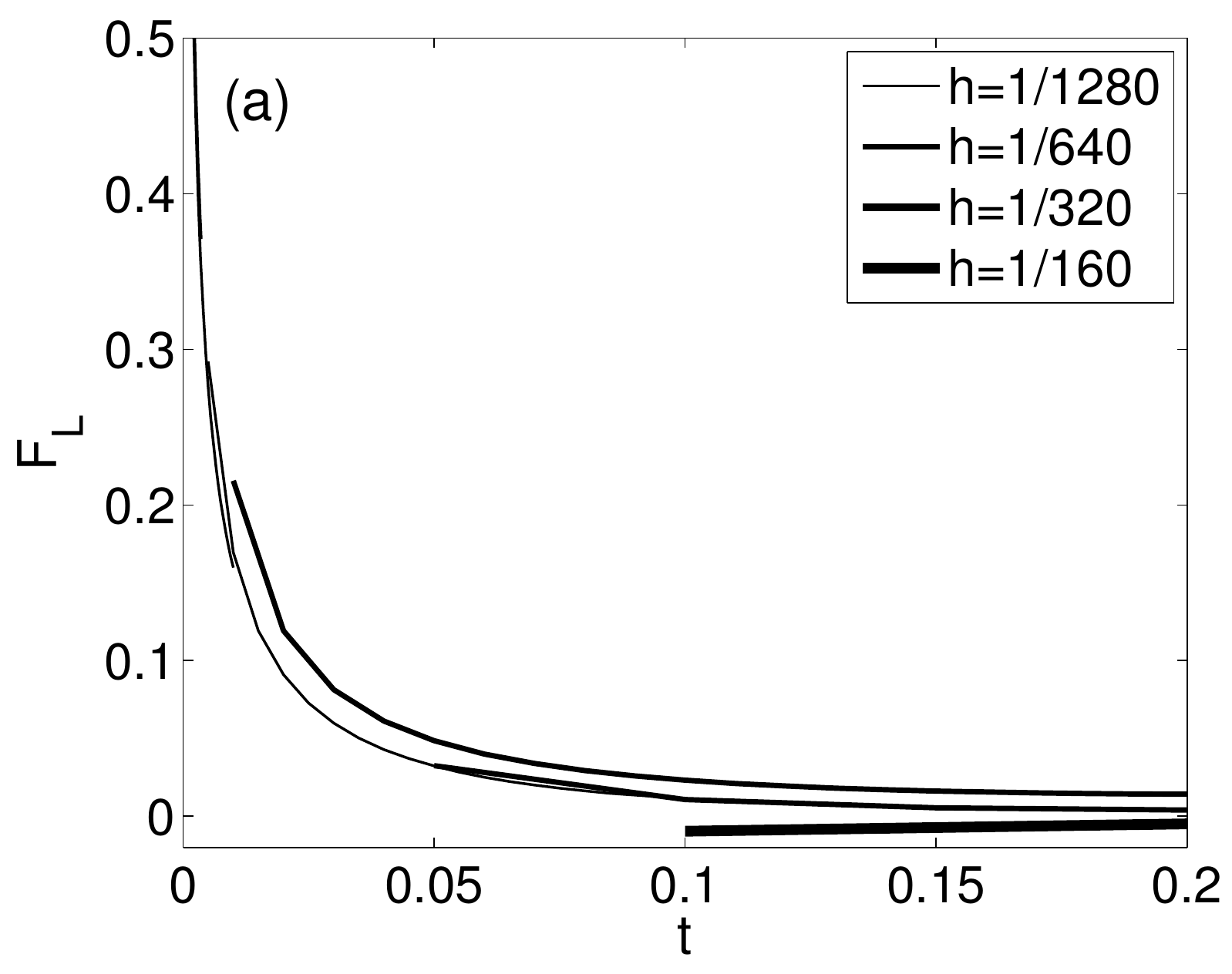}
\includegraphics[height=0.37\textwidth]{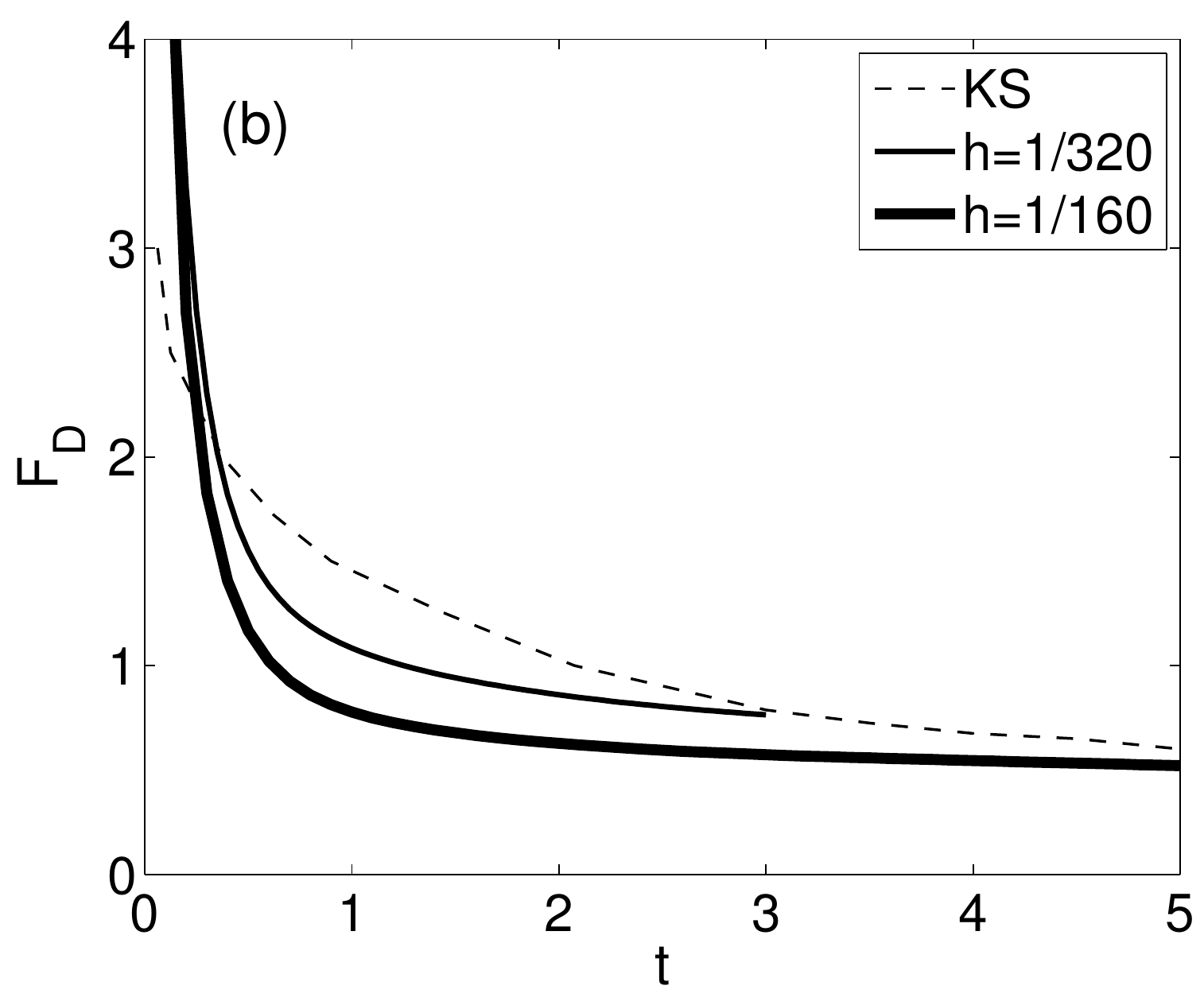}
\caption{
Forces (a) $F_L$ and (b) $F_D$ for $Re=1000$,
computed with the indicated values of $h$.
For comparison, (b) includes 
the data by Koumoutsakos \& Shiels (1996, denoted by KS).  }
\label{F:drag1000}
\end{figure}

Computing the forces 
using formulation (\ref{E:forces})
is sensitive to discretization errors,
since the formulation depends on values 
of the vorticity and its derivatives on the plate wall. 
These values are large and difficult to compute accurately,
specially near the tip.
To illustrate, figure \ref{F:drag1000}
plots the computed lift and drag for $Re=1000$ for various 
values of $h$ used and shows that the convergence in $h$ is slow.
An alternative formulation for the drag force is used
by \cite{koushiels96} (KS), who compute drag as the time
derivative of an area integral, see their equations (40-41).
The value of $F_D$ given by (\ref{E:forces}b) corresponds to
the variable $c_D$ plotted in KS in their figure 13. 
Figure \ref{F:drag1000}(b) in this paper compares 
the drag computed here 
with the values computed by KS, for $Re=1000$.
The figure shows that the values are in fairly good agreement, 
although differences exist, mainly in the decay rate at early times. 

Figures \ref{F:force}(a,b) plot the lift and drag force on a logarithmic
scale, for all Reynolds numbers computed here, and for all values of $h$ used.
They show that the results at early times collapse quite well 
onto a common curve. 
Figure \ref{F:force}(a) plots the lift force versus a scaled time $Re\cdot t$.
In these variables, the data at early times collapses onto a curve
that decreases in time and in $Re$ approximately as 
\begin{equation}
F_L(t)\approx (Re\cdot t)^{-{1/2}}~.
\end{equation}

The drag, on the other hand, plotted in figure \ref{F:force}(b) 
versus time $t$, appears to be almost independent of $Re$,
and is approximately given by 
\begin{equation}
F_D(t)\approx 6 t^{-1/2}~.
\end{equation}
These results suggest that at a fixed time, the lift
decays significantly faster, as $1/Re^{1/2}$, than the drag, 
which remains almost constant in $Re$ for early times.
In view of equation (\ref{E:forces}), this in turn indicates that the wall vorticity grows as
$Re^{1/2}$, while the wall vorticity gradients grow faster, almost
linearly in $Re$.
At later times the drag force decreases as $Re$ increases, consistent
with the results shown by Dennis et al (1993) and by KS.

\begin{figure}
 \centering
\includegraphics[height=0.39\textwidth]{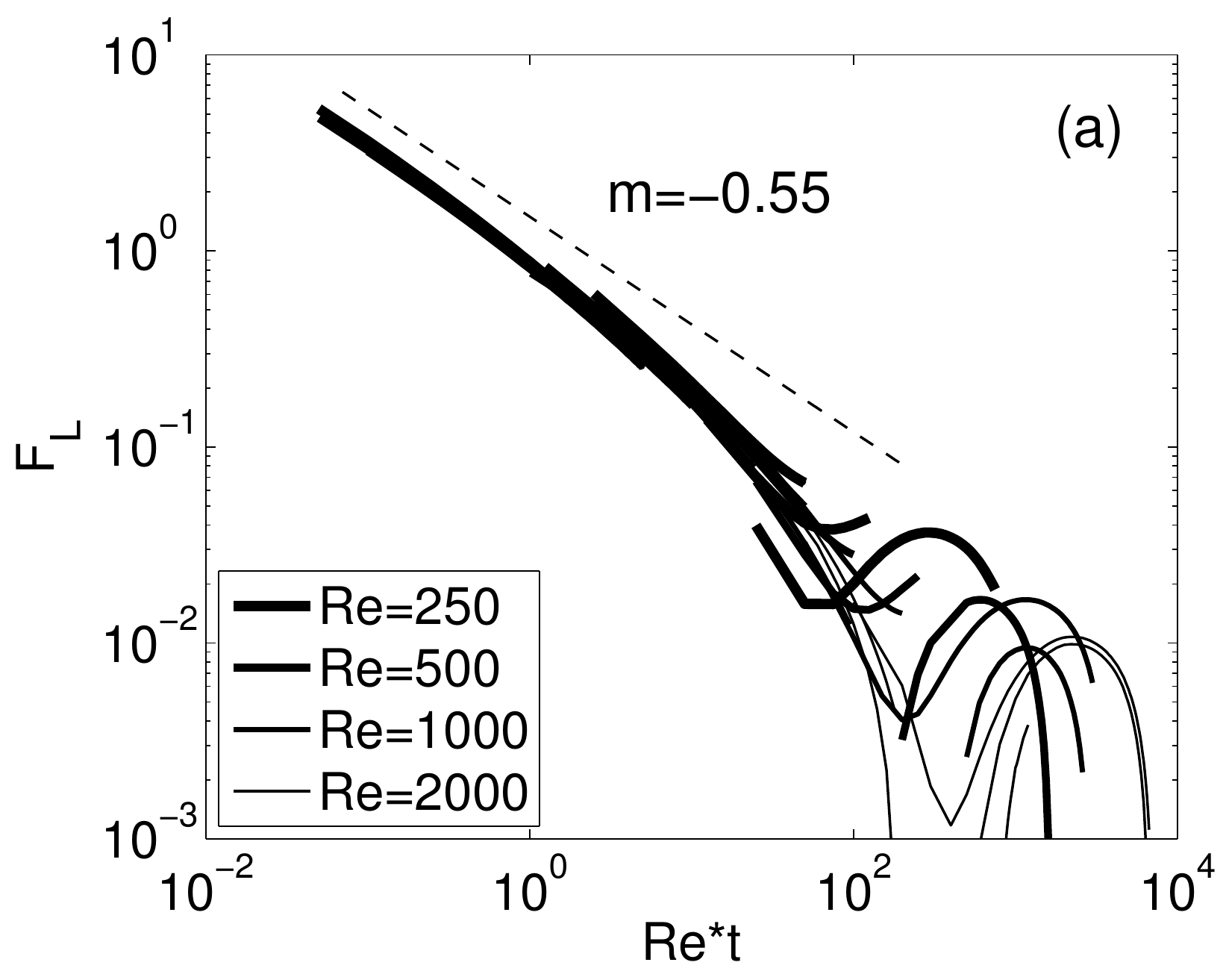}
\includegraphics[height=0.39\textwidth]{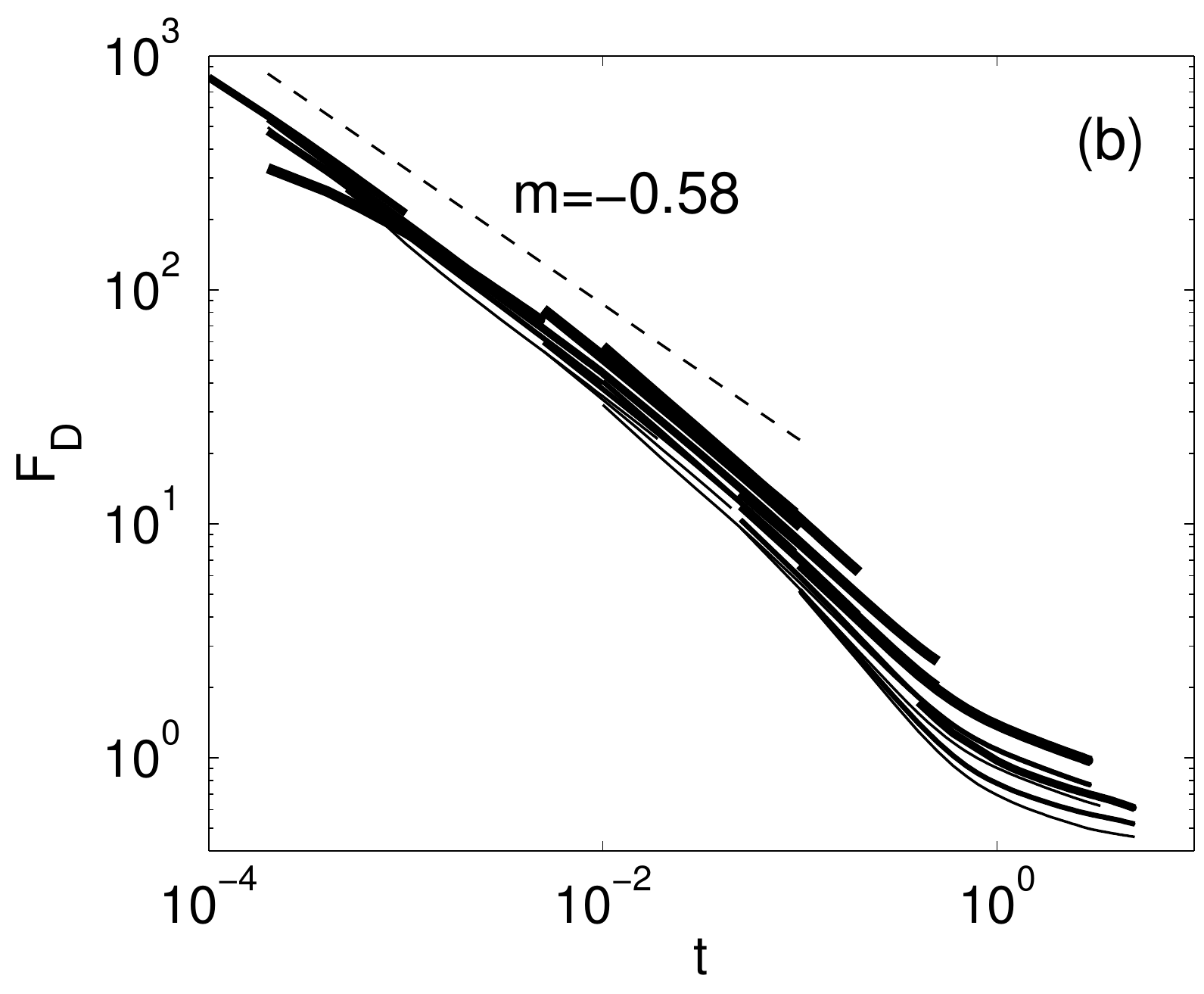}
\caption{
(a) Lift force \vs $Re\cdot t$.
(b) Drag force \vs $t$.
Results are shown for $Re=250,500,1000,2000$, as indicated 
in the legend in figure (a).
The dashed lines have the indicated slopes.
}
\label{F:force}
\end{figure}

\section{Summary}
\label{sec:conc}
Viscous flow past a flat plate of zero thickness which
is impulsively started in direction normal to itself is studied using
highly resolved numerical simulations for a range of Reynolds numbers $Re\in[250,2000]$.
Several features of the flow are revealed by the computations.

The evolution of the vorticity profiles shows the growth of 
the leading vortex emanating from the boundary layer
during the early starting flow, the growth of the region 
of opposite-signed vorticity on the plate, and the evolution 
of the plate vorticity and vorticity gradients.
Increasing Reynolds numbers leads to 
thinning of boundary layer and shear layer thicknesses,
and to increasingly many and tighter spiral streakline 
turns in the vortex center. 

Scaling behaviours that capture the dependence on 
time and on $Re$ were found for 
several quantities in the flow,
over several decades in time.
Some quantities clearly depend strongly on $Re$, such as
the core vorticity, the boundary layer thickness, and 
the lift force over the half-plate, with
\begin{subequations}
\begin{equation}
\omega_c(t)\approx 0.335 \Big({\tiny Re\over t}\Big)^{3/4}~,\quad 
\delta(t)\sim \Big({\tiny t\over Re}\Big)^{1/2}~,\quad
F_L(t)\approx (Re\cdot t)^{-1/2} ~,
\end{equation}
all as $t\to 0$.
Other quantities of the early starting flow 
are largely independent on $Re$, such as
the core trajectory, the shed circulation, 
the integral negative vorticity, and the drag force over the half-plate,
with 
\begin{eqnarray}
y_c(t)&\approx& 0.37 t^{2/3}~,\quad
x_c(t)-0.5\approx -0.1t^{2/3}~,\quad\\
\Gamma(t)&\approx& 2t^{1/3}~, \qquad
F_D(t) \sim t^{-1/2}~.
\end{eqnarray}
\label{E:scalsumm}
\end{subequations}
The scaling for $x_c$, $y_c$ and $\Gamma$ are in excellent agreement
with the inviscid scaling laws for self-similar roll-up.

One of the main contributions is to define
and compute the viscous shed circulation, specially at
early times, when the shed vorticity is not clearly
separated from the boundary layer vorticity.
Our definition is validated by plots of the circulation 
shedding rates across various
portions of the boundary defining the shed vorticity, 
which show that practically no vorticity enters
or leaves the boundary by convection except near the tip,
where it is convected into the vortex by the separating
boundary layers.
With this definition, the shed circulation satisfies the 
self-similar scaling laws for more than three decades in time.
%

The presentation of several components of the circulation shedding 
rate gives insight into the effect of viscosity on the total circulation.
The largest component 
is the gain of circulation due to convection of 
vorticity from the upstream boundary layer into the leading vortex. 
This component is highly independent of the Reynolds number.
It is offset by loss of circulation due to 
viscous diffusion of vorticity out 
of the boundary, which is of opposite sign but also 
significant in magnitude.
Interestingly, this diffusive component also 
depends little on $Re$ at early times,
consistent with the fact that the overall shed circulation
is basically independent of $Re$ at these times.
This observation suggests that that as $Re$ increases, vorticity
gradients responsible for viscous diffusion grow in such a way
that the quotient $\nabla\omega/Re$ changes little. 
We conclude that the effect of viscous diffusion on the overall shed circulation
is significant, but its contribution depends little on the value of $Re$.
%
%
%
%


\section*{Appendix}
In this appendix we compare the scaling laws observed in this paper using two alternate
nondimensionalizations. For clarity, in this appendix only, let all variables
without a hat or double hat, such as $x,t,\Gamma,\omega, F_D$, denote 
the original {\it dimensional} variables.
Let all variables with one hat 
denote variables nondimensionalized by length and time scales
$L$ and $L/U$. For example, 
$$
\xhat={x\over L}~,\quad \that={Ut\over L}~,\quad\gamhat={\Gamma\over LU}~,\quad 
\what={L\omega\over U}~,\quad
\fhat={\FD\over {1\over2} LU^2}~.
\eqno(A1)$$
These are the nondimensional variables used throughout this paper, 
where for simplicity, the hats were dropped.
Let all variables with two hats
denote variables nondimensionalized by length and time scales
$\nu/U$ and $\nu/U^2$. For example, 
$$
\xhhat={Ux\over \nu}~,\quad \thhat={U^2t\over \nu}~,
\quad\gamhhat={\Gamma\over \nu}~,\quad 
\whhat={\nu\omega\over U^2}~,\quad
\fhhat={F_D\over {1\over2} \nu U}~.
\eqno(A2)
$$

It follows that 
$$
      \xhhat   =Re\cdot\xhat ~,
\quad \thhat   =Re\cdot\that ~,
\quad \gamhhat =Re\cdot\gamhat~,
\quad \whhat   ={\what\over Re} ~,
\quad \fhhat   =Re\cdot\fhat ~.
\eqno(A3)
$$
Therefore, the observed scaling laws, given in equations (\ref{E:scalsumm})
in the single-hat variables, are given in the double-hat variables
as 
$$\whhat_c\approx\, 0.335\, {Re^{1/2}~\thhat{~ }^{-{3/4}}}~,\eqno(A.3a)$$
$$\delhhat\sim \thhat{ }^{~{1/2}}~,\eqno(A.3b)$$
$$\flhhat\approx\, Re\cdot \thhat{ }^{~-{1/2}} ~,\eqno(A.3c)$$
$$\yhhat_c\approx\, 0.37\, Re^{1/3}~\thhat{ }^{~{2/3}}~,\eqno(A.3d)$$
$$\xchhat\approx\, -0.1\,Re^{1/3}~\thhat{ }^{~{2/3}}~,\eqno(A.3e)$$
$$\gamhhat\approx \,2\,Re^{2/3}~\thhat{ }^{~{1/3}}~,\eqno(A.3f)$$ 
$$\fhhat \sim (Re/\,\thhat ~)^{{1/2}}~,\eqno(A.3g)$$
all as $\thhat\to 0$.
Notice that of all these, one variable, namely the boundary layer thickness
$\delhhat$, scales as a function of $\thhat$ independent of the value of $Re$. One
may therefore suggest that the double-hat nondimensionalization is more
natural for this variable. Using the same argument one may say that
the single-hat nondimensionalization is more natural to describe the vortex
coordinates and circulation, as in equations (5.1bc).
The essence is that the variables depend on $L,U,\nu$ as described by
either equations (A.3) or (5.1).

\end{document}